\documentclass[12pt]{article}

\usepackage{amsmath}
\usepackage{amsfonts}
\usepackage{amsthm}
\usepackage{latexsym}
\usepackage{mathrsfs}

\allowdisplaybreaks[1] 

\newcommand{\amsatop}[2]{\genfrac{}{}{0pt}{}{#1}{#2}}
\newcommand{\twovect}[2]{\left( \amsatop{#1}{#2} \right)}
\newcommand{\twomat}[1]{\left( \begin{array}{cc} #1 \end{array} \right)}

\newcommand{\mytable}[3]{\begin{table}
\rule{\textwidth}{1pt}\\
#1
\vspace*{-5ex}\rule{0pt}{1pt}\\
\rule{\textwidth}{0.5pt}\\[-5ex]
\caption{#2}
\vspace*{-1.5ex}
\rule{\textwidth}{1pt}
\label{#3}
\end{table}
}

\def\A{{\mathscr A}}
\def\e{{\rm e}}

\def\l{\langle}
\def\r{\rangle}
\def\da{{\dot\alpha}}
\def\be{\beta}

\renewcommand{\S}{{\mathscr S}}
\renewcommand{\O}{{\cal O}}

\newcommand{\M}{{\mathscr M}}
\newcommand{\G}{{\mathscr G}}
\newcommand{\W}{{\mathscr W}}
\newcommand{\w}{{\mathscr W}_{\rm local}}
\newcommand{\D}{{\mathscr D}}
\newcommand{\poly}{{\mathscr P}}
\newcommand{\F}{{\mathscr F}}
\newcommand{\dimIR}{{\rm dim}_{\rm IR}}
\newcommand{\dimUV}{{\rm dim}_{\rm UV}}
\newcommand{\au}{A_1}
\newcommand{\aub}{\bar{A}_1}
\newcommand{\av}{A_2}
\newcommand{\avb}{\bar{A}_2}
\newcommand{\f}{f}

\newcommand{\vecv}{\vec v}

\newcommand{\Geff}{\Gamma_{\rm eff}}
\newcommand{\Gcl}{{\Gamma_{\rm cl}}}
\newcommand{\Ggf}{\Gamma_{\rm g.f.}}
\newcommand{\Gext}{\Gamma_{\rm ext}}
\newcommand{\Ginv}{\Gamma_{\rm ct,inv}}
\newcommand{\Ginva}{\Gamma_{\rm ct,inv,1a}}
\newcommand{\Ginvb}{\Gamma_{\rm ct,inv,1b}}
\newcommand{\Ginvc}{\Gamma_{\rm ct,inv,finite}}
\newcommand{\Grest}{\Gamma_{\rm ct,restore}}
\newcommand{\dS}{\!\!{\rm d}^6z\,}
\newcommand{\dV}{\!\!{\rm d}^8z\,}
\newcommand{\dx}{\!\!{\rm d}^4x\,}
\newcommand{\al}{\alpha}
\newcommand{\eps}{\varepsilon}
\newcommand{\mbv}{\mathbf{v}}
\renewcommand{\i}{{\rm i}}

\newcommand{\T}{{\rm T}}
\newcommand{\s}{\mbox{\bf s}\,}
\newcommand{\B}{\S_\Gcl}
\newcommand{\tr}{{\rm tr }}

\newcommand{\tfr}[2]{{\textstyle \frac{#1}{#2}}}
\newcommand{\fdq}[2]{\frac{\delta #1}{\delta #2}}
\newcommand{\pdq}[2]{\frac{\partial #1}{\partial #2}}

\newcommand{\wl}{w^{\lambda'}}

\newcommand{\lambdabar}{{\bar\lambda}}

\newcommand{\alphadot}{{\dot\alpha}}

\newcommand{\cbar}{\bar{c}}
\newcommand{\Z}{R}
\newcommand{\ZW}{z_V}
\newcommand{\ZV}{Z_V}

\newcommand{\Zc}{{Z_c}}
\newcommand{\Zcbar}{{Z_{\bar{c}}}}

\newcommand{\Rsu}{\Z_{\tilde{u}}}
\newcommand{\Rsd}{\Z_{\tilde{d}}}
\newcommand{\Rse}{\Z_{\tilde{e}}}

\newcommand{\VL}{\left( \begin{array}{c}}
\newcommand{\VR}{\end{array} \right)}
\newcommand{\ML}{\left( \begin{array}{cc}}
\newcommand{\MLv}{\left( \begin{array}{cccc}}
\newcommand{\MR}{\end{array} \right)}
\newcommand{\deltan}{\delta^{(n)}}

\def\dfrac#1#2{\frac{\delta{#1}}{\delta{#2}}}

\def\pslash#1{{\setbox0=\hbox{$#1$}
  \rlap{\ifdim\wd0>.7em\kern.22\wd0\else\kern.1\wd0\fi /}#1}}
\def\psl{\pslash p}

\def\GG#1{{\Gamma_{#1}}}

\begin{document}

\thispagestyle{empty}
\vspace*{-5mm}
\begin{flushright}
BN-TH-2002-03\\
DESY-02-054\\
LU-ITP 2002/006\\
KA-TP-9-2002\\
hep-ph/0204350
\end{flushright}

\vspace{0.5cm}
\begin{center}
{\Large \bf Renormalization of the Minimal Supersymmetric Standard
  Model%
\footnote{Supported in part by the European Community's Human Potential
  Programme under contract HPRN-CT-2000-00149 ``Physics at Colliders''.}
}

\vspace{0.5cm}

Wolfgang Hollik${}^{a,b}$,
Elisabeth Kraus${}^c$,
Markus Roth${}^a$, \\
Christian Rupp${}^d$, 
Klaus Sibold${}^e$,
Dominik St{\"o}ckinger${}^f$
\\[1cm]
${}^a$ Institut f{\"u}r Theoretische Physik, Universit{\"a}t Karlsruhe, 
Germany\\
${}^b$ Max-Planck-Institut f\"ur Physik, M\"unchen, Germany\\
${}^c$ Institut f{\"u}r Theoretische Physik, Universit{\"a}t Bonn, Germany\\
${}^d$ Institut f{\"u}r Theoretische Physik, Universit{\"a}t Bern, 
Switzerland\\
${}^e$ Institut f{\"u}r Theoretische Physik, Universit{\"a}t Leipzig, Germany\\
${}^f$  Deutsches Elektronen-Synchrotron DESY, Hamburg, Germany
\vspace{0.3cm}

\end{center}

\vspace{1cm}

\begin{center}
\parbox{12cm}{
\centerline{\small \bf Abstract}
\small \noindent 
The renormalization of the Minimal Supersymmetric Standard Model of 
electroweak interactions is presented to all orders of perturbation theory 
using the algebraic method. Special attention is directed to the issues 
of soft supersymmetry breaking, gauge fixing, and infrared finiteness. 
We discuss the implications of $\hbar$-dependent field parametrizations on 
the counterterm structure and provide a complete set of on-shell 
normalization conditions.}
\end{center}


\newpage

\section{Introduction}

\noindent
Independent of 
the success of the Standard Model (SM) \cite{SM}, manifested in more than a 
decade of precision tests \cite{SMtests}, 
the quest for a more comprehensive description
of the fundamental interactions has brought forward the concept of
supersymmetry as the favored extension of the SM.
The Minimal Supersymmetric Standard Model (MSSM)~\cite{MSSM} 
is the theoretically best motivated and conceptually most elaborated 
and predictive framework beyond the SM.
The possible existence of a light Higgs boson, in a mass range
consistent with the
electroweak precision analysis~\cite{SMtests} and with the direct
searches~\cite{Hanson:2001cq}, would find a natural explanation 
within the MSSM. Precision analyses of the MSSM~\cite{MSSMtests}, 
performed in analogy to those of the SM, 
show a similar quality for the overall fits as in the SM.
They are presently based on one-loop
calculations~\cite{susy1,susydelr,susy3} with specific
two-loop improvements~\cite{drosusy} on the $\rho$ parameter.

With the increase of the experimental precision at future colliders,
decisive precision tests of both the SM and the MSSM will become
possible (see e.g.\ the reports 
\cite{Altarelli:2000ye,Aguilar-Saavedra:2001rg}).  
On the theoretical side, a thorough control of the quantization and the
renormalization of the MSSM as a supersymmetric gauge theory, with
spontaneously broken gauge symmetry and softly broken supersymmetry,
is required. This is not only a theoretical question for 
establishing a solid and consistent basis 
but also a matter of practical importance for
concrete higher-order calculations, where
the quantum contributions to the Green functions have to fulfil
the symmetry properties of the underlying theory.
An increasing number of
phenomenological applications has been carried out in the 
Wess-Zumino gauge where the  number of unphysical degrees 
of freedom is minimal, but 
where supersymmetry is no longer manifest.
Moreover, a manifestly supersymmetric and gauge-invariant 
regularization for divergent loop integrals is missing.
Hence, renormalization and the structure of counterterms
have to be adapted by exploiting the basic symmetries expressed
in terms of the supersymmetric BRS transformations 
\cite{White:ai,Maggiore:1996gg}.
An additional complication in the conventional approach
assuming an invariant regularization scheme, 
however, arises from the 
modification of the symmetry transformations themselves by
higher-order terms.    

The method of algebraic renormalization, applied 
in~\cite{Kraus:1997bi,Grassi:1999nb}
to the electroweak SM, avoids the difficulties of
the conventional approach. The theory is defined 
at the classical as well as the quantum level  
by the particle content and by the basic symmetries.
The essential features of the algebraic method are the 
combination of all symmetries into the BRS transformations 
leading to the Slavnov-Taylor (ST) identity. 
In this way, the theory is defined 
by symmetry requirements that have to be satisfied after renormalization
in all orders of perturbation theory. In the case of symmetry violation
during the explicit calculation of vertex functions in a given order,
additional non-invariant counterterms would be uniquely determined
to restore the symmetry, besides the invariant counterterms
needed for absorbing the divergences and for the normalization of fields and
parameters. 
Examples are given in~\cite{Hollik:1999xh,Hollik:2001cz} for supersymmetric 
QED and QCD and in~\cite{Grassi:algmeth} for the SM case.

In this article we adopt the algebraic method to define the (electroweak part 
of the) MSSM  as a quantum theory to all orders, including a complete
discussion of renormalization. The fundamental symmetries are
established in a functional way introducing  the ST and Ward operators,
and the invariant and non-invariant counterterms for the renormalization
are discussed, together with a proof of the infrared (IR) finiteness of
the renormalized theory.
In section 2 we define  the foundations  of the MSSM
specifying the field multiplets, the classical action, and the
gauge and soft symmetry breakings. Section 3 is devoted to the 
discussion of quantization and renormalization based on
the symmetry requirements, and of the IR finiteness.
In section 4 the on-shell renormalization is formulated and it is 
shown that the complete set of on-shell conditions can be fulfilled.

The rather involved structure of the MSSM leads to a series of
specific difficulties which had to be overcome and which will be
addressed in the paper:

\begin{itemize}
\item In the MSSM
supersymmetry is softly broken. Since our approach of
algebraic renormalization relies substantially on 
symmetries, we couple the soft breaking terms to external spurion fields
such that the combined system is again supersymmetric. Two different
methods which accomplish this have been discussed 
\cite{Maggiore:1996gg,Hollik:2000pa}; here
we use a synthesis of both methods which avoids unphysical
counterterms (section \ref{sec:spurion}). 
\item For gauge fixing, we use a variant of the $R_\xi$ gauge, which breaks 
  rigid gauge invariance. In order to have a rigid Ward identity, external
  fields have to be introduced that 
  restore the invariance. The particular
  structure of the MSSM Higgs sector requires two external tensor fields
  $\Phi_{1}^{a\, i}$, $\Phi_{2}^{a\, i}$ transforming as ${\rm ad}\otimes T$, 
  where $T$ is the
  representation of the Higgs fields (section \ref{sec:gaugefixing}).
\item Since the U(1) gauge field transforms linearly as opposed to its
  superpartner, algebraic consistency conditions have to be
  established. Nilpotency of the linearized Slavnov-Taylor (ST) operator holds
  only on a suitably defined subspace (section \ref{sec:AlgebraicRelations}).
  In particular, it has to be clarified to what extent
  the usual cohomology structure is affected (section \ref{sec:strategy}). 
\item The model has to be formulated in terms of physical fields describing
  mass eigenstates. However, the relation of physical and
  symmetric fields is modified by loop graphs which implies that the Ward
  and ST operators --- formulated in terms of physical fields --- become
  $\hbar$ dependent. 
  The field reparametrization corresponds to 
  an additional type of invariant counterterms (section \ref{sec:CTs}). 
\item Several field monomials would give rise to
  IR-singular insertions and must not be used for the cancellation of 
  breaking terms. We have to ensure that such counterterms can indeed be 
  avoided (section \ref{sec:IR}).
\item Since the model possesses many symmetries, there are not enough free
  parameters to establish all relevant on-shell normalization conditions. 
  Instead it has to be shown that some of the normalization conditions are 
  fulfilled automatically due to the ST and Ward identities 
  (section \ref{sec:EvaluationNormCond}).
\item The interpretation of the on-shell conditions for the Higgs fields is 
  complicated by the mixing of the physical $A^0$ boson and the 
  unphysical longitudinal gauge-boson degrees of freedom. It has to be 
  clarified whether the on-shell condition for the mass $M_{A^0}$ really 
  corresponds to the pole of the $A^0$ propagator 
  (section \ref{sec:A0longmixing}).    
\end{itemize}

\section{Foundations of the MSSM}
\label{sec:foundations}
\setcounter{equation}{0}

The MSSM of electroweak interactions is a supersymmetric SU(2)$\times$U(1) 
gauge theory where supersymmetry is softly broken by Girardello-Grisaru terms.
Its field content comprises the following superfields

\begin{tabular}{ll}
$\hat V$ & real SU(2) gauge superfield, $\hat V=\hat V^a T^a$,\\ 
$\hat V'$ & real U(1) gauge superfield,\\
$\hat H_1$, $\hat H_2$ & two  SU(2) doublets of chiral Higgs multiplets,\\
$\hat L$ & SU(2) doublet of the  chiral left-handed lepton multiplet,\\
$\hat R$ &  singlet of the   chiral right-handed electron multiplet,\\
$\hat Q$ &  SU(2) doublet of the  chiral left-handed quark multiplet,\\
$\hat U$, $\hat D$ &  singlets of  chiral right-handed up and down
quark multiplets.
\end{tabular}

We restrict our analysis to one generation of matter fields. As
required by anomaly cancellation, the quark fields appear in three
colours, but colour indices are suppressed throughout. To be close to
phenomenological applications we work in the Wess-Zumino gauge, where
the superfield expansions in terms of component fields read
\begin{subequations}
\begin{align}
\hat{V}&=\theta\sigma^\mu \bar\theta V_\mu+
\bar\theta^2 \theta^\alpha \lambda_\alpha+
\theta^2 \bar\theta_{\dot{\alpha}} \bar\lambda^{\dot{\alpha}}+
\frac{1}{2} \theta^2 \bar\theta^2 D_V,\\
\hat{V}^\prime&=\theta\sigma^\mu \bar\theta V_\mu^\prime+
\bar\theta^2 \theta^\alpha \lambda^\prime_\alpha+
\theta^2 \bar\theta_{\dot{\alpha}} \bar\lambda^{\prime \dot{\alpha}}+
\frac{1}{2} \theta^2 \bar\theta^2 D_V^\prime,\\
\hat{H}_i&=\e^{-\i \theta \sigma^\mu \bar\theta \partial_\mu}
\left(H_i+\sqrt{2}\theta^\alpha h_{i \alpha}+\theta^2 F_i\right),\qquad 
i=1,2,\\
\hat{L}&=\e^{-\i \theta \sigma^\mu \bar\theta \partial_\mu}
\left(L+\sqrt{2}\theta^\alpha l_\alpha+\theta^2 F_L\right),\\
\hat{R}&=\e^{-\i \theta \sigma^\mu \bar\theta \partial_\mu}
\left(R+\sqrt{2}\theta^\alpha r_\alpha+\theta^2 F_R\right),\\
\hat{Q}&=\e^{-\i \theta \sigma^\mu \bar\theta \partial_\mu}
\left(Q+\sqrt{2}\theta^\alpha q_\alpha+\theta^2 F_Q\right),\\
\hat{U}&=\e^{-\i \theta \sigma^\mu \bar\theta \partial_\mu}
\left(U+\sqrt{2}\theta^\alpha u_\alpha+\theta^2 F_U\right),\\
\hat{D}&=\e^{-\i \theta \sigma^\mu \bar\theta \partial_\mu}
\left(D+\sqrt{2}\theta^\alpha d_\alpha+\theta^2 F_D\right).
\end{align}
\label{eq:fields}
\end{subequations}
An overview of the component fields is given in table \ref{ta:fields},
and our conventions for superspace and gauge-group generators can be
found in appendix \ref{ap:conventions}.

\mytable{
\begin{center}
\vspace*{-12mm}\rule{0pt}{1pt}
\begin{tabular*}{\textwidth}{@{\extracolsep{5mm}}lccccc}
 & Bosonic fields & Fermionic fields & Isospin & Hypercharge \\
\hline
Gauge & $V^a$ & $\lambda^a$ & triplet & $0$ \\  
multiplets & $V^\prime$ & $\lambda^\prime$ & singlet & $0$ \\
\hline
Higgs fields  
 & $H_1=(H_1^1,H_1^2)^T$ & $h_1=(h_1^1,h_1^2)^T$ & doublet & $-1$ \\  
 & $H_2=(H_2^1,H_2^2)^T$ & $h_2=(h_2^1,h_2^2)^T$ & doublet & $1$ \\
\hline  
Leptons & $L=(L^1,L^2)^T$ & $l=(l^1,l^2)^T$ & doublet & $-1$ \\  
 &  $R$ & $r$ & singlet & $2$ \\
\hline  
Quarks & $Q=(Q^1,Q^2)^T$ & $q=(q^1,q^2)^T$ & doublet & $1/3$ \\  
 &  $U$ & $u$ & singlet & $-4/3$ \\
 &  $D$ & $d$ & singlet & $2/3$  \\
\end{tabular*}
\end{center}
}{Multiplets and field content of the theory}{ta:fields}

The defining symmetries of the 
MSSM are 
 SU(2)$\times$U(1) gauge invariance and  supersymmetry.
 The transformation laws of both 
symmetries are formulated in the BRS form \cite{BRS} with ghosts $c=c^aT^a$
for  
local SU(2) gauge symmetry (weak isospin), $c'$ for local U(1) gauge symmetry
(hypercharge), $\eps^\al$ for rigid supersymmetry, and $\xi^\mu$ for
rigid translational transformations. $\eps^\al$ is a constant complex
bosonic and $\xi^\mu$ a constant imaginary fermionic ghost (see
\cite{Hollik:1999xh,Maggiore:1995gr,White:ai,Rupp:2000db}
for details on the BRS formulation of supersymmetry). The BRS transformation 
of a field $\phi$ out of (\ref{eq:fields}) reads
\begin{equation}
\s \phi=(\delta_c+\delta_{c'}+\eps^\alpha Q_\alpha+
\bar{Q}_{\dot{\alpha}}\bar\eps^{\dot{\alpha}}-\i\xi^\mu\partial_\mu)\phi,
\end{equation}
where $\delta_c$, $\delta_{c'}$ are local gauge transformations with ghosts 
$c$, $c'$ used as transformation parameters, and where $Q_\alpha$, 
$\bar{Q}_\alphadot$ are the supersymmetry generators. The explicit form of all 
BRS transformations can be found in appendix~\ref{app:BRS}.
Unless we eliminate the auxiliary $D$ and $F$ fields, the BRS transformations 
are nilpotent:
\begin{align}
\s^2 \phi&=0 & \mbox{for all fields $\phi$}.
\end{align}

Although BRS invariance includes a substitute for local gauge invariance, rigid
gauge invariance does not necessarily follow from BRS invariance in higher 
orders and has to be established in addition. 
The SU(2)$\times$U(1) gauge 
transformations $\delta_\omega$, $\delta_{\omega'}$ read
\begin{subequations}
\label{eq:gaugetransf}
\begin{align}
\delta_\omega \phi &= -\i g \omega \phi && 
\text{for isospin doublets,}\\
\delta_\omega \phi &= 0 && \text{for isospin singlets,}\\
\delta_\omega \phi &= - \i g [ \omega , \phi ] && 
\text{for fields in the adjoint representation,} \\
\delta_{\omega'} \phi&= -\i g' \omega' \frac{Y}{2} \phi &&
\text{for all fields}
\end{align}
\end{subequations}
with constant transformation parameters $\omega=\omega^a T^a$, $\omega'$. The 
hypercharges $Y$ of the fields are given in table \ref{ta:fields} and are
related to the 
electric charge  by the Gell-Mann--Nishijima relation
\begin{equation}
\label{eq:Qem}
Q_{\rm em}=T^3+\frac{Y}{2}.
\end{equation}

Furthermore, we assume lepton- and baryon-number conservation, CP invariance, 
and continuous R symmetry. The CP transformations are given in table 
\ref{ta:CP}. The continuous R transformation with transformation parameter 
$\alpha$ reads ($\hat \phi$ stands for an arbitrary superfield):
\begin{equation}
\hat{\phi}(x, \theta, \bar\theta) \to \e^{2 \i n \alpha} 
\hat{\phi}(x, \e^{-\i \alpha} \theta, \e^{\i \alpha} \bar\theta). 
\end{equation}
For the R weights $n$ of chiral multiplets we choose 
$n=1/2$ for $\hat{H}_1, \hat{H}_2$, 
$n=1/4$ for $\hat{L}, \hat{R}, \hat{Q}, \hat{U}, \hat{D}$. The R weights
of the other multiplets are zero.

\mytable{
\begin{center}
\vspace*{-12mm}\rule{0pt}{1pt}
\begin{tabular*}{\textwidth}{lr@{\,}c@{\,}ll}
Type & \multicolumn{3}{l}{CP transformation} & Fields \\ 
\hline
Scalar &
$\phi$ & $\xrightarrow{\rm CP}$ & $ \bar{\phi}$ &
$H_1^i,H_2^i,L^i,R,Q^i,U,D,A,G^\pm,H^\pm,H,h,\au,\av$ \\
Real pseudo-scalar \hspace*{-0.4cm} &
$\phi$ & $\xrightarrow{\rm CP}$ & $-\phi$ &
$c,c^\prime,\bar{c},\bar{c}^\prime,B,B^\prime,G^0,A^0$ \\
Vector boson & 
$\phi^\mu$ & $\xrightarrow{\rm CP}$ & $- \bar{\phi}_\mu$ &
$V^\mu, V^{\prime\mu},A^\mu, Z^\mu, W^{\pm \mu}$ \\
Weyl spinor & 
$\phi_\alpha$ & $\xrightarrow{\rm CP}$ & 
$-\e^{-\i \omega_{\rm CP}}\bar{\phi}^{\dot{\alpha}}$ & 
$\lambda_\alpha, \lambda^\prime_\alpha, h_{1 \alpha}^i, h_{2 \alpha}^i,
l^i_\alpha,r_\alpha,q^i_\alpha,u_\alpha,d_\alpha,a_\alpha,\chi^\pm_\alpha,
\chi^0_\alpha$\\
& $\bar{\phi}^{\dot{\alpha}}$ & $\xrightarrow{\rm CP}$ 
& $\e^{-\i \omega_{\rm CP}}\phi_\alpha$ &
\end{tabular*}
\end{center}
}{CP transformations ($\omega_{\rm CP}$ is an arbitrary phase)}{ta:CP}

\subsection{Classical action in superfield formulation}

To construct the most general classical action, it is convenient to start 
from the superspace formulation, which is more condensed and
incorporates supersymmetry manifestly. The supersymmetric kinetic and
interaction part of the classical action reads
\begin{eqnarray} 
\Gamma_{\rm susy}&=&\frac{1}{16} \sum\limits_{i=1,2} \int \dV 
\bar{\hat{H}}_i e^{2 g \hat{V}+ g^\prime Y \hat{V}^\prime} \hat{H}_i
+\frac{1}{16}\int \dV
\bar{\hat{L}} e^{2 g \hat{V}+ g^\prime Y \hat{V}^\prime} \hat{L}
\nonumber \\ && {}
+\frac{1}{16}\int \dV
\bar{\hat{R}} e^{g^\prime Y \hat{V}^\prime} \hat{R}
+ \frac{1}{16} \int \dV
\bar{\hat{Q}} e^{2 g \hat{V}+ g^\prime Y \hat{V}^\prime} \hat{Q}
\nonumber \\ && {}
+\frac{1}{16}\int \dV \bar{\hat{U}} e^{g^\prime Y \hat{V}^\prime} \hat{U}
+\frac{1}{16}\int \dV \bar{\hat{D}} e^{g^\prime Y \hat{V}^\prime} \hat{D}
\nonumber \\ && {}
-\frac{1}{512 g^2} \int \dS 2 \tr[ \hat{F}^\alpha \hat{F}_\alpha]
-\frac{1}{128 g^{\prime 2}} \int \dS \hat{F}^{\prime \alpha} 
\hat{F}^\prime_{\alpha}
-\frac{v^\prime}{8}\int \dV \hat{V}^\prime
\nonumber \\ && {} 
+\bigg[
-\frac{f_R}{4} \int \dS \hat{H}_1^T (\i \sigma_2) \hat{L} \hat{R}
-\frac{f_U}{4} \int \dS \hat{H}_2^T (\i \sigma_2) \hat{Q} \hat{U}
\nonumber \\ && {} \qquad
-\frac{f_D}{4} \int \dS \hat{H}_1^T (\i \sigma_2) \hat{Q} \hat{D}
+\frac{\mu}{4} \int \dS \hat{H}_1^T (\i \sigma_2) \hat{H}_2
+{\rm c.c.}\bigg]
\label{eq:susy}
\end{eqnarray}
with
\begin{align}
\hat{F}^\alpha&=\bar \D^2 \e^{-2 g \hat{V}} \D^\alpha \e^{2 g \hat{V}}, &
\hat{F}^{\prime \alpha}&=
\bar \D^2 \e^{-g^\prime \hat{V}^\prime} \D^\alpha 
\e^{g^\prime \hat{V}^\prime}.
\end{align}
Note that we added a Fayet-Iliopoulos term involving $v'$ to the action.
Explicit results of the classical action in terms of component fields can be 
found in appendix \ref{ap:explicit}. 
 
\subsection{Soft supersymmetry breaking} 
\label{sec:spurion}

Since so far no supersymmetric partners of the Standard Model particles have 
been experimentally observed, supersymmetry has to be broken in such a way 
that the superpartners acquire a higher mass. Explicit supersymmetry breaking 
is also necessary to trigger the spontaneous breakdown of gauge symmetry by 
the Higgs mechanism. 
The MSSM restricts the breaking terms to the soft 
breaking terms found by Girardello and Grisaru \cite{Girardello:1981wz}.
These comprise mass terms for scalar fields and gauginos and up to 
trilinear interactions between the scalar components of either purely
chiral or purely antichiral multiplets. They are generally believed to be 
a sufficient starting point for supersymmetry phenomenology.

In order to motivate our strategy how soft supersymmetry breaking can be 
introduced in the MSSM, we consider slepton and gaugino mass terms as 
examples. They can be written as 
\begin{align}
\int\dV\theta^2 \bar\theta^2 \bar{\hat L}\e^{2g\hat V+g'Y\hat V'}\hat L 
\sim \int \dx \bar L  L,\\
\int \dS \theta^2 \hat{F}'{}^\al \hat{F}'_\al \sim \int \dx \lambda'{}^\al \lambda'{}_\al,
\end{align}

A customary tool to control symmetry-breaking terms is to couple them
to external fields in such a way that the total action is again invariant. 
These external fields must have inhomogeneous transformation laws such that 
the breaking terms reappear for vanishing external fields.
For this purpose we choose a chiral scalar spurion field $\hat A$ as 
proposed  in \cite{Girardello:1981wz}:
\begin{equation}
\hat{A}=\e^{-\i \theta \sigma^\mu \bar\theta \partial_\mu}
\left(A+\sqrt{2}\theta^\alpha a_\alpha+\theta^2 F_A\right).
\end{equation}
By shifting the $\theta^2$ component of this field, 
$F_A \to F_A+v_A$, the desired inhomogeneity is introduced 
into the transformation law, and for $\hat A\to \theta^2 v_A$ 
the above breaking terms 
are produced when the field is coupled as
\begin{eqnarray}
\int \dV \hat A \bar{\hat A} \bar{\hat L} 
\e^{2g\hat V + g' Y \hat V'} \hat L 
&\stackrel{{\hat A\to \theta^2 v_A}}{\longrightarrow}&
v_A^2 \int \dV \theta^2 \bar \theta^2 \bar{\hat L} \e^{2g\hat V+g'Y\hat V'} 
\hat L,\\
\int \dS \hat A \hat{F}'{}^\al \hat{F}'_\al
&\stackrel{{\hat A\to \theta^2 v_A}}{\longrightarrow}&
v_A \int \dS \theta^2 F^\al F_\al.
\end{eqnarray}
The BRS transformations of the spurion multiplet are given by
\begin{subequations}
\begin{align}
\s A&=\sqrt{2} \eps^\alpha a_\alpha-\i \xi^\mu \partial_\mu A,\\
\s a_\alpha &=\sqrt{2} \eps_\alpha (F_A+v_A)+\i \sqrt{2} 
\sigma^\mu_{\alpha \dot{\alpha}}\bar\eps^{\dot{\alpha}}\partial_\mu A
-\i \xi^\mu \partial_\mu a_\alpha,\\
\s F_A&=\i \sqrt{2} \partial_\mu a^\alpha \sigma^\mu_{\alpha \dot{\alpha}} 
\bar\eps^{\dot{\alpha}}-\i \xi^\mu \partial_\mu F_A.
\end{align}
\end{subequations}

The spurion fields provide an adequate characterization of soft 
supersymmetry breaking 
versus soft breaking of gauge invariance. However, 
since the spurion field is dimensionless, it can in principle appear
in arbitrary powers in the action. In this way infinitely many new
renormalization parameters could appear in higher orders of
perturbation theory, but it has been shown in \cite{Hollik:2000pa}
that the infinite towers of terms in the action are irrelevant
for physical processes and 
it suffices to include terms up to first order 
in each $\hat A$, $\bar{\hat A}$ and $\hat A\bar{\hat A}$. 

An alternative approach to soft supersymmetry breaking
has been presented in \cite{Maggiore:1996gg}, using no superfield 
but two ordinary scalar fields of dimension 1 which form a BRS doublet. 
Because of their dimension, only finitely many terms can be formed, and 
since BRS doublets cannot contribute to the cohomology, it is clear that no 
new anomaly can emerge. We can pass over from the chiral spurion field 
$\hat A$ to the BRS doublet $\au$, $\av$ (originally named $u$, $v$) of 
\cite{Maggiore:1996gg}  by identifying
\begin{equation}
A=0 ,\qquad 
a^\alpha = \sqrt{2} \eps^\alpha \au ,\qquad
F_A = \av. 
\label{spurionexpansion}
\end{equation}
The transformation law for the BRS doublet%
\footnote{Strictly speaking, the BRS doublet is given by
  $\au$ and $\av'\equiv \av+v_A-\i \xi^\mu \partial_\mu \au$, since $\s
  \au=\av'$, $\s A'_2=0$.\label{BRSdoubletfootnote}}%
$\au$, $\av$ is 
\begin{subequations}
\begin{align}
\s \au&= \av+v_A-\i \xi^\mu \partial_\mu \au,\\
\s \av&= \i 2 \eps^\alpha \sigma^\mu_{\alpha \dot{\alpha}} 
\bar\eps^{\dot{\alpha}} \partial_\mu \au-\i \xi^\mu \partial_\mu \av.
\end{align}
\end{subequations}
From now on we consider a restricted classical action (and later on the
vertex functional) where the spurion multiplet is always parametrized
as in (\ref{spurionexpansion}). Equivalently, it depends on $\au$ only
through the combination $a^\al=\sqrt{2}\eps^\al \au$. 

Since we have put $A=0$ in (\ref{spurionexpansion}), powers of $\hat A$ 
vanish and the most general real and chiral polynomials in $\hat A$, 
$\bar{\hat A}$ are given by
\begin{equation}
\poly_{\rm real}=c_{\rm real,1}(\hat A+\bar{\hat A})+c_{\rm real,2} \hat A
\bar{\hat A}, \qquad
\poly_{\rm chiral} =c_{\rm chiral} \hat A.
\label{Apolydef}
\end{equation}
By inserting polynomials of the form (\ref{Apolydef}) in $\Gamma$
wherever possible, we obtain
\begin{eqnarray}
\Gamma_{\rm soft}&=&
\frac{1}{16} \sum\limits_{i=1,2} \int \dV \poly_{H_i}  
\bar{\hat{H}}_i e^{2 g \hat{V}+ g^\prime Y \hat{V}^\prime} \hat{H}_i
+\frac{1}{16}\int \dV \poly_L 
\bar{\hat{L}} e^{2 g \hat{V}+ g^\prime Y \hat{V}^\prime} \hat{L}
\nonumber \\ && {} 
+\frac{1}{16}\int \dV \poly_R 
\bar{\hat{R}} e^{g^\prime Y \hat{V}^\prime} \hat{R}
+ \frac{1}{16} \int \dV \poly_Q 
\bar{\hat{Q}} e^{2 g \hat{V}+ g^\prime Y \hat{V}^\prime} \hat{Q}
\nonumber \\ && {}
+\frac{1}{16}\int \dV \poly_U  
\bar{\hat{U}} e^{g^\prime Y \hat{V}^\prime} \hat{U}
+\frac{1}{16}\int \dV \poly_D 
\bar{\hat{D}} e^{g^\prime Y \hat{V}^\prime} \hat{D}
\nonumber \\ && {}
+\bigg[
-\frac{c_{V}}{512 g^2} \int \dS \hat{A} 2 \tr[ \hat{F}^\alpha \hat{F}_\alpha]
-\frac{c_{V^\prime}}{128 g^{\prime 2}} 
\int \dS \hat{A} \hat{F}^{\prime \alpha} \hat{F}^\prime_{\alpha}
\nonumber \\ && {} \qquad
-\frac{c_{HLR}}{4} \int \dS \hat{A} \hat{H}_1^T (\i \sigma_2) \hat{L} \hat{R}
-\frac{c_{HQU}}{4} \int \dS \hat{A} \hat{H}_2^T (\i \sigma_2) \hat{Q} \hat{U}
\nonumber \\ && {} \qquad
-\frac{c_{HQD}}{4} \int \dS \hat{A} \hat{H}_1^T (\i \sigma_2) \hat{Q} \hat{D}
+\frac{c_{HH}}{4} \int \dS \hat{A} \hat{H}_1^T (\i \sigma_2) \hat{H}_2
+{\rm c.c.}\bigg]
\label{eq:soft}
\end{eqnarray}
with
\begin{align}
\poly_\phi&=c_{\phi 1} (\hat{A}+\bar{\hat{A}})
+c_{\phi 2} \bar{\hat{A}}\hat{A},&\phi&=H_1,H_2,L,R,Q,U,D
\end{align}
and $c_{\phi 1}$, $c_{\phi 2}$ real.

\subsection{External fields and elimination of auxiliary fields}
\label{sec:externalfields}

The BRS transformations 
of most of the fields [see eqs.~(\ref{BRS_V})--(\ref{BRS_epsdot})] 
are non-linear in the propagating fields, i.e.\ they correspond to 
composite operators in the quantum theory. Therefore, we couple the
non-linear BRS transformations to external source fields $Y$, $y$ in
such a way that differentiation of the action with respect to these
external fields yields the desired non-linear field expressions. 
This yields the external field part of the classical action:
\begin{eqnarray}
\Gext&=& \int \dx \Big[
\, 2 \tr \left(Y_V^{\mu} \s V_\mu +y_\lambda^{\alpha} \s \lambda_\alpha
+\bar{y}_{\lambda \dot{\alpha}} \s \bar{\lambda}^{\dot{\alpha}} 
+Y_c \s c \right)
+y_{\lambda^\prime}^{\alpha} \s \lambda^\prime_\alpha
+\bar{y}_{\lambda^\prime \dot{\alpha}}\s \bar{\lambda}^{\prime\dot{\alpha}} 
\nonumber \\ && {}
+Y_L^T \s L+\bar{Y}_L \s \bar{L}^T+y_l^{\alpha T} \s l_\alpha
+\bar{y}_{l \dot{\alpha}} \s \bar{l}^{\dot{\alpha} T} 
+Y_R \s R+\bar{Y}_R \s \bar{R}+y_r^\alpha \s r_\alpha
+\bar{y}_{r \dot{\alpha}} \s \bar{r}^{\dot{\alpha}} 
\Big]
\nonumber \\ && {}
+\mbox{analogous quark and Higgs terms.}
\label{eq:ext}
\end{eqnarray}
We do not introduce $Y$ fields corresponding to the $D$  and $F$ fields
as these fields will be eliminated later. The $Y$ fields are invariant 
under BRS transformations:
\begin{align}
\s Y&=0,& \s y&=0.
\end{align} 
Since $\s^2=0$, $\Gext$ is BRS invariant, and we obtain
\begin{equation}
\s (\Gamma_{\rm susy}+\Gamma_{\rm soft}+\Gext)=0.
\end{equation} 
The definition of the $Y$ fields is such that $\Gext$ is bosonic 
and real.\footnote{In our conventions $y_\lambda$ is a bosonic spinor and
its complex conjugate is $\bar{y}_\lambda$. Conversely,
$Y_L$ is a fermionic scalar and its complex conjugate is $Y_L^*=-\bar{Y}_L$.}
The gauge transformations of the external fields 
$Y$ are such that $\Gext$ is gauge invariant.

The auxiliary fields can be eliminated from the action and the 
BRS transformations by using the equations of motion. The resulting
replacements are
\begin{subequations}
\label{eq:elimresults}
\begin{eqnarray}
D_V&=&{}
-g\T^a \left[\bar{H}_1 \T^a H_1
 +\bar{H}_2 \T^a H_2+\bar{L} \T^a L+\bar{Q} \T^a Q \right]
-y_\lambda^\alpha \eps_\alpha
+\bar{y}_{\lambda \dot{\alpha}} \bar{\eps}^{\dot{\alpha}}
\nonumber \\ && {}
+c_{V} \au \eps^\alpha \lambda_\alpha 
+c_{V} \aub \bar\eps_{\dot{\alpha}} \bar\lambda^{\dot{\alpha}}, 
\qquad\\
D_V^\prime&=&v^\prime
-\frac{g^\prime}{2}\left[
  \bar{H_1} Y H_1+\bar{H_2} Y H_2+\bar{L} Y L+\bar{R} Y R
 +\bar{Q} Y Q+\bar{U} Y U+\bar{D} Y D
\right]
\nonumber \\ && {}
-y_{\lambda^\prime}^{\alpha} \eps_\alpha
+\bar{y}_{\lambda^\prime \dot{\alpha}} \bar{\eps}^{\dot{\alpha}}
+c_{V^\prime} \au \eps^{\alpha} 
 \lambda_\alpha^\prime 
+c_{V^\prime} \aub \bar\eps_{\dot{\alpha}} 
 \bar\lambda^{\prime \dot{\alpha}}, \\
F_1&=&-\f_R (\i \sigma_2) L^* \bar{R}
-\f_D (\i \sigma_2) Q^* \bar{D}
+\mu (\i \sigma_2) H_2^*
+\sqrt{2} \bar{y}_{h_1 \dot{\alpha}} \bar{\eps}^{\dot{\alpha}}
\nonumber \\ && {}
-c_{H_1 1}\left[(\av+v_A)H_1+\sqrt{2} 
 \au \eps^\alpha h_{1\alpha} \right], \\
F_2&=&-\f_U (\i \sigma_2) Q^* \bar{U}
-\mu (\i \sigma_2) H_1^*
+\sqrt{2} \bar{y}_{h_2 \dot{\alpha}} \bar{\eps}^{\dot{\alpha}}
\nonumber \\ && {}
-c_{H_2 1}\left[(\av+v_A)H_2+\sqrt{2} 
 \au \eps^\alpha h_{2\alpha} \right], \\
F_L&=&\f_R (\i \sigma_2) H_1^* \bar{R}
+\sqrt{2} \bar{y}_{l \dot{\alpha}} \bar{\eps}^{\dot{\alpha}}
-c_{L1}\left[(\av+v_A)L+\sqrt{2} \au \eps^\alpha l_\alpha \right], \\
F_R&=&{}-\f_R \bar{H}_1 (\i \sigma_2) L^*
+\sqrt{2} \bar{y}_{r \dot{\alpha}} \bar{\eps}^{\dot{\alpha}}
-c_{R1}\left[(\av+v_A)R+\sqrt{2} \au \eps^\alpha r_\alpha \right], \\
F_Q&=&\f_D (\i \sigma_2) H_1^* \bar{D}
+\f_U (\i \sigma_2) H_2^* \bar{U}
+\sqrt{2} \bar{y}_{q \dot{\alpha}} \bar{\eps}^{\dot{\alpha}}
\nonumber \\ && {}
-c_{Q1}\left[(\av+v_A)Q+\sqrt{2} \au \eps^\alpha q_\alpha \right], \\
F_U&=&{}-\f_U \bar{H}_2 (\i \sigma_2) Q^*
+\sqrt{2} \bar{y}_{u \dot{\alpha}} \bar{\eps}^{\dot{\alpha}}
-c_{U1}\left[(\av+v_A)U+\sqrt{2} \au \eps^\alpha u_\alpha \right],\\
F_D&=&{}-\f_D \bar{H}_1 (\i \sigma_2) Q^*
+\sqrt{2} \bar{y}_{d \dot{\alpha}} \bar{\eps}^{\dot{\alpha}}
-c_{D1}\left[(\av+v_A)D+\sqrt{2} \au \eps^\alpha d_\alpha \right].
\end{eqnarray}
\end{subequations}
After eliminating the auxiliary fields, the nilpotency is respected only
on shell, i.e.\
\begin{eqnarray}
\s^2 \phi&=&0 + \mbox{equations of motion}.
\label{eq:nilpotency}
\end{eqnarray}
Since the action is stationary with respect to the auxiliary fields after their
elimination, the invariance 
$\s(\Gamma_{\rm susy}+\Gamma_{\rm soft}+\Gext)=0$
still holds if the results (\ref{eq:elimresults}) are used in the BRS
transformations (\ref{BRS_V})--(\ref{BRS_epsdot}). Furthermore, the
bilinear terms in the external fields introduced in
\cite{Maggiore:1996gg,Hollik:1999xh,Hollik:2000pa,Maggiore:1995gr,White:ai}
are automatically included in $\Gext$ after inserting (\ref{eq:elimresults}).

\subsection{Soft supersymmetry-breaking parameters}
\label{sec:softsusybreaking}

After elimination of the auxiliary fields, we can evaluate the physical terms
induced by the shift in the spurion field. We obtain the following
contributions in the limit $\hat{A}=\theta^2 v_A$ and $Y=0$:
\begin{eqnarray}
\Gamma_{\rm soft}|_{\hat{A}=Y=0}&=&{}
-\int \dx \left(m_1^2 |H_1|^2+m_2^2 |H_2|^2 +m_{\tilde{l}}^2 |L|^2
+m_{\tilde{e}}^2  |R|^2\right) 
\nonumber \\ && {}
-\int \dx \tfr{1}{2}\left[M_2 2 \tr\left(\lambda^\alpha\lambda_\alpha\right)
+M_1 \lambda^{\prime\alpha}\lambda'_\alpha +{\rm c.c.}\right]
\nonumber \\ && {}
- \int\dx\left[m_3^2 H_1^T (\i \sigma_2) H_2
+f_R A_e H_1^T (\i \sigma_2) L R+{\rm c.c.}\right]
\nonumber \\ && {}
+ \mbox{ analogous quark terms}.
\end{eqnarray}
with the explicit results
\begin{subequations}
\label{physspurionterms}
\begin{align}
m_i^2&=v_A^2 \left(c_{H_i1}^2-c_{H_i2} \right)+Y\tfr{g'}{2} v',
\qquad i=1,2,\\
m_{\tilde{l}}^2&=v_A^2 \left(c_{L 1}^2-c_{L 2}\right)+Y\tfr{g'}{2}v',\\ 
m_{\tilde{r}}^2&=v_A^2 \left(c_{R 1}^2-c_{R 2}\right)+Y\tfr{g'}{2}v',\\
m_3^2&=c_{HH}v_A-\mu v_A (c_{H_1 1}+c_{H_2 1}),\\
M_1^2&=-2 v_A c'_V, \\
M_2^2&=-2 v_A c_V,\\
f_R A_e&=v_A(c_{L1} f_R+c_{R1} v_A+c_{H_11} f_R-c_{HLR}),
\end{align}
\end{subequations}
where $Y$ symbolizes the hypercharge of the corresponding field as defined 
in table \ref{ta:fields}. The equations (\ref{physspurionterms}) show
that only certain combinations of the spurion parameters are physical
parameters of the MSSM, namely $m_1^2$, $m_2^2$, $m_{\tilde{l}}^2$,
$m_{\tilde{e}}^2$, $M_1$, $M_2$, $m_3^2$, $A_e$, and similar
parameters $m_{\tilde{q}}^2$, $m_{\tilde{u}}^2$, $m_{\tilde{d}}^2$,
$A_u$, $A_d$ for the quark fields. The remaining parameters are
irrelevant for physical observables \cite{Hollik:2000pa}. 

\subsection{Field parametrization}
\label{sec:fieldparam}

The question of field parametrization is a central aspect of spontaneously 
broken gauge theories. Up to now we have introduced fields which are 
multiplets under the SU(2)$\times$U(1) gauge group ({\em symmetric} fields). 
They do not correspond directly to mass eigenstates. However, we are still 
free to choose a basis of independent {\em physical} fields in terms of
which the vertex functional and the symmetry operators are parametrized. 

In this section we introduce a set of fields that correspond to mass 
eigenstates. Schematically, we define the physical fields as general linear 
combinations of the symmetric fields with $\hbar$-dependent coefficients: 
\begin{equation}
\phi_i^{\rm sym} = \Z_{ij} \phi_j^{\rm phys}, \qquad
\Z_{ij}=\sum_{n=0}^\infty \Z_{ij}^{(n)} \hbar^n. 
\label{Phisymphys}
\end{equation}
Later in section \ref{sec:renscheme}, it will be shown that by adjusting the
$\Z_{ij}^{(n)}$ the mass matrix can be diagonalized%
\footnote{To be precise, this diagonalization is not possible 
  for the mixings between Higgs fields and the unphysical longitudinal
  gauge bosons and $B$ fields.}
to all orders in $\hbar$, i.e.
\begin{align}
{\rm Re} \Gamma_{\phi_i^{\rm phys} \phi_j^{\rm phys}} \Bigr|_{p^2=m_i^2}=0,
\end{align}
where $m_i$ is the mass of the field $\phi^{\rm phys}_i$. In this 
way, the $\phi_i^{\rm phys}$ can be identified with mass eigenstates up to 
effects due to non-vanishing imaginary parts of self energies.

From the phenomenological point of view it is certainly desirable to work with
physical fields and on-shell normalization conditions. In some cases, we are 
also forced to use a specific parametrization due to IR finiteness as we 
will see in section \ref{sec:IR}. 
Classically, $\Z^{(0)}$ can be chosen as a rotation matrix. Loop corrections 
lead in general to remixing of the classically diagonal fields so that the 
matrix $\Z$ has to be a general non-singular matrix. 
Since mixing is only possible between fields with matching quantum
numbers, the general demixing matrix $\Z$ can be 
decomposed into several submatrices $\Z_\phi$ for the various kinds of fields. 

The mass eigenstates of the gauge bosons are the $Z$ boson, $W^\pm$ boson, and 
the photon. We write
\begin{align}
\label{eq:Vectorparam}
\VL V'_\mu \\ V^3_\mu \VR &= \Z_V \VL A_\mu \\ Z_\mu \VR, &
\frac{1}{\sqrt2}(V^1_\mu\mp\i V^2_\mu)&=  W^\pm_\mu.
\end{align}
Here, $\Z_V$ is a real $2\times 2$ matrix.
Similarly, we write for ghosts and antighosts
\begin{subequations}
\begin{align}
\VL c' \\ c^3 \VR &= \Z_c \VL c_A \\ c_Z \VR, &
\frac{1}{\sqrt2}(c^1\mp\i c^2)&= c^\pm, 
\\
\VL \bar{c}' \\ \bar{c}^3 \VR &= 
\Z_{\bar{c}} \VL \bar{c}_A \\ \bar{c}_Z \VR, &
\frac{1}{\sqrt2}(\bar{c}^1\mp\i\bar{c}^2)&= \bar{c}^\pm
\end{align}
\end{subequations}
with real $2\times 2$ matrices $\Z_c$, $\Z_{\bar{c}}$.
As usual, we will couple the auxiliary fields $B^{(\prime)}$ to linear
gauge-fixing functions
${\cal F}^{(\prime)} = \partial^\mu V^{(\prime)}_\mu + \cdots$
 in the gauge-fixing term (see section 
\ref{sec:gaugefixing}). For linear gauge fixing the $B$ fields have no
interaction vertices, and they are auxiliary fields.
Thus, independent mixing matrices for $B$ fields are not required and  we 
choose
\begin{align}
\VL B'\\B_3\VR & = \Z_V^{-1T} \VL B_A\\B_Z\VR, &
\frac{1}{\sqrt2}(B^1\mp\i B^2) &= B^\pm.
\end{align}

For the scalar Higgs fields, we have three real $2\times 2$ matrices 
$\Z_{H^0}$, $\Z_{H^\pm}$, and $\Z_{A^0}$ for the neutral CP-even, CP-odd, and 
charged sector, respectively. With the help of the parametrization
\begin{align}
\label{eq:Higgs}
\bar{H}_1 (\i \sigma_2)&=
\VL \phi_1^+ \\[0.5ex] \frac{1}{\sqrt2}(\phi_1^0 + \i\rho_1^0) \VR, &
H_2 &= \VL \phi_2^+ \\[0.5ex] \frac{1}{\sqrt2}(\phi_2^0 + \i\rho_2^0) \VR 
\end{align}
we introduce the matrices $\Z_{H^0}$, $\Z_{A^0}$, $\Z_{H^\pm}$ as follows:
\begin{align}
\VL \phi_1^0 \\ \phi_2^0 \VR &=
\Z_{H^0}\VL H^0 \\ h^0 \VR,&
\VL \rho_1^0 \\ \rho_2^0 \VR &=
\Z_{A^0} \VL G^0 \\ A^0 \VR,&
\VL \phi_1^{\pm} \\ \phi_2^{\pm} \VR &=
\Z_{H^\pm} \VL G^{\pm} \\ H^{\pm} \VR.
\end{align}
The electric charges of the fields are indicated in the superscript. The 
neutral fields are real and the charged ones complex with $G^-=\bar G^+$, 
$H^-=\bar H^+$. Furthermore, the fields $A^0$ and $G^0$ are CP odd. The 
notation anticipates that the normalization conditions will be such that the 
$G^{0,\pm}$ correspond to the unphysical Goldstone degrees of freedom. 

Gauginos and Higgsinos also mix. This requires two complex $2\times 2$
matrices for the charged sector and one complex $4\times 4$ matrix for the
neutral sector:
\begin{align}
\twovect{\lambda^+_\alpha}{h_{2 \alpha}^1} &= 
\Z_{\chi^+} \twovect{\chi_{1 \alpha}^+}{\chi_{2 \alpha}^+},&
\twovect{\lambda^-_\alpha}{h_{1\alpha}^2} &= 
\Z_{\chi^-} \twovect{\chi_{1 \alpha}^-}{\chi_{2 \alpha}^-},&
\left(\! \begin{array}{c} \lambda'_\alpha \\ \lambda^3_\alpha \\ 
h_{1\alpha}^1 \\ h_{2\alpha}^2 \end{array}\!\right) 
&= \Z_{\chi^0} \left( \!\begin{array}{c} \chi_{1 \alpha}^0 \\  
\chi_{2 \alpha}^0 \\ \chi_{3 \alpha}^0 \\\chi_{4 \alpha}^0 
\end{array}\!\right).
\end{align}

In order to respect all normalization conditions in the matter sector 
of the MSSM (see section \ref{sec:renscheme}), 
we introduce normalization factors for the left-handed fermions:
\begin{align}
q^1_\alpha & = \Z_{u_L} u_{L\alpha},&
q^2_\alpha & = \Z_{d_L} d_{L\alpha},\\
l^1_\alpha & = \Z_{\nu_L} \nu_{L\alpha},&
l^2_\alpha & = \Z_{e_L} e_{L\alpha}.
\end{align}

Since we consider only one generation, there is no mixing within quarks and 
leptons. However, there is in general a mixing of the superpartners of the 
left- and right-handed sfermions:
\begin{align}
\VL L^2\\ \bar{R} \VR & = 
\Rse \VL \tilde{e}_1 \\ \tilde{e}_2\VR,&
L^1 & = \tilde{\nu}_1, \\
\VL Q^1\\ \bar{U} \VR & = 
\Rsu \VL \tilde{u}_1 \\ \tilde{u}_2\VR,&
\VL Q^2\\ \bar{D} \VR & =  
\Rsd \VL \tilde{d}_1 \\ \tilde{d}_2\VR.
\end{align}

\subsection{Gauge fixing}
\label{sec:gaugefixing}

To quantize the theory, gauge fixing is necessary. We use a gauge-fixing 
and ghost term of the general form 
\begin{eqnarray}
\label{eq:gaugefixgeneral}
\Ggf &=& \int \dx \s \left[2 \tr 
\left( \bar c \F + \tfr{1}{2} \zeta \bar c B \right) 
+ \bar c' \F' + \tfr{1}{2} \zeta' \bar c' B'\right] 
\nonumber\\
&=&\int \dx 2\tr \left( B \F + \tfr{1}{2} \zeta B^2 -\bar c \s\F 
-\i \xi^\mu \partial_\mu \bar c \F 
-\i\zeta\eps^\al\sigma^\mu_{\al \da} \bar\eps^\da \bar c\partial_\mu 
\bar c\right) 
\nonumber \\ && {}
+ \int \dx \left( B' \F' +\tfr{1}{2} \zeta' {B'}^2 -\bar c' \s\F' 
-\i \xi^\mu \partial_\mu \bar c'   \F' 
-\i\zeta'\eps^\al\sigma^\mu_{\al \da}\bar\eps^\da \bar c'\partial_\mu\bar c' 
\right)
\end{eqnarray}
with gauge-fixing functions $\F=\F^a T^a$, $\F'$. The auxiliary fields
$B$, $B'$ are introduced as the BRS transformations of the antighosts
$\bar c$, $\bar c'$ (see appendix \ref{app:BRS}). In this way,
$\s^2=0$ holds also on these fields 
and the gauge-fixing and ghost action (\ref{eq:gaugefixgeneral}) is
BRS invariant, $\s \Ggf=0$. For perturbative calculations, the
$B$ fields can be readily eliminated using their equations
of motion, yielding the usual form of the gauge-fixing term
$-2\tr(\F^2)/(2\zeta) - (\F')^2/(2 \zeta')$.

Analogously to the decomposition of the symmetric gauge-boson fields into mass 
eigenstates (\ref{eq:Vectorparam}) we rewrite the gauge-fixing functions 
as
\begin{align}
\label{eq:Fphys}
\VL \F' \\ \F^3 \VR &= \Z_V \VL \F_A \\ \F_Z \VR, &
\frac{1}{2}(\F^1 \mp\i \F^2) &=   \F^\pm.
\end{align}
For the gauge-fixing functions $\F^\pm$, $\F_A$, and $\F_Z$ we choose a 
linear function in propagating fields. Especially useful for phenomenological 
calculations are the gauge-fixing functions
\begin{subequations}
\label{eq:gaugefix1}
\begin{align}
\F^\pm & =  \partial^\mu W^\pm_\mu \pm\i M_W(\zeta^{G^\pm} G^\pm 
+ \zeta^{H^\pm} H^\pm), \\
\VL \F_A\\ \F_Z\VR
 & =  \VL \partial^\mu A_\mu \\ \partial^\mu Z_\mu \VR + M_Z 
\twomat{\zeta^{A G^0} & \zeta^{A A^0} \\
        \zeta^{Z G^0} & \zeta^{Z A^0}} 
\VL G^0 \\ A^0 \VR ,
\end{align}
\end{subequations}
where 
$M_W$ and $M_Z$ are the (real) masses of the $W^\pm$ and $Z$ bosons. 

Such a gauge-fixing condition gives masses to the unphysical Goldstone bosons, 
and it allows to eliminate mixing terms between Goldstone bosons and 
longitudinal gauge bosons from the action by a suitable choice of the gauge 
parameters $\zeta^{G^\pm}$, $\zeta^{AG^0}$, and $\zeta^{ZG^0}$. Furthermore, 
the physical Higgs bosons $H^\pm$ and $A^0$ can be excluded from the 
gauge-fixing functions by defining $\zeta^{H^\pm}=\zeta^{AA^0}=\zeta^{ZA^0}=0$.
In the actual renormalization analysis of section \ref{sec:Renormalization} 
we stay general and treat all $\zeta$ parameters as independent.

The gauge-fixing functions (\ref{eq:gaugefix1}) exhibit a severe problem: they
are not gauge covariant, and the gauge-fixing term (\ref{eq:gaugefixgeneral}) 
is, therefore, not invariant with respect to {\em rigid} gauge transformations.
Since rigid invariance is an important ingredient of the definition of the 
MSSM, it has to be restored.

The violation of rigid gauge invariance is particularly apparent if
$\F'$, $\F^a$ are rewritten in terms of the original Higgs fields 
\begin{subequations}
\label{eq:gaugefix2}
\begin{align}
\F^a &= \partial^\mu V^a_\mu-(\i\bar\mbv^a_1H_1+\i\bar\mbv^a_2H_2
+{\rm h.c.}), \\
\F' &= \partial^\mu V'_\mu-(\i\bar\mbv'_1H_1+\i\bar\mbv'_2H_2
+{\rm h.c.})
\end{align}
\end{subequations}
with fixed isospin doublets $\mbv_{1,2}'$, $\mbv_{1,2}^{a}$. From
comparing (\ref{eq:gaugefix2}) with (\ref{eq:gaugefix1}) we get
\begin{subequations}
\begin{align}
 \mbv_1^1 &= \twovect{0}{(\mbv_1^1)^2} , & 
 \mbv_1^2 &= \twovect{0}{\i(\mbv_1^1)^2} , & 
 \mbv_1^3 &= \twovect{(\mbv_1^3)^1}{0}, & 
 \mbv_1' &= \twovect{(\mbv_1')^1}{0}, \\
 \mbv_2^1 &= \twovect{(\mbv_2^1)^1}{0}, &
 \mbv_2^2 &= \twovect{-\i(\mbv_2^1)^1}{0}, &
 \mbv_2^3 &= \twovect{0}{(\mbv_2^3)^2}, &
 \mbv_2' &= \twovect{0}{(\mbv_2')^2}. 
\end{align}
\end{subequations}
where the vanishing components are in agreement with electric charge 
conservation. The six non-vanishing components are given by
\begin{subequations}
\begin{align}
\twovect{(\mbv_1^1)^2}{(\mbv_2^1)^1}&=
\frac{M_W}{\sqrt2} \Z_{H^\pm}^{-1T}  \twovect{\zeta^{G^\pm}}{\zeta^{H^\pm}},\\
\twomat{-(\mbv_1')^1 & (\mbv_2')^2 \\
        -(\mbv_1^3)^1 & (\mbv_2^3)^2}&=
\frac{M_Z}{\sqrt2} \Z_{A^0}^{-1} 
\twomat{\zeta^{A G^0} & \zeta^{A A^0} \\
        \zeta^{Z G^0} & \zeta^{Z A^0}}
\Z_V .
\end{align}
\label{eq:Goldstren}
\end{subequations}

The solution of the non-covariance problem is to make the isovectors $\mbv_i$ 
{\em local}, i.e.\ to introduce eight external isospin doublets 
$(\Phi^a_i+\mbv^a_i)$,  $(\Phi'_i+\mbv'_i)$ where the $\mbv_i$ appear as 
constant shifts.
Using these external fields, we redefine the gauge-fixing functions
into the final form,
\begin{subequations}
\label{eq:gaugefixingfct}
\begin{align}
\F^a &= \partial^\mu V^a_\mu
-[\i(\bar\Phi^a_1+\bar\mbv^a_1)H_1+\i(\bar\Phi^a_2+\bar\mbv^a_2)H_2
+{\rm h.c.}], \\
\F' &= \partial^\mu V'_\mu
-[\i(\bar\Phi'_1+\bar\mbv'_1)H_1+\i(\bar\Phi'_2+\bar\mbv'_2)H_2
+{\rm h.c.}],
\end{align}
\end{subequations}
which is used from now on. The point in the replacement $\mbv_i\to
(\Phi_i+\mbv_i)$ is that non-trivial gauge transformation laws can be
assigned to the external fields $\Phi_i$. With the choice
\begin{subequations}
\label{eq:rigidgaugaPhi}
\begin{align}
\delta_{\omega'} \Phi_i^a &= - \i g'\omega'\frac{Y}{2}(\Phi_i^a + \mbv_i^a), &
\delta_{\omega'} \Phi'_i &= -\i g' \omega'\frac{Y}{2}(\Phi'_i + \mbv'_i), \\
\delta_\omega \Phi_i^a &= - \i g \omega ( \Phi_i^a +\mbv_i^a) 
  - \i g   (\i \epsilon^{bca}) \omega^b ( \Phi_i^c + \mbv_i^c ), &
\delta_\omega \Phi'_i &= -\i g\, \omega (\Phi'_i + \mbv'_i),
\end{align}
\end{subequations}
the two 8-component objects $(\Phi_i+\mbv_i)$ for $i=1,2$ transform according 
to the product of the adjoint and the fundamental representation of 
SU(2)$\times$U(1), where the hypercharges in (\ref{eq:rigidgaugaPhi}) are 
defined as those of the corresponding Higgs fields (see table \ref{ta:fields}).
As a consequence, the gauge-fixing term (\ref{eq:gaugefixgeneral}) becomes 
invariant under rigid gauge transformations. For $\Phi_i=0$ the gauge-fixing 
functions reduce to (\ref{eq:gaugefix1}).

For the BRS transformations of the $\Phi_i$, we introduce further
external fields $\Psi_i^a$, 
$\Psi'_i$ with $Q_{\Phi\Pi}=1$ which form BRS doublets%
\footnote{Footnote \ref{BRSdoubletfootnote} applies also here.}
together with $\Phi_i^a$, $\Phi_i'$. The BRS transformations of the fields 
$\Phi_i$, $\Psi_i$ are defined as follows
\begin{subequations}
\begin{align}
\s \Phi_i^a &= \Psi_i^a -\i \xi^\mu \partial_\mu \Phi_i^a ,&
\s \Psi_i^a &= 2 \i \eps^\al \sigma^\mu_{\al \da} \bar\eps^\da\partial_\mu\Phi_i^a 
-\i \xi^\mu \partial_\mu \Psi_i^a , \\
\s \Phi'_i &= \Psi'_i -\i \xi^\mu \partial_\mu \Phi'_i ,&
\s \Psi'_i &= 2 \i \eps^\al \sigma^\mu_{\al \da} \bar\eps^\da \partial_\mu \Phi'_i 
-\i \xi^\mu \partial_\mu \Psi'_i.
\end{align}
\end{subequations}
The rigid gauge transformations and hypercharges of $\Psi_i$ are 
defined analogously to the ones of $\Phi_i$ but without $\mbv_i$.

\subsection{Classical action}
\label{sec:GammaCl}

With the gauge fixing introduced in the above section \ref{sec:gaugefixing}
we can complete the discussion of the classical approximation of the MSSM. 
Its building blocks are the manifestly supersymmetric part, 
$\Gamma_{\rm susy}$ (\ref{eq:susy}), the soft-supersymmetry breaking part, 
$\Gamma_{\rm soft}$ (\ref{eq:soft}), the external-field part, 
$\Gext$ (\ref{eq:ext}), and the gauge-fixing term, $\Ggf$ 
(\ref{eq:gaugefixgeneral}). Hence, a classical action satisfying all 
symmetries reads 
\begin{equation}
\Gcl=\Gamma_{\rm susy}+\Gamma_{\rm soft}+\Gext+\Ggf.
\label{eq:GclwithoutVEV}
\end{equation}

Although spontaneous symmetry breaking has already been taken into account in 
the definition of the gauge-fixing term, the finite vacuum expectation values 
of the Higgs bosons have not been introduced so far. Therefore, no
physical particle masses are generated in (\ref{eq:GclwithoutVEV}). In the
present framework, where rigid gauge invariance is
ensured by using the external fields $\Phi_i$ in the gauge-fixing
term, the vacuum expectation values arise along with a new class of
free parameters. The situation is similar to the one in
\cite{Kraus:1997bi,Kraus:1995jk}, where, however, the multiplet structure of 
the external fields $\Phi_i$ is simpler.

From the tensor fields ${\Phi}^a_i$, ${\Psi}^a_i$ we can form SU(2) doublets 
by contraction with the generators
\begin{align}
\hat\Phi_i+\hat{\mbv}_i &= 2 T^a (\Phi_i^a+\mbv_i^a),&
\hat\Psi &= 2 T^a \Psi_i^a.
\label{eq:hatPhidef}
\end{align}
The external fields $(\hat\Phi_i+\hat\mbv_i)$, $(\Phi'_i+\mbv'_i)$ have 
the same quantum numbers and rigid gauge-transformation laws as the physical 
Higgs fields $H_i$. However, the BRS transformations of the $\Phi$ fields 
prevent them from appearing arbitrarily in the classical action. 
A possibility to introduce them into the classical action consists in 
redefining the BRS transformation laws of the Higgs fields
\begin{equation}
\label{eq:HPhiredef}
\s H_i \to (\s H_i)|_{H_i\to H^{\rm eff}_i}
- x_i (\hat \Psi_i -\i \xi^\mu \partial_\mu \hat\Phi_i) 
- x'_i (\Psi'_i -\i \xi^\mu \partial_\mu \Phi'_i)
\end{equation}
and simultaneously substituting 
\begin{equation}
\label{eq:HPeff}
H_i \to H_i^{\rm eff} = 
H_i + x_i (\hat\Phi_i+\hat\mbv_i) + x'_i(\Phi'_i+\mbv')
\end{equation}
everywhere in the classical action apart from the gauge-fixing term.
The new parameters $x_i$, $x'_i$ are arbitrary and have to be fixed by 
normalization conditions. The $H_i^{\rm eff}$ obey the BRS transformation law of 
the original Higgs field. In this way, the complete vacuum expectation value 
of the effective Higgs fields $H^{\rm eff}_i$ is generated via the fields
$\hat\Phi_i+\hat\mbv_i$, $\Phi'_i+\mbv'_i$:
\begin{equation}
\langle H^{\rm eff}_i\rangle = H^{\rm eff}_i|_{\phi\to0}
= 2 x_i T^a \mbv_i^a + x_i' \mbv'_i \equiv \vecv_i.
\label{eq:decomposition}
\end{equation}
According to the form of $\mbv_i^a$, $\mbv'_i$ in (\ref{eq:Goldstren}),
the vacuum expectation values of the effective Higgs bosons read 
\begin{align}
\label{eq:VEVs}
\vecv_1&=\twovect{v_1}{0}, & \vecv_2&=\twovect{0}{v_2},
\end{align}
in agreement with electric charge conservation.
CP invariance restricts $v_1$, $v_2$ to be real.
When we perform the replacements (\ref{eq:HPhiredef}) and (\ref{eq:HPeff}) 
in $\Gcl$, we obtain another solution of the classical action obeying all 
symmetries. It turns out that this solution contains all parameters
needed for the renormalization. Hence, we adopt this solution and 
define the classical action $\Gcl$ in this way. $x_1$, $x_2$, $x'_1$, $x'_2$ 
are additional free parameters. Owing to the decomposition 
(\ref{eq:decomposition}) of $\vecv_1$, $\vecv_2$,  it is equivalent to 
eliminate $x'_1$, $x'_2$ in favour of $v_1$, $v_2$ and choose
\begin{align}
v_1, \quad v_2, \quad x_1, \quad x_2 
\end{align}
instead of $x'_1$, $x'_2$, $x_1$, $x_2$ as free parameters of the 
theory. 

The appearance of $\vecv_i$ in $H^{\rm eff}_i$ accounts for the spontaneous 
breaking of gauge invariance. In the following, we sketch the resulting 
bilinear part of $\Gcl$. The corresponding calculations are well-known 
in the literature (see  e.g.\ \cite{MSSM,Gunion:1989we}). 
Hence we just quote the results. The minimization conditions yield
\begin{subequations}
\label{eq:Min}
\begin{align}
-\dfrac{\Gcl}{{\rm Re}H_1^1}\bigg|_{\phi=0}
&=2\bigg[(m_1^2+\mu^2)v_1+m_3^2 v_2
+\frac{g^2+g'{}^2}{4}v_1(v_1^2-v_2^2)\bigg]\equiv\sqrt2 \, t_1
\stackrel{!}{=}0,\\
-\dfrac{\Gcl}{{\rm Re} H_2^2}\bigg|_{\phi=0}
&=2\bigg[(m_2^2+\mu^2)v_2+m_3^2 v_1
-\frac{g^2+g'{}^2}{4}v_2(v_1^2-v_2^2)\bigg]\equiv\sqrt2 \, t_2
\stackrel{!}{=}0.
\end{align}
\end{subequations}
The minimization of the Higgs potential is equivalent to vanishing
tadpole contributions $t_1$, $t_2$.

\subsubsection*{Higgs and gauge-boson sector}

The bilinear part of $\Gcl$ contains the mass matrices for the symmetric 
fields. These mass matrices have to be diagonalized by the demixing matrices 
$\Z_\phi$. In the Higgs and gauge-boson sector, the classical demixing 
matrices are rotation matrices with three angles $\theta_W$, $\beta$, 
$\alpha$ defined by
\begin{subequations}
\begin{align}
\tan \beta&=\frac{v_2}{v_1},&
\tan 2 \alpha &= \tan 2 \beta \frac{M_{A^0}^2+M_Z^2}{M_{A^0}^2-M_Z^2},&
\cos \theta_W &=\frac{M_W}{M_Z}
\label{eq:ThetaWDef}
\end{align}
\end{subequations}
with $-\pi/2<\alpha<0$. The shorthands $s_\alpha=\sin \alpha$, 
$s_\be=\sin \be$, $s_W=\sin \theta_W$ (and similar for cosine) are used in 
the following. Classically, the demixing matrices read
\begin{subequations}
\begin{align}
\Z^{(0)}_{A^0,H^\pm}&=\twomat{c_\be & -s_\be \\ s_\be & c_\be}, &
\Z^{(0)}_{H^0}&= \twomat{c_\al & -s_\al \\ s_\al & c_\al}, &
\Z^{(0)}_V&=
 \twomat{c_W & -s_W \\ s_W & c_W}.
\end{align}
\end{subequations}
After diagonalizing the mass matrices, the 
masses of the $W^\pm$ boson, $Z$ boson, and the $A^0$ field result in
\begin{subequations}
\begin{align}
M_W^2&=\frac{1}{2}g^2(v_1^2+v_2^2), \\
M_Z^2&=\frac{1}{2}(g^2+g^{\prime 2})(v_1^2+v_2^2),\\
M_{A^0}^2&=\sin^2\beta\frac{t_1}{\sqrt2 v_1}
+\cos^2\beta\frac{t_2}{\sqrt2 v_2}-m_3^2(\tan\beta+\cot\beta).
\end{align}
\end{subequations}
We consider $M_W^2, M_Z^2$, and $M^2_{A^0}$ 
as independent parameters in the following. Then
the masses of the remaining Higgs bosons are dependent parameters and
are determined in lowest order by
\begin{subequations}
\label{eq:MHDef}
\begin{align}
\label{eq:MHpmDef}
M_{H^\pm}^2{}^{(0)}&=M_{A^0}^2+M_W^2,\\
M_{H^0,h^0}^2{}^{(0)}&=\frac{1}{2}\left(M_{A^0}^2+M_Z^2\pm 
\sqrt{(M_{A^0}^2+M_Z^2)^2 - 4M_Z^2 M_{A^0}^2 \cos^2(2\beta)}\right).
\end{align}
\end{subequations}
As indicated by the superscript $^{(0)}$, these mass relations
are only valid in lowest order. In general, the mass relations 
(\ref{eq:MHDef}) are corrected by loop contributions.

In the chargino and neutralino sector, the lowest-order demixing matrices are 
conventionally denoted by
\begin{align}
\label{eq:UVNDef}
\Z^{(0)}_{\chi^+} & = {\cal V}^{-1},&
\Z^{(0)}_{\chi^-} & = {\cal U}^{-1},&
\Z^{(0)}_{\chi^0} & = {\cal N}^{-1},
\end{align}
where $\cal V$, $\cal U$, $\cal N$ are unitary matrices and satisfy
\begin{align}
{\cal N}^{ *}{\cal Y} {\cal N}^\dagger&=
\mbox{diag}(m_{\chi^0_1},\dots,m_{\chi^0_4}),&
{\cal U}^* {\cal X} {\cal V}^\dagger&=\mbox{diag}(m_{\chi^+_1},m_{\chi^+_2})
\end{align}
with the neutralino and chargino mass matrices 
\begin{subequations}
\begin{align}
{\cal Y}&= 
\left(\begin{array}{rrrr}
M_1                & 0                & -M_Z s_W  c_\beta & M_Z  s_W s_\beta \\
0                  & M_2              & M_Z c_W  c_\beta  & -M_Zc_W s_\beta \\
-M_Z s_W  c_\beta  & M_Z c_W  c_\beta & 0                 &  -\mu \\
M_Z  s_W s_\beta   & -M_Zc_W s_\beta  & -\mu              & 0
\end{array}\right) 
\label{YDef},\\
{\cal X}&=\left(\begin{array}{cc} 
M_2                  & M_W \sqrt{2} s_\beta \\
M_W \sqrt{2} c_\beta & \mu
\end{array}\right). 
\label{XDef}
\end{align}
\end{subequations}

After defining the Majorana spinors of the neutralinos and the
Dirac spinors of the charginos
(with $\bar\chi^\pm=\overline{\chi^\mp}$)
\begin{align}
\chi^0_i&={\chi^0_{i\alpha} \choose \bar\chi^{0\alphadot}_i},&
\chi^\pm_i&={\chi^\pm_{i\alpha} \choose \bar\chi^{\pm\alphadot}_i},
\end{align}
the bilinear part of the classical action contributing to the Higgs and 
gauge-boson sector reads
\begin{eqnarray}
\lefteqn{\Gamma_{\rm cl}|_{\rm bil,Higgs,gauge} =\int\dx \bigg\{
|\partial^\mu H^+|^2+|\partial^\mu G^+|^2-M_{H^\pm}^{2(0)} |H^+|^2}
\nonumber\\&&{}
+\frac12\left[(\partial^\mu H^0)^2+(\partial^\mu h^0)^2 +(\partial^\mu G^0)^2 
+(\partial^\mu A^0)^2-M_{H^0}^{2(0)} (H^0)^2-M_{h^0}^{2(0)} (h^0)^2
-M_{A^0}^2 (A^0)^2 \right]
\nonumber \\ && {} 
-\frac14 \left(A^{\mu\nu}A_{\mu\nu} +Z^{\mu\nu}Z_{\mu\nu} +
2W^{+\mu\nu}W^-_{\mu\nu}\right) 
+\frac12 M_Z^2 Z^\mu Z_\mu+M_W^2 W^{+\mu} W^-_\mu
\nonumber \\ && {} 
-M_Z Z^\mu \partial_\mu G^0
+\i M_W \left(W^{+\mu} \partial_\mu G^--W^{-\mu} \partial_\mu G^+ \right)
\nonumber \\ && {}
+\frac12 \sum_{i=1}^4\left(\bar\chi_i^0\i\gamma^\mu\partial_\mu\chi_i^0 
-m_{\chi_i^0}\bar\chi^0_i\chi^0_i \right)
+\sum_{i=1}^2\left(\bar\chi_i^-\i\gamma^\mu\partial_\mu\chi_i^+ 
-m_{\chi_i^+}\bar\chi^-_i\chi^+_i\right)\bigg\},
\label{eq:GclBil}
\end{eqnarray}
where the linear parts of the field strength tensors are given by 
$V^{\mu\nu}=\partial^\mu V^\nu-\partial^\nu V^\mu$ with $V=A,Z,W^\pm$.
The masses $M_{H^0}^{(0)}$, $M_{h^0}^{(0)}$, and 
$M_{H^\pm}^{(0)}$ and three of the neutralino and chargino masses are not 
considered as independent parameters of the MSSM, but they are functions of 
the other masses and couplings. 

\subsubsection*{Matter sector}

In the matter sector, the quark and lepton masses are given by
\begin{align}
m_u&={}-f_U v_2,& m_d&=f_D v_1,& m_e&=f_R v_1.
\end{align}
The mass matrices of the up- and down-type squarks in the basis 
$(Q^1,\bar{U})$, $(Q^2,\bar{D})$ have the following form
\begin{subequations}
\begin{align}
M_{\tilde{u}}^2 &= 
\left(\begin{array}{rr}
    m_u^2+m_{\tilde{q}}^2 + M_Z^2 c_{2\beta} (T_3-Q_{\rm em} s_W^2) 
 &  m_u\left(-\mu/\tan\beta + A_u \right) \\ 
    m_u\left(-\mu/\tan\beta + A_u \right) 
 &  m_u^2+m_{\tilde{u}}^2 + M_Z^2  c_{2\beta} Q_{\rm em} s_W^2 
\end{array}\right) ,\\
M_{\tilde{d}}^2 &= 
\left(\begin{array}{rr}
    m_d^2+m_{\tilde{q}}^2 + M_Z^2 c_{2\beta} (T_3-Q_{\rm em} s_W^2) 
 &  m_d\left(-\mu\tan\beta + A_d \right) \\ 
    m_d\left(-\mu\tan\beta + A_d \right) 
 &  m_d^2+m_{\tilde{d}}^2 + M_Z^2  c_{2\beta} Q_{\rm em} s_W^2 
\end{array}\right)
\label{eq:MSqDef}
\end{align}
\end{subequations}
with the electric charge $Q_{\rm em}$ and $c_{2\beta}=\cos(2\beta)$. 
The corresponding mass matrix for the selectron in the basis $(L^2,\bar{R})$ 
can be derived from (\ref{eq:MSqDef}) by replacing $d\to e$ and
$\tilde{d}\to\tilde{e}$. The demixing matrices $\Z_{\tilde{f}}$ have
to be chosen in such a way that they satisfy
\begin{equation}
\Z_{\tilde{f}}^{(0) -1} M_{\tilde{f}}^2 \Z_{\tilde{f}}^{(0)}=
\mbox{diag}(m_{\tilde{f}_1}^2,m_{\tilde{f}_2}^2)
\end{equation}
with $\tilde{f}=\tilde{u},\tilde{d},\tilde{e}$. Since there is no sneutrino 
mixing the sneutrino mass is given by
\begin{equation}
m_{\tilde{\nu}}^2=m_{\tilde{l}}^2 + \frac{1}{2} M_Z^2 \cos(2\beta). 
\end{equation}
With the definition of the Dirac spinors 
\begin{align}
e&={l^1_\alpha \choose \bar{r}^\alphadot},&
\nu&={l^2_\alpha \choose 0},&
u&={q^1_\alpha \choose \bar{u}^\alphadot},& 
d&={q^2_\alpha \choose \bar{d}^\alphadot},
\end{align}
and the lowest-order values
$\Z_{\nu_L}^{(0)}=\Z_{e_L}^{(0)}=\Z_{u_L}^{(0)}=\Z_{d_L}^{(0)}=1$,
the bilinear part of the matter sector can be written as
\begin{eqnarray}
\Gcl|_{\rm bil,matter}&=&\int\dx \Big[
\bar{f}\left(\i\gamma^\mu\partial_\mu - m_f\right)f
+\sum_{i=1,2}|\partial^\mu \tilde{f}_i|^2 -m_{\tilde{f}_i}^2 |\tilde{f}_i|^2 
\Big],
\end{eqnarray}
where $f$, $\tilde{f}$ run over all matter fields, i.e.\ $f=e,\nu,u,d$ and 
$\tilde{f}_i=\tilde{\nu}_i,\tilde{e}_i,\tilde{u}_i,\tilde{d}_i$.

\subsubsection*{Gauge-fixing sector}

We come back to the gauge-fixing and ghost terms already discussed in 
section \ref{sec:gaugefixing} in order to evaluate the mass eigenstates of 
ghosts. For vanishing $\Phi$ fields the gauge-fixing functions of 
(\ref{eq:gaugefixingfct}) reduce to the functions $\F^\pm$, $\F_A$, and 
$\F_Z$ defined in (\ref{eq:gaugefix1}). Classically, evaluating the
bilinear ghost terms in (\ref{eq:gaugefixgeneral}) yields
\begin{eqnarray}
\Gcl|_{\rm bil,gh}&=&\int\dx\Bigg\{-(\cbar_A,\cbar_Z) 
 \left[\Box+M_Z^2 \twomat{0 & \zeta^{AG^0} \\ 0 & \zeta^{ZG^0}} \right]
 \twovect{c_A}{c_Z}
\nonumber \\ && {}
-(\cbar^-,\cbar^+)\left[\Box + M_W^2 \zeta^{G^\pm}\right] 
 \twovect{c^+}{c^-}\Bigg\}
\end{eqnarray}
with $\Box=\partial^\mu \partial_\mu$ and where the lowest-order ghost
and antighost demixing matrices are given by
\begin{align}
\Z_c^{(0)}=\Z_{\cbar}^{(0)}=\Z_V^{(0)}.
\end{align}

An especially convenient choice is the $R_\xi$ gauge which corresponds to
\begin{align}
\zeta^{G^\pm}&=\zeta^{ZG^0}=\zeta'=\zeta\equiv\xi,&
\zeta^{H^\pm}&=\zeta^{ZA^0}=\zeta^{AG^0}=\zeta^{AA^0}=0.
\label{eq:Rxigauge}
\end{align}
In the $R_\xi$ gauge the gauge-boson--Higgs mixing in (\ref{eq:GclBil}) 
cancels when the auxiliary $B$ fields are eliminated by their equations of 
motion. Furthermore, the ghost and Goldstone-boson masses are
proportional to the masses of the corresponding gauge bosons:
\begin{subequations}
\begin{align}
M_{c_A}^2{}^{(0)}&=0,\\
M_{c_Z}^2{}^{(0)}&=M_{G^0}^2{}^{(0)}=\xi M_Z^2,\\
M_{c^\pm}^2{}^{(0)}&=M_{G^\pm}^2{}^{(0)}=\xi M_W^2.
\end{align}
\end{subequations}

\subsection{Free parameters of the MSSM}
\label{sec:physicalparameters}

The free parameters of the MSSM can be summarized as follows:
\begin{subequations}
\begin{align}
\label{eq:physparametersStart}
\text{coupling constants} \qquad & g, g', f_R, f_U, f_D, \mu, \\
\text{soft parameters} \qquad & M_1, M_2, m_1, m_2, m_3, m_{\tilde{l}},
m_{\tilde{e}}, m_{\tilde{q}}, m_{\tilde{u}}, m_{\tilde{d}}, A_e, A_u, A_d,
\\
\text{Higgs/$\Phi$ parameters}\qquad & 
v_1, v_2, x_1, x_2,
\label{eq:physparametersEnd}
\\
\text{gauge parameters}\qquad & 
\zeta, \zeta', \zeta^{G^\pm},\zeta^{H^\pm}, \zeta^{ZG^0},
\zeta^{ZA^0}, \zeta^{AG^0}, \zeta^{AA^0}.
\label{eq:gaugeparameters}
\end{align}  
\end{subequations}
In section \ref{sec:GammaCl} some of these parameters were related to a 
set of parameters with a more direct physical interpretation. 
Schematically, the relations between the two sets of parameters are
\begin{eqnarray}
g, g', v_1, v_2 & \leftrightarrow & e, M_Z, M_W, \tan\beta,\\
m_1, m_2, m_3 & \leftrightarrow & M_{A^0}, t_1, t_2, \\
\mu, M_1, M_2 & \leftrightarrow & m_{\chi^\pm_1}, m_{\chi^\pm_2}, m_{\chi^0_1},\\
f_R, f_U, f_D & \leftrightarrow & m_e, m_u, m_d,\\
m_{\tilde{l}}, m_{\tilde{e}}, m_{\tilde{q}}, m_{\tilde{d}}, m_{\tilde{u}} & 
\leftrightarrow & m_{\tilde{\nu}_1},  m_{\tilde{e}_{1}},  m_{\tilde{u}_{1}},  
m_{\tilde{u}_{2}},  m_{\tilde{d}_{1}},
\end{eqnarray}
where the electric charge is defined by
\begin{eqnarray}
e & = & g\sin\theta_W = g^\prime\cos\theta_W.
\end{eqnarray}
Of course, the choice of the independent parameters in the neutralino 
and sfermion sectors is not unique. 

We close with a remark on the mass relations present in the MSSM. In the 
MSSM not all masses are independent. The masses of $H^\pm$, $H^0$, $h^0$, 
$\chi^0_{2,3,4}$, and of $\tilde{e}_2$, $\tilde{d}_2$ are functions 
of the free parameters of the MSSM. Accordingly, there is a distinction
between the notions $M_{H^\pm}^2{}^{(0)}$ and 
$M_{H^\pm}^2$ (and similar for the other dependent masses). 
$M_{H^\pm}^2{}^{(0)}$ denotes the result of the lowest-order formula 
(\ref{eq:MHpmDef}) and is thus always to be treated as an abbreviation for 
$M_{A^0}^2+M_W^2$. In particular, in the renormalization
transformation 
$M_{H^\pm}^2{}^{(0)}\to M_{H^\pm}^2{}^{(0)}+\delta M_{H^\pm}^2$, 
used in section \ref{sec:renscheme}, $\delta M_{H^\pm}^2$ is a 
substitute of $\delta M_{A^0}^2+\delta M_W^2$. On the other hand, 
$M_{H^\pm}^2$ denotes the physical mass of the charged Higgs boson, which can 
be calculated order by order as a function of the independent
parameters of the MSSM.

\section{Renormalization} 
\label{sec:Renormalization}
\setcounter{equation}{0}

In this section general aspects of the renormalization of the MSSM are 
presented. After establishing the symmetries in a functional way by 
introducing the ST and Ward operators, all defining symmetries
of the MSSM are summarized, and all invariant and symmetry-restoring 
counterterms required for the renormalization are discussed. The section is
completed with a proof of the IR finiteness of the theory.

\subsection{Functional characterization of the symmetries}
\label{sec:symmetries}

In order to characterize the MSSM in a regularization-scheme
independent way, it is convenient to formulate the fundamental
symmetries in terms of a set of functional equations comprising the ST
identity for BRS invariance and the Ward identities for rigid gauge and
R invariance. 

\subsubsection*{ST identity}

BRS invariance is expressed in form of a ST identity 
\begin{equation}
\S(\Gamma)=0
\end{equation}
with the ST operator
\begin{eqnarray}
\label{eq:STOp}
\S (\Gamma)&=&\int \dx \bigg\{
2 \tr \left[
 \frac{\delta \Gamma}{\delta Y_V^{\mu}} 
 \frac{\delta \Gamma}{\delta V_\mu} 
+\frac{\delta \Gamma}{\delta y_{\lambda \alpha}} 
 \frac{\delta \Gamma}{\delta \lambda^{\alpha}}
+\frac{\delta \Gamma}{\delta \bar{y}_{\lambda}^{\dot{\alpha}}} 
 \frac{\delta \Gamma}{\delta \bar{\lambda}_{\dot{\alpha}}} 
+\frac{\delta \Gamma}{\delta Y_c} 
 \frac{\delta \Gamma}{\delta c} 
+\s \bar{c} \frac{\delta \Gamma}{\delta \bar{c}} 
+\s B \frac{\delta \Gamma}{\delta B} 
\right]
\nonumber \\ && {}
+\s  V^\prime_\mu
 \frac{\delta \Gamma}{\delta V^\prime_\mu} 
+\frac{\delta \Gamma}{\delta y_{\lambda^\prime \alpha}} 
 \frac{\delta \Gamma}{\delta \lambda^{\prime \alpha}}
+\frac{\delta \Gamma}{\delta \bar{y}_{\lambda^\prime}^{\dot{\alpha}}} 
 \frac{\delta \Gamma}{\delta \bar{\lambda}^{\prime}_{\dot{\alpha}}} 
+\s c^\prime \frac{\delta \Gamma}{\delta c^\prime} 
+\s \bar{c}^\prime \frac{\delta \Gamma}{\delta \bar{c}^\prime} 
+\s B^\prime \frac{\delta \Gamma}{\delta B^\prime} 
\nonumber \\ && {}
+\s \au \frac{\delta \Gamma}{\delta \au}
+\s \aub \frac{\delta \Gamma}{\delta \aub}
+\s \av \frac{\delta \Gamma}{\delta \av}
+\s \avb \frac{\delta \Gamma}{\delta \avb}
\nonumber \\ && {}
+\s \Phi_i^{aT} \frac{\delta \Gamma}{\delta \Phi_i^{aT}}
+\s \Phi_i^{\prime T} \frac{\delta \Gamma}{\delta \Phi_i^{\prime T}}
+\s \Psi_i^{aT} \frac{\delta \Gamma}{\delta \Psi_i^{aT}}
+\s \Psi_i^{\prime T} \frac{\delta \Gamma}{\delta \Psi_i^{\prime T}}
\nonumber \\ && {}
+\s \bar\Phi_i^a \frac{\delta \Gamma}{\delta \bar\Phi_i^a}
+\s \bar\Phi_i' \frac{\delta \Gamma}{\delta \bar\Phi_i'}
+\s \bar\Psi_i^a \frac{\delta \Gamma}{\delta \bar\Psi_i^a}
+\s \bar\Psi_i' \frac{\delta \Gamma}{\delta \bar\Psi_i'}
\nonumber \\ && {}
+\frac{\delta \Gamma}{\delta Y_{L}} 
 \frac{\delta \Gamma}{\delta L^T} 
+\frac{\delta \Gamma}{\delta \bar{Y}_{L}^T} 
 \frac{\delta \Gamma}{\delta \bar{L}} 
+\frac{\delta \Gamma}{\delta y_{l \alpha}} 
 \frac{\delta \Gamma}{\delta l^{\alpha T}}
+\frac{\delta \Gamma}{\delta \bar{y}_l^{\dot{\alpha} T}} 
 \frac{\delta \Gamma}{\delta \bar{l}_{\dot{\alpha}}} 
\nonumber \\ && {}
+\frac{\delta \Gamma}{\delta Y_{R}} 
 \frac{\delta \Gamma}{\delta R} 
+\frac{\delta \Gamma}{\delta \bar{Y}_{R}} 
 \frac{\delta \Gamma}{\delta \bar{R}} 
+\frac{\delta \Gamma}{\delta y_{r \alpha}} 
 \frac{\delta \Gamma}{\delta r^{\alpha}}
+\frac{\delta \Gamma}{\delta \bar{y}_{r}^{\dot{\alpha}}} 
 \frac{\delta \Gamma}{\delta \bar{r}_{\dot{\alpha}}} 
\bigg\}
+\s \xi^\mu \frac{\partial \Gamma}{\partial \xi^\mu}
\nonumber \\ && {}
+\mbox{analogous quark and Higgs terms}.
\end{eqnarray}
For subsequent discussions we introduce the condensed notation
\begin{equation}
\S(\Gamma) = \sum_a \fdq{\Gamma}{Y_a} \fdq{\Gamma}{\phi_a} + \sum_b \s\phi_b'
\fdq{\Gamma}{\phi_b'},
\label{STopdef}
\end{equation}
where $\phi_a$ runs over all non-linearly transforming fields and $\phi'_b$
over all linearly transforming ones. Owing to the construction of $\Gcl$, 
the ST identity $\S(\Gcl)=0$ holds at the classical level. At higher orders, 
$\S(\Gamma)=0$ will be imposed as a condition on the vertex functional 
$\Gamma$. 

In addition to the ST operator, we define the linearized ST operator 
\begin{equation}
\S_\gamma =\sum_a \left(\fdq{\gamma}{Y_a} \fdq{}{\phi_a}+
  \fdq{\gamma}{\phi_a}\fdq{}{Y_a} \right) + \sum_b \s\phi_b' \fdq{}{\phi_b'},
\label{STlinopdef}
\end{equation}
which obeys the relation
\begin{equation}
\S(\gamma+\Delta) = \S(\gamma) + \S_\gamma \Delta + \O(\Delta^2),
\end{equation}
where $\gamma$, $\Delta$ are arbitrary field polynomials.

\label{sec:AlgebraicRelations}

For the algebraic characterization of possible anomalies the nilpotency 
relations of the ST operator
\begin{subequations}
\label{eq:nilpot}
\begin{align}
\label{eq:mixednilpot}
\S_\gamma \S(\gamma) &=0 \qquad \text{for all $\gamma$,} \\
\label{eq:Bnilpot}
\S_\Gcl \S_\Gcl &=0, 
\end{align}
\end{subequations}
have to be realized. These nilpotency relations are important
ingredients for the algebraic proof of the renormalizability discussed
later in section \ref{sec:strategy}.

However, a short calculation shows that (\ref{eq:mixednilpot}) is in general 
not true:
\begin{equation}
\label{eq:nilpotviolation}
\S_\gamma \S(\gamma) = \sum_b (\S_\gamma \s\phi_b') \fdq{\gamma}{\phi_b'} =
(\S_\gamma \s V_\mu') \fdq{\gamma}{V_\mu'}.
\end{equation}
The non-vanishing r.h.s.~of (\ref{eq:nilpotviolation}) originates from the 
different treatment of linear and non-linear BRS transformations in the ST 
operator (\ref{eq:STOp}). BRS transformations of the linearly
transforming fields are "hard coded" in  
$\S$ while those of the non-linearly transforming ones are coupled to source 
fields $Y$, $y$ and, therefore, obtain higher-order corrections. We ensure 
(\ref{eq:mixednilpot}) for $\gamma=\Gamma$ by imposing the requirement 
\begin{equation}
\S_\Gamma \s V_\mu'=0, 
\label{eq:Vlambda_consistency}
\end{equation}
which holds in particular for $\Gamma=\Gcl$.
In terms of the linear functional 
differential operator 
\begin{equation}
\wl_\mu = \eps^\al \sigma_{\mu\al\da} \fdq{}{\bar y_{\lambda' \da}} 
-\fdq{}{y_{\lambda' \al}} \sigma_{\mu\al\da} \bar \eps^\da,
\label{eq:wldef}
\end{equation}
the requirement (\ref{eq:Vlambda_consistency}) can be rewritten as
the linear equation
\begin{equation}
\wl_\mu \Gamma = 2 \i \eps^\alpha \sigma^\nu_{\alpha \dot{\alpha}} 
\bar{\eps}^{\dot{\alpha}} F'_{\mu\nu} + \i \xi^\nu
\partial_\nu \left( \eps^\al \sigma_{\mu \al \da} \bar {\lambda'}^\da 
+ {\lambda'}^\al \sigma_{\mu \al \da}  \bar \eps^\da 
\right). 
\label{eq:Vlambda_consistency2}
\end{equation}

Also the second nilpotency condition is in general violated.
Using $\S(\Gcl)=0$ and (\ref{eq:Vlambda_consistency}), the square of 
$\B$ yields the non-vanishing result
\begin{equation}
\S_\Gcl \S_\Gcl = 
- \fdq{(\s V_\mu')}{\lambda'{}^\al} \fdq{\Gcl}{V_\mu'}
  \fdq{}{y_{\lambda'}{}^\al}
+ \fdq{(\s V_\mu')}{\bar \lambda'{}^\da}\fdq{\Gcl}{V_\mu'}
  \fdq{}{\bar y_{\lambda'\da}}.
\label{eq:Sgcl2}
\end{equation}
However, it is possible to define a space of constrained functionals on which 
the linearized ST operator is nilpotent. Since the r.h.s.\ of 
(\ref{eq:Sgcl2}) is proportional to $\wl_\mu$, the operator $\S_\Gcl$
is nilpotent on the kernel of $\wl_\mu$, i.e.
\begin{equation}
\S_\Gcl^2 \gamma =0\qquad \text{if $\wl_\mu \gamma=0$}.
\end{equation}
This is a weaker but sufficient form of (\ref{eq:Bnilpot}). The constraint 
$\wl_\mu \gamma=0$ means simply that $\gamma$ depends not arbitrarily on 
$y_\lambda'$, $\bar y_\lambda'$ but only on a certain combination, 
\begin{equation}
\gamma(y_{\lambda'}, \bar y_{\lambda'}) = \gamma(\Upsilon), \qquad \Upsilon =
\eps^\al y_\lambda'{}_\al - \bar \eps_\da \bar y_\lambda'{}^\da.
\label{eq:Upsilondef}
\end{equation}

\subsubsection*{Ward identities}

Rigid gauge invariance is formulated in the form of Ward identities
\begin{align}
\W(\omega) \Gamma&=0, \qquad \W'(\omega') \Gamma=0,
\end{align}
where the Ward operators for rigid SU(2), $\W$, and for rigid U(1) gauge 
invariance, $\W'$, are defined by
\begin{subequations}
\label{eq:Wardoperators}
\begin{align}
\W(\omega)&=\int \dx \left[
 \sum_a \delta_\omega \phi_a^\prime \frac{\delta}{\delta \phi_a^\prime} 
+\sum_b \left(\delta_\omega \phi_b \frac{\delta}{\delta \phi_b} 
+\delta_\omega Y_b \frac{\delta}{\delta Y_b}\right)\right], \\ 
\label{eq:WIAbelian}
\W'(\omega')&=\int \dx \left[
 \sum_a \delta_{\omega^\prime} \phi_a^\prime \frac{\delta}{\delta
 \phi_a^\prime} 
+\sum_b \left(\delta_{\omega^\prime} \phi_b \frac{\delta}{\delta \phi_b}
+\delta_{\omega^\prime} Y_b \frac{\delta}{\delta Y_b}\right) \right].
\end{align}
\end{subequations}
The Ward operators for lepton-number conservation, $\W_L$, for quark-number 
conservation, $\W_Q$, and for R symmetry, $\W^R$ are defined in an obvious way.
Explicit expressions for the Ward operators are given in appendix 
\ref{ap:operators}. 

Since the gauge group of the MSSM is non-semisimple not all couplings
of the classical action
are fixed by the algebra of the gauge group. The abelian couplings,
i.e.~the hypercharges of matter couplings to $V'_\mu$,
are not determined by (\ref{eq:WIAbelian}) but have to be determined
by an additional  symmetry, either by the local U(1) 
Ward identity or by the corresponding ghost equation
\cite{Kraus:1997bi,BBBC,Grassi:1997mc}.
For the classical action
 the hypercharges of the MSSM are determined 
 by the Gell-Mann--Nishijima relation 
(\ref{eq:Qem}).  At higher orders, the functional analog of the
Gell-Mann--Nishijima relation holds for the Ward operators:
\begin{equation}
\label{eq:Wem}
\frac{1}{e}\W_{\rm em}(\omega_{\rm em}) = 
  \left[\frac{1}{g} \W(T^3 \omega_{\rm em})  
+ \frac{1}{g'} \W'(\omega_{\rm em}) \right],
\end{equation}
where the Ward operator of electromagnetic symmetry
$\W_{\rm em}$ is defined similar to (\ref{eq:Wardoperators}) with
$\delta_{\omega_{\rm em}}=-\i e \omega_{\rm em}Q_{\rm em}$. 

We choose to determine the abelian couplings by the local Ward
identity corresponding to the abelian operator $\W'$, which can be 
derived also in supersymmetric models \cite{Hollik:1999xh}:
\begin{equation}
\label{eq:localWI}
\w' \Gamma=\Box (B' + \i \xi^\mu \partial_\mu \bar{c}')
\end{equation}
with the Ward operator
\begin{equation}
\w'=\frac{1}{g'}\left[\sum_{a}
 \delta_{\omega'} \phi_a^\prime \frac{\delta}{\delta \phi_a^\prime}
+\sum_b \left(\delta_{\omega'} \phi_b \frac{\delta}{\delta \phi_b}
+\delta_{\omega'} Y_b \frac{\delta}{\delta Y_b}\right) \right]
-\partial^\mu \fdq{}{{V'}^\mu}.
\end{equation}
If (\ref{eq:localWI}) is established to all orders, the hypercharges 
are fixed in such a way
that electromagnetic symmetry is gauged  and that the Adler-Bardeen
anomaly is cancelled to all orders. 

Due to parity non-conservation and due to supersymmetry,  invariant
regularization schemes of the MSSM probably do not exist. 
Hence one has to construct the MSSM by algebraic renormalization
restoring the defining symmetries by 
(scheme-dependent) non-invariant counterterms order by order in perturbation
theory. For this purpose one needs a complete algebraic
characterization of invariant counterterms and breaking terms, which
is governed by the algebra
of defining symmetry operators.
 Hence, we conclude this section by 
listing the commutation relations of the ST and Ward operators:
\begin{subequations}
\begin{align}
[\W(\omega_1), \W(\omega_2)]&=\W([\omega_1,\omega_2]),\\
[\W_1,\W_2]&=0, \\
[\W(\omega),\W_1]&=0, \\
[\w',\W_1]&=0,\\
\label{cons_WST}
\W(\omega) \S(\Gamma)-\S_\Gamma \W(\omega) \Gamma &=0,\\
\label{cons_WprimeST}
\W_1 \S(\Gamma)-\S_\Gamma \W_1 \Gamma &=0,\\
\w' \S(\Gamma)-\S_\Gamma \w' \Gamma &=
-2 \i \eps^\alpha \sigma^\mu_{\alpha \dot{\alpha}} 
 \bar\eps^{\dot{\alpha}} \partial_\mu \frac{\delta \Gamma}{\delta c'}
+\i \xi^\nu \partial_\nu \partial^\mu\frac{\delta \Gamma}{\delta V'{}^\mu},
\label{eq:commutelocalWO}
\end{align}
\end{subequations}
where $\W_1$, $\W_2$ stand for the Ward operators $\W'(\omega')$, $\W_L$, 
$\W_Q$, $\W^R$.

\subsection{Defining symmetries} 
\label{sec:symmetryrequirements}

To define a quantum theory in the framework of algebraic renormalization, 
a suitable set of requirements has to be imposed on the vertex
functional $\Gamma$. $\Gamma$ is calculated order by order in perturbation 
theory starting from the classical action. In higher orders,
intermediate results for regularized loop diagrams and counterterms
are regularization-scheme dependent, but the final results for
observables are regularization-scheme independent and unique if the
requirements are chosen appropriately.  

A complete set of conditions for the MSSM is given in the following, where
symmetric fields are understood to be expressed by the physical
fields. We will see that this set fixes $\Gamma$ uniquely up to
the usual freedom of mass, coupling-constant, and wave-function
renormalizations. 

The conditions express the symmetries, the minimization of the
Higgs potential as well as constraints for the gauge fixing and the
$\xi^\mu$ dependence.
\begin{itemize}
\item CP invariance (see table \ref{ta:CP}).
\item $\Gamma$ depends on $\au$ only through the combination $\eps_\al \au$.
\item ST identity with the ST operator (\ref{eq:STOp}):
\begin{equation}
\label{eq:STidentity}
\S(\Gamma)=0.
\end{equation}
\item Nilpotency condition for the ST operator:
\begin{equation}
\wl_\mu \Gamma = 2 \i \eps^\al \sigma^\nu_{\al \da} \bar \eps^\da 
  F'_{\mu\nu} + \i \xi^\nu \partial_\nu 
  \left( \eps^\al \sigma_{\mu \al \da} \bar {\lambda'}^\da 
+ {\lambda'}^\al \sigma_{\mu \al \da} \bar \eps^\da \right),
\label{eq:NilpotencyReq}
\end{equation}
where $\wl_\mu$ is defined in (\ref{eq:wldef}).
\item Rigid SU(2)$\times$U(1) gauge invariance:
\begin{align}
\W \Gamma &=0,\qquad \W' \Gamma=0.
\end{align}
\item Lepton- and quark-number conservation:
\begin{align}
\W_L \Gamma&=0,\qquad \W_Q \Gamma=0.
\end{align}
\item Continuous R symmetry:
\begin{equation}
\W^R \Gamma=0.
\end{equation}
\item Local U(1) gauge invariance:
\begin{equation}
\label{eq:LocalWIDef}
\w' \Gamma=\Box (B' + \i \xi^\mu \partial_\mu \bar{c}').
\end{equation}
\item Gauge-fixing conditions:
\begin{align}
\fdq{\Gamma}{B^a}&=\F^a+\zeta B^a,\qquad \fdq{\Gamma}{B'}=\F'+\zeta' B',
\label{eq:gaugecondition}
\end{align}
where $\F^a$, $\F'$ are defined in (\ref{eq:gaugefixingfct}).
\item Translational-ghost equation:
\begin{equation}
\fdq{\Gamma}{\xi_\mu} = - \i \sum_a Y_a \partial^\mu \phi_a.
\label{eq:ghotsequation}
\end{equation}
\item Minimum requirement for the Higgs potential:
\begin{equation}
\int \dx \left. \fdq{\Gamma}{H_i(x)} \right|_{\phi\to 0}=0.
\end{equation}
\end{itemize}

\subsection{Cohomology and symmetry-restoring counterterms} 
\label{sec:strategy}

The general strategy of calculating radiative corrections to 
a given order in perturbation theory for a renormalizable model 
consists of three steps:
\begin{itemize}
\item In the first step all loop diagrams of order $\hbar^n$ are calculated within a chosen regularization scheme.
\item In general the result does not satisfy the defining symmetry
properties. To restore the symmetry identities symmetry-restoring 
counterterms $\Grest^{(n)}$ have to be added. One contribution to
$\Grest$ originates from the non-invariance of the regularization scheme. 
Another contribution to $\Grest$ is given by products of lower-order 
counterterms involving for instance $\delta g^{(n-m)}\delta z^{(m)}$ 
with $0<m<n$.
\item Finally, invariant counterterms $\Ginv^{(n)}$ are added to cancel
UV singularities and to establish the normalization conditions.
\end{itemize}

The classical action, the invariant counterterms $\Ginv$ (see section 
\ref{sec:CTs}), and the symmetry-restoring counterterms $\Grest$ add up to 
\begin{equation}
\label{eq:defGeff}
\Geff = \Gcl+\sum_{n=1}^\infty \left(\Ginv^{(n)}+\Grest^{(n)}\right)
\end{equation}
from which the Feynman rules are derived. 
The full vertex functional $\Gamma$ differs from $\Geff$ by non-local 
loop contributions.

In the following, we give a few comments to the symmetry-restoring 
counterterms and possible anomalies. The third 
step is treated in detail in section \ref{sec:CTs}.

We consider the ST identity (\ref{eq:STidentity}) and the nilpotency 
requirement (\ref{eq:NilpotencyReq}) as an example. 
Assuming that all symmetry identities have already been established up to 
order $\hbar^{n-1}$ the symmetry identities yield
\begin{align}
\wl_\mu\Gamma - \mbox{r.h.s.\ of (\ref{eq:NilpotencyReq})}&=
\hbar^n\Delta_{\wl}{}_\mu+\O(\hbar^{n+1}),
\label{eq:Nilbreak}\\
\S(\Gamma) &= \hbar^n\Delta_{\rm ST}+\O(\hbar^{n+1}).
\end{align}
Owing to the quantum action principle \cite{QAP}, $\Delta_{\wl}$, 
$\Delta_{\rm ST}$ are 
local functionals. They have to be absorbed by adding symmetry-restoring 
counterterms of order $\hbar^n$ to $\Geff$. In this way, the symmetry 
properties can be restored in the given order.

Since the nilpotency condition (\ref{eq:NilpotencyReq}) is the basis for the
restoration of the ST identity, (\ref{eq:Nilbreak}) has to be recovered as 
a first step. It can be shown that $\Delta_{\wl}$ can be written as a variation
\begin{equation}
\Delta_{\wl}{}_\mu = \wl_\mu\hat{\Delta}_{\wl}
\end{equation}
with a local, power-counting renormalizable functional $\hat\Delta_{\wl}$,
we can add $-\hat\Delta_{\wl}$ to $\Gamma$ and recover (\ref{eq:Nilbreak})
in order $\hbar^n$.

Similarly, the breaking of the ST identity can be absorbed, if 
$\Delta_{\rm ST}$ can be written as 
\begin{equation}
\label{eq:cohomology}
\Delta_{\rm ST}=\B\hat{\Delta}_{\rm ST}.
\end{equation} 
The question whether any $\Delta_{\rm ST}$ can be written in the form of 
(\ref{eq:cohomology}) can be reduced to the algebraic problem to find the 
cohomology of $\B$. In our case, the nilpotency of the ST operator $\B$ 
has to be established by restricting $\Delta_{\rm ST}$ and 
$\hat\Delta_{\rm ST}$ to the kernel of $\wl_\mu$ or equivalently to 
functionals depending on $y_{\lambda'}$, $\bar{y}_{\lambda'}$ only 
through $\Upsilon$ defined in (\ref{eq:Upsilondef}). Indeed, if the 
nilpotency condition (\ref{eq:NilpotencyReq}) holds we find
\begin{align}
0=\wl_\mu \S(\Gamma)&=\wl_\mu \Delta_{\rm ST}+\O(\hbar^{n+1}),\\
0=\S_\Gamma \S(\Gamma)&=\S_\Gcl \Delta_{\rm ST}+\O(\hbar^{n+1}),
\label{nilpotconsist}
\end{align}
which implies that $\Delta_{\rm ST}$ lies in the kernel of $\wl_\mu$ 
and satisfies the consistency condition for possible anomalies:
\begin{align}
\B\Delta_{\rm ST}&=0.
\label{eq:conscond}
\end{align}
Since $\B^2 \gamma =0$ for $\wl_\mu\gamma=0$ as shown in section
\ref{sec:AlgebraicRelations}, the most general solution of the
consistency condition can be written as follows
\begin{align}
\Delta_{\rm ST} &= \sum_i r_i \A_i +\B \hat \Delta_{\rm ST} &&  
\text{with } 
\wl_\mu \hat \Delta_{\rm ST}=0,  \label{eq:anomalies}
\end{align}
where $\B \hat \Delta_{\rm ST}$ is a trivial solution of
(\ref{eq:conscond}). The breaking terms $\A_i$ satisfy $\B\A_i=0$, but 
cannot be written as $\B\hat{\A}_i$, i.e.\ they span the cohomology 
of $\B$ on the kernel of $\wl_\mu$. Hence, $\A_i$ constitute possible 
anomalies. 

We assume that the only possible anomaly is the Adler-Bardeen anomaly
in its supersymmetric version, whose coefficient $r_i$ vanishes in the
MSSM owing to the choice of the matter field representations. So we are left 
with the breaking terms $\B\hat\Delta_{\rm ST}$. $\Delta_{\rm ST}$ can be 
cancelled by adding the counterterms $-\hat\Delta_{\rm ST}$ to $\Gamma$. 
Because of (\ref{eq:anomalies}) they do not destroy the nilpotency relation 
and both symmetry requirements can be satisfied simultaneously.
The remaining symmetries can be restored in a similar manner. 

\subsection{Invariant counterterms}
\label{sec:CTs}

Having established the defining symmetries of the model by adding suitable
symmetry-restoring counterterms, we are still free to add invariant 
counterterms which are in agreement with all symmetry properties. There are 
two fundamental types of invariant counterterms: the first type is required 
for the cancellation of UV singularities, while the second type comprises 
finite field reparametrizations as discussed in section \ref{sec:fieldparam}. 

Invariant counterterms of the first type have to respect all symmetry 
properties of the MSSM (see section \ref{sec:symmetryrequirements}). In 
particular, the counterterms of order $\hbar^n$ are restricted by
\begin{subequations}
\label{eq:GinvConditions}
\begin{align}
\B\Ginv^{(n)}&=0,&
\wl_\mu\Ginv^{(n)}&=0,&
W\Ginv^{(n)}&=0, \\
\frac{\delta\Ginv^{(n)}}{\delta B}&=0,&
\frac{\delta\Ginv^{(n)}}{\delta B'}&=0,&
\frac{\delta\Ginv^{(n)}}{\delta \xi^\mu}&=0,
\end{align}
\end{subequations}
where the Ward operator $W$ runs over $\W,\W',\W_L,\W_Q,\W^R$. 
If the vertex functional $\Gamma$ satisfies all symmetry identities, 
$\Gamma+\Ginv$ satisfy the symmetries as well. Since $\B$ is nilpotent 
on the kernel of $\wl_\mu$, the counterterms $\Ginv^{(n)}$ can be rewritten 
into the form
\begin{equation}
\Ginv^{(n)} = \Ginva^{(n)} + \B\hat\Gamma_{\rm ct,inv,1b}^{(n)},
\end{equation}
where $\B\Ginva^{(n)}=0$ but $\Ginva^{(n)}$ cannot be written as a
$\B$ variation. 

The strategy to find all invariant counterterms of the first type is the
following: we first neglect the soft and spontaneous symmetry breaking and 
determine the most general contribution to $\Ginva$. Then it is shown that the 
soft and spontaneous symmetry breaking contribute only to $\Ginvb$. 
In a second step, the most general form of $\Ginvb$ is constructed.

The structure of $\Ginva^{(n)}$ can be derived from the supersymmetric 
SU(2)$\times$U(1) gauge theory without spontaneous and soft-symmetry breaking, 
where the spurion fields and the $\Phi$ fields are absent. The invariant 
counterterms $\Ginva^{(n)}$ correspond to the free parameters of 
$\Gamma_{\rm susy}$ (\ref{eq:susy}) \cite{Maggiore:1995gr}. In the MSSM, 
$\Ginva^{(n)}$ can be written as a combination of parameter 
renormalizations of the supersymmetric and gauge-invariant couplings 
$g,f_{R,U,D},\mu,v'$ and of a combined field and parameter renormalization of 
$V'{}^\mu,\lambda',y_{\lambda'},c',\cbar',B',\zeta',g'$ as in 
\cite{Hollik:1999xh}. The counterterm for $g'$ is not independent but 
governed by the field renormalization of $V'{}^\mu$ owing to the requirement 
of the local U(1) Ward identity (\ref{eq:LocalWIDef}). Note that $V'{}^\mu$ 
is the only propagating, linearly transforming field of the MSSM. The field 
renormalizations of the propagating, non-linearly transforming fields are 
included in $\Ginvb$ due to the presence of $\Gext$. Hence, $\Ginva^{(n)}$ 
takes the form
\begin{eqnarray}
\Ginva^{(n)} &=&
\bigg(\deltan g \pdq{}{g}
 + \deltan f_R \pdq{}{f_R} + \deltan f_U \pdq{}{f_U}
 + \deltan f_D \pdq{}{f_D}
 + \deltan \mu \pdq{}{\mu} + \deltan v' \pdq{}{v'}
\bigg)\Gcl
\nonumber\\&& {}
+\frac12\deltan z_{V'}\bigg[
\int\dx\bigg(
V'{}^\mu\fdq{}{V'{}^\mu}
+\lambda'{}^\alpha\fdq{}{\lambda'{}^\alpha}
-y_{\lambda'}^\alpha\fdq{}{y_{\lambda'}^\alpha}
+\lambdabar'_\alphadot\fdq{}{\lambdabar'_\alphadot}
-\bar{y}_{\lambda'}{}_\alphadot\fdq{}{\bar{y}_{\lambda'}{}_\alphadot}
\nonumber\\&&{}
+c'\fdq{}{c'}-\cbar'\fdq{}{\cbar'}-B'\fdq{}{B'}
\bigg)
-g'\pdq{}{g'}+2\zeta'\pdq{}{\zeta'}\bigg]\Gcl.
\end{eqnarray}

Special attention has to be drawn to the spurion parameters. These parameters
are introduced in section \ref{sec:spurion} with the help of the spurion 
superfield, and the spurion superfield is replaced later by the BRS doublet 
$(A_1,A_2+v_A)$ 
[c.f.\ (\ref{spurionexpansion})]. This replacement does not invalidate the
conditions (\ref{eq:GinvConditions}), but it excludes certain
counterterms involving $A,a_\alpha$. As a special property of BRS doublets, 
all counterterms involving BRS doublets are $\B$ variations and thus 
included in $\Ginvb^{(n)}$, which follows from the well-known theorems 
(see e.g.\ section 5.2 of \cite{PiSo}) implying that the cohomology of 
$\B$ is isomorphic to a subspace of the cohomology of 
$\B|_{\text{doublets=0}}$ in the space where the doublets are set to zero. 

Similar arguments hold also for the invariant counterterms 
involving the other BRS doublets $(\Phi_i+\mbv_i,\Psi)$, 
$(\Phi_i'+\mbv_i',\Psi')$, $(\bar{c},B)$, $(\bar{c}',B')$. In particular, 
the invariant counterterms including these fields contribute to $\Ginvb$.

In a second step, the most general form of 
$\Ginvb^{(n)}=\B\hat\Gamma_{\rm ct,inv,1b}^{(n)}$ 
is obtained by writing down all possible local field polynomials with 
dimension 4 and $Q_{\Phi\Pi}=-1$ whose $\B$ variation satisfies all 
symmetry requirements. Apart from the global symmetries, the requirement 
that $\Ginv^{(n)}$ depends on $\au$ only through the combination 
$a_\alpha=\sqrt2\epsilon_\alpha \au$ is important and reduces the number of 
possible spurion counterterms strongly.

As shown in appendix \ref{app:invCT}, one part of the resulting invariant
counterterms can be identified with the renormalization of parameters
and fields in $\Gcl$. This part of $\Ginvb$ comprises the field 
renormalization of all non-linearly transforming fields, the counterterms 
to all soft-breaking parameters, and the renormalization of $x_1,x_2,v_1,v_2$. 
Due to the presence of $\Gext$ and $\Ggf$, the field renormalizations 
must be compensated by similar renormalizations for the $Y$ fields, 
$B$ fields, and gauge parameters $\zeta,\zeta'$. The remaining part of 
the counterterms in $\Ginvb$ is irrelevant for physical quantities and 
discarded in the following.

Combining the results for $\Ginva^{(n)}$ and $\Ginvb^{(n)}$ shows 
that the relevant counterterms of the first type are generated by 
renormalization of the fields and parameters in $\Gcl$. The parameter 
renormalization in $\Ginv$ can be obtained by 
applying the following renormalization transformations on $\Gcl$
\begin{equation}
g_i \to g_i+\sum_n \deltan g_i
\end{equation}
for the free parameters $g_i$ of the MSSM listed in 
(\ref{eq:physparametersStart})--(\ref{eq:physparametersEnd}). 
The general renormalization transformations for the fields are given by
\begin{eqnarray}
\{V^\mu,Y_{V}^\mu,B,\cbar,\zeta\} & \to & 
\{\sqrt{z_V} V^\mu, \sqrt{z_V}^{-1}Y_{V},
\sqrt{z_V}^{-1}B,\sqrt{z_V}^{-1}\bar{c}, z_V\zeta\},
\nonumber\\
\{\lambda, y_{\lambda}\} & \to &
\{\sqrt{z_\lambda}\lambda, \sqrt{z_\lambda}^{-1}y_{\lambda}\},
\nonumber\\
\{c, Y_c\} & \to & 
\{\sqrt{z_c} c, \sqrt{z_c}^{-1} Y_c\},
\nonumber\\
\{V'{}^\mu,\lambda',y_{\lambda'},c',B',\cbar',\zeta',g'\}
& \to & 
\left\{\begin{array}{l}\sqrt{z_{V'}}V'{}^\mu,\sqrt{z_{V'}}\lambda',
\sqrt{z_{V'}}^{-1}y_{\lambda'}, \\
\sqrt{z_{V'}}c',
\sqrt{z_{V'}}^{-1}B',\sqrt{z_{V'}}^{-1}\cbar',z_{V'}\zeta',
\sqrt{z_{V'}}^{-1}g'
\end{array}
\right\},
\nonumber\\
\{H_i,Y_{H_i}\} & \to &
\{\sqrt{z_{H_i}}H_i,\sqrt{z_{H_i}}^{-1}Y_{H_i}\},
\nonumber\\
\{h_i,y_{h_i}\} & \to &
\{\sqrt{z_{h_i}}h_i,\sqrt{z_{h_i}}^{-1}y_{h_i}\},
\nonumber\\
({\Phi}^a_i+\mbv^a_i,\Psi^a_i) & \to &
 \sqrt{{z_V}}\sqrt{z_{H_i}}^{-1} ({\Phi}^a_i+\mbv^a_i,\Psi^a_i),
\nonumber\\
({\Phi}'_i+\mbv'_i,\Psi'_i) & \to &
 \sqrt{z_{V'}}\sqrt{z_{H_i}}^{-1} ({\Phi}'_i+\mbv'_i,\Psi'_i),
\nonumber\\
\{L, Y_L\} & \to & \{\sqrt{z_{L}} L, \sqrt{z_{L}}^{-1}Y_L\},
\nonumber\\&&
(\mbox{and analogously for the sfermions $R,Q,U,D$}),
\nonumber\\
\{l, y_l\} & \to & \{\sqrt{z_{l}} l, \sqrt{z_{l}}^{-1}Y_l\},
\nonumber\\&&
(\mbox{and analogously for the fermions $r,q,u,d$}).
\label{eq:RT}
\end{eqnarray}

The second type of invariant counterterms differs substantially from the
first type. From the algebraic point of view, $\Ginv$ is sufficient to
cancel all UV singularities. However, we are still free to choose a basis of 
independent {\em physical} fields in terms of which the vertex functional 
is parametrized. 

This parametrization is defined by the choice of finite demixing matrices 
$\Z$. Indeed, changing these matrices $\Z=\Z^{(0)}(1+\sum_n \deltan \Z)$ 
simultaneously in the vertex functional $\Gamma$ and the ST and Ward 
identities preserves all symmetries. 
Since the validity of the symmetries is not affected, these new contributions 
satisfy
\begin{subequations}
\label{eq:symreqtypethree}
\begin{align} 
  \deltan\S(\Gcl) + \S_\Gcl\ \Ginvc^{(n)} &=0, \\
  \deltan W \Gcl + W^{(0)}\ \Ginvc^{(n)}  &= 0 .
\end{align}
\end{subequations}
In accordance with (\ref{Phisymphys}), the invariant counterterms 
$\Ginvc^{(n)}$ and the  counterterms of the symmetry operators $\deltan\S$, 
$\deltan W$ are equivalently generated by the replacement 
\begin{equation}
\phi^{\rm phys}_{j} \to \deltan\Z_{ji} \phi^{\rm phys}_i.
\end{equation}
Symbolically, $\Ginvc^{(n)}$ is given by
\begin{equation}
\Ginvc^{(n)} = \deltan \Z_{ji} \,\left( \int\dx \phi^{\rm phys}_i
   \fdq{}{\phi^{\rm phys}_j} \right) \Gcl,
\label{geffbasis}
\end{equation}
where $\phi^{\rm phys}_i$ runs over all propagating fields. The matrix $\Z$ 
has a block structure since mixing is only possible between fields with equal 
quantum numbers. The different submatrices $\Z_\phi^{(0)}$ are already 
introduced in section \ref{sec:fieldparam}.

The net result for the invariant counterterms is very simple. The invariant
counterterms of order $\hbar^n$ correspond to the renormalization of all
free parameters and fields of the classical action. They can be generated
if the following transformations and replacements are performed
in the classical action:
\begin{subequations}
\label{eq:rentransform}
\begin{align}
g_i & \to g_i+\sum_n \deltan g_i,\\
\phi^{\rm sym} & \to \sqrt{z} \, \phi^{\rm sym},\\
\phi^{\rm sym} & = \Z \, \phi^{\rm phys},
\end{align}
\end{subequations}
and by extracting the pure $\deltan$ part.
This shows that the classical action we started with was complete and that the
theory is multiplicatively renormalizable.

\subsection{Masslessness of the photon} 
\label{sec:photonmassless}

The MSSM contains exactly one unbroken gauge symmetry, and this
implies that there is exactly one massless gauge boson.
For completeness, we will give a simple proof of this well-known
statement, relying only on the ST identity. In addition, we will
show that there is one massless Faddeev-Popov ghost and that
there are three massless directions of the Higgs potential.

The results of this section hold in all orders of perturbation
theory and are important for the physical interpretation as well as
for the theoretical considerations of the following sections. 
They will be of particular importance for the discussion of the IR 
finiteness and of the renormalization scheme. 

For convenience, we use the condensed notation 
$V^a=(V^1,V^2,V^3,V')$ and similarly for $c^a$, $\cbar^a$, $B^a$.
Field derivatives are denoted by subscripts, i.e.\ 
$\Gamma_\phi=\delta\Gamma/\delta \phi$ etc.

Using $\delta^2 \S(\Gamma)/(\delta c^a\delta H_l^j)=0$ we obtain
\begin{align}
\label{eq:dSsHdc}
 0 & =   \Gamma_{c^a Y_i^k}\Gamma_{H_l^j H_i^k}
       + \Gamma_{c^a \bar Y_i^k} \Gamma_{H_l^j \bar H_i^k}
       + \Gamma_{c^a Y_V^{b\mu}} \Gamma_{H_l^j V^b_\mu},
\end{align}
where the last term vanishes for $p=0$ 
since both factors are proportional to $p^\mu$.
Combining (\ref{eq:dSsHdc}) with a similar identity involving 
$\bar H_l^j$ we derive the following identity at $p=0$: 
\begin{align}
 0 = (\Gamma_{c^a Y_i^k}, \Gamma_{c^a \bar Y_i^k})
 \twomat{\Gamma_{H_l^j H_i^k}&\Gamma_{\bar H_l^j H_i^k}
 \\  \Gamma_{H_l^j \bar H_i^k} &\Gamma_{\bar H_l^j \bar H_i^k}}.
\end{align}
If the vector $(\Gamma_{c^a Y_i^k}, \Gamma_{c^a \bar Y_i^k})$ is 
non-vanishing for fixed $a$, it constitutes an eigenvector of the Higgs 
self-energy matrix to the eigenvalue zero. Thus, to each non-vanishing  
$(\Gamma_{c^a Y_i^k}, \Gamma_{c^a \bar Y_i^k})$ there is a massless 
direction of the Higgs self-energy matrix. 

In the case of gauge bosons, the ST identity 
$\delta^2 \S(\Gamma)/(\delta c^a\delta V^{b\mu})$ reads
\begin{align}
 0  =   \Gamma_{c^a Y_i^k}\Gamma_{V^{b\mu} H_i^k}
      + \Gamma_{c^a \bar Y_i^k}\Gamma_{V^{b\mu} \bar H_i^k}
      + \Gamma_{c^a Y^c_{V\nu}}\Gamma_{V^{b\mu} V^{c\nu}}.
\label{Vector}
\end{align}
Suppose there is a non-trivial linear combination of the four vectors
$(\Gamma_{c^a Y_i^k}, \Gamma_{c^a \bar Y_i^k})$ that satisfies
\begin{align}
 \sum_{a=1,2,3,\prime} b_a (\Gamma_{c^a Y_i^k}, \Gamma_{c^a \bar Y_i^k}) = 0.
\label{NullVector}
\end{align}
After contracting with $b_a$ and performing the derivative 
$\partial/\partial p^\mu$ at $p=0$, (\ref{Vector}) reduces to 
\begin{align}
\label{MasslessVector}
0 = \sum_{a,c=1,2,3,\prime} b_a \, D_{ac} \, \Gamma^1_{V^bV^c},
\end{align}
where the Lorentz decompositions $\Gamma_{c^a Y^c_{V\nu}}=p_\nu D_{ac}$ 
and $\Gamma_{V^{b\mu} V^{c\nu}}=g_{\mu\nu} \Gamma_{V^b V^c}^1+p_\mu p_\nu 
\Gamma_{V^b V^c}^2$ are used. Since $D_{ac}$ is invertible, 
$\Gamma_{V^b V^c}^1\propto p^2$. After defining the transverse and 
longitudinal parts of the gauge-boson self energy 
\begin{equation}
\Gamma_{V^{b\mu} V^{c\nu}}=
-\left(g_{\mu \nu}-\frac{p_\mu p_\nu}{p^2}\right)\Gamma_{V^b V^c}^T
-\frac{p_\mu p_\nu}{p^2} \Gamma_{V^b V^c}^L,
\label{eq:transverse}
\end{equation}
we find $\Gamma_{V^b V^c}^T=\Gamma_{V^b V^c}^1\propto p^2$. This proves 
that there is a massless gauge boson corresponding to the linear combination 
(\ref{NullVector}). 

Similarly, contracting the identity 
$\delta^2 \S(\Gamma)/(\delta c^a\delta B^b)$ with $b_a$ yields
\begin{align}
0 = \sum_{a=1,2,3,\prime}b_a \, \Gamma_{c^a \cbar^b},
\end{align}
proving that there is a massless ghost.

As a result, each non-vanishing vector 
$(\Gamma_{c^a Y_k^i}, \Gamma_{c^a \bar Y_k^i})$
corresponds to a broken symmetry and implies the existence of a
massless Goldstone direction of the Higgs potential. Furthermore, a 
vanishing linear combination (\ref{NullVector}) corresponds to an unbroken 
symmetry and implies the existence of a massless gauge boson and a massless 
ghost.

In the MSSM, the five physical Higgs bosons and the $W^\pm$ and $Z$ bosons 
are massive in lowest order and thus also in higher orders. 
Therefore, the four eigenvectors
$(\Gamma_{c^a Y_k^i}, \Gamma_{c^a \bar Y_k^i})$, $a=1,2,3,\prime$
of the Higgs self-energy matrix are linearly dependent, and there
exists at least one vanishing, non-trivial linear combination 
(\ref{NullVector}) and at least one unbroken symmetry.
On the other hand, there is at most one massless gauge boson and thus
at most one unbroken symmetry in higher orders. 

Combining both results we find that to all orders there is exactly one
unbroken and three broken symmetries, and there are one massless
gauge boson, one massless ghost, and three massless would-be Goldstone modes.

\subsection{Off-shell infrared finiteness} 
\label{sec:IR}

Since our model contains massless fields, infrared (IR) 
singularities can appear in loop diagrams. Partly, these IR singularities 
are of physical origin and well understood \cite{IRdiv}. They 
originate from soft photons and cancel in combination with the corresponding 
contributions from the bremsstrahlung processes. On the other hand, additional 
artificial IR singularities without any  physical 
meaning can appear in principle. Such IR singularities would arise
already in off-shell amplitudes and would destroy the mathematical
consistency of the model. The purpose of the present section is to
prove the absence of such unphysical IR singularities to all
orders. In the following, we restrict our discussion only to these  
artificial IR singularities.

While the appearance or non-appearance of IR singularities is not specific to 
any regularization scheme, we adopt Lowenstein's generalization of the BPHZ 
subtraction scheme \cite{BPHZL} because it provides a very systematic 
way to deal with IR singularities. This BPHZL scheme assigns to massless fields
an auxiliary mass $M (s-1)$ that regularizes all IR singularities in 
intermediate steps. At the end the limit $s\to 1$ is taken. The Bogoliubov 
$R$ operation for a renormalization part $\gamma$ reads
\begin{equation}
R \gamma = \big( 1- t_{p,s-1}^{\rho(\gamma)-1}\big) 
  \big(1- t_{p,s}^{\delta(\gamma)}\big) \gamma,
\label{eq:subtractions}
\end{equation} 
where $t_x^d$ is the Taylor expansion operator in the variable $x$ up to
degree $d$. The subtractions are performed in the external momenta $p$ as
well as in the parameter $s$ (resp.\ $s-1$). $\rho(\gamma)$, $\delta(\gamma)$ 
are the IR resp.\ UV degrees of divergence of the diagram $\gamma$, given by 
the usual power-counting law in terms of the IR and UV dimensions of the 
fields. While we use the canonical dimensions for the UV dimensions, 
the freedom in assigning IR degrees has to be used to meet some basic
requirements. 

The two subtractions in (\ref{eq:subtractions}) are compatible if the 
following inequalities are satisfied:
\begin{subequations}
\begin{align}
\dimIR \phi_i &\ge \dimUV \phi_i, \label{IRcond1}\\
\dimIR \phi_i +\dimIR \phi_j &\le 4+r_{ij}, 
\label{IRcond2}
\end{align}
\label{IRcond}
\end{subequations}
where $r_{ij}$ is the scaling exponent of the propagator 
$\Delta_{ij}=\l \phi_i \phi_j\r$,
\begin{equation}
\Delta_{ij} (\lambda p, \lambda (s-1)) \sim \lambda^{r_{ij}}
\qquad\text{for $\lambda\to 0$, $s\to 1$}.
\end{equation}

The BPHZL procedure then yields finite results 
at non-exceptional momenta
for all diagrams  that
contain an arbitrary number of integrated vertices of IR dimension $\dimIR
\ge 4$. In addition, a diagram is allowed to contain one single integrated 
vertex of dimension $4> \dimIR \ge 5/2$. As a consequence, if all interaction 
vertices in $\Geff$ --- the classical action plus counterterms ---
have $\dimIR \ge 4$, then all diagrams contributing to $\Gamma$ are IR
finite. There is an exception: the IR 
degree of terms that are linear in propagating fields is irrelevant as such terms 
cannot occur in one-particle-irreducible (1PI) loop diagrams.

In addition to the IR finiteness of $\Gamma$, there is a more subtle
requirement. The application of Ward 
and ST operators on the vertex functional yields an insertion which has to be 
off-shell IR finite, since otherwise the notion of  ``symmetry 
transformation'' would be meaningless. If $\Geff$ has $\dimIR=4$, an 
operator $\int \dx \phi_i \delta / \delta \phi_j$ leads to an insertion with 
$\dimIR=4+\dimIR \phi_i-\dimIR \phi_j$. The insertions resulting from rigid 
symmetry operators must have $\dimIR\ge 5/2$. Again, there is the exception 
that linear contributions to an insertion are allowed to have arbitrarily low 
IR dimension. Moreover, local Ward operators yield non-integrated
insertions and can have arbitrary IR dimensions.

In order to prove the IR finiteness, we have to find a choice of IR dimensions 
that meets the requirements (\ref{IRcond}). Then it has to be shown that only 
the IR-safe terms listed above appear in $\Geff$ and the Ward and ST operators. 
For this proof it is important to exploit the freedom in assigning the
IR dimensions in such a way that all IR dimensions are as high as
possible, otherwise forbidden terms in $\Geff$ or the symmetry
operators cannot be avoided.

\mytable{
\begin{center}
\vspace*{-12mm}\rule{0pt}{1pt}
\begin{tabular*}{\textwidth}{@{\extracolsep{3mm}}lcccccccccccc}
Field    & $A^\mu$ & $c_A$ & $B_A$ & $\bar c_A$ & $c_Z$ & $B_Z$ & $\bar
c_Z$ & $c_\pm$ & $B_\pm$ & $\bar c_\pm$ & $l^1_\alpha$ \\
$\dimIR$ & 1       & 0     & 2     & 2          & 1     & 2     & 3
& 1 & 2 & 3 & 3/2
\end{tabular*}
\end{center}
}{Exceptional IR dimensions of propagating fields}{tab:IRdim}
A possible choice of IR dimensions is as follows. The dynamical fields
\begin{equation}
W^{\pm\mu}, Z^\mu, G^0, G^\pm, A^0, H^\pm, H^0, h^0, \chi_\al^\pm, \chi_\al^0,
L, R, l^2_\alpha, r_\alpha, Q, U, D, q_\alpha, u_\alpha, d_\alpha
\end{equation}
have $\dimIR=2$ as usual for massive fields. The assignments for the remaining
propagating fields are collected in table \ref{tab:IRdim}.
For the constant ghosts $\xi^\mu,\eps_\alpha$ we assign the UV and IR 
dimension zero, and we set the ultraviolet dimensions of all
non-constant external fields to their canonical dimension. 
To the external source fields $Y,y$ we assign IR dimensions that equal their 
canonical dimensions. For the external fields $\Phi$, $A_2$ we distinguish 
between the components appearing with a shift in some Ward or ST operator and 
the remaining ones. In the SU(2) Ward identity there are three shifted 
components, which can be identified in the spirit of external Goldstone 
modes as
\begin{equation}
\sum_{i=1,2}
-\i g\fdq{}{\Phi_i^a}\left(T^b\mbv^a_i+\i\epsilon^{bca}\mbv_i^c \right)
-\i g'\fdq{}{\Phi_i'} T^b \mbv'_i+{\rm c.c.}\sim\fdq{}{\G^b}.
\label{BackgrGoldstDef}
\end{equation}
In the same way we define an external Goldstone mode $\G^R$ corresponding to R
symmetry by requiring that this field carries the whole shift contribution
of the R Ward identity,
\begin{equation}
\sum_{i=1,2}\i\fdq{}{\Phi_i^a}\mbv_i^a+\i\fdq{}{\Phi_i'}\mbv_i'
-2\i\fdq{}{\av}v_A+{\rm c.c.}\sim\fdq{}{\G^R}.
\end{equation}
We consider $\G^1$, $\G^2$, $\G^3$, $\G^R$, $\av$ and 28 independent
components of $\Phi$, $\Phi'$ as our basis fields. The $\G$ modes have
$\dimIR=1$, the remaining fields have $\dimIR=2$.
These exceptional assignments as well as the IR dimensions for the constant 
ghosts can be found in table \ref{tab:IRnonprop}.

Using these assignments we will show in the following that the vertex 
functional as well as the insertions produced by Ward and ST operators are 
off-shell IR finite. We begin by decomposing $\Geff$ into a part
$\Geff^{\rm dyn}$ that is at
least quadratic in propagating fields and an at most
linear part $\Geff^{\rm triv}$. By requiring $\dimIR\Geff^{\rm dyn}=4$ we 
ensure IR finiteness of all Green functions. $\Geff^{\rm dyn}$ leads to a 
vertex functional $\Gamma^{\rm dyn}$, and the full vertex functional is given 
by
\begin{equation}
\Gamma=\Gamma^{\rm dyn}+\Geff^{\rm triv},
\end{equation}
since the terms in $\Geff^{\rm triv}$ cannot contribute to non-trivial 1PI
diagrams. Applying a Ward operator yields
\begin{equation}
\W \Gamma = \W \Gamma^{\rm dyn} + \W \Geff^{\rm triv}
= [\W \Geff] \cdot \Gamma^{\rm dyn}+ \W \Geff^{\rm triv}
= [\W \Geff] \cdot \Gamma+ \W \Geff^{\rm triv}.
\end{equation}
Clearly, the IR dimension of $\W \Geff$ must be $\dimIR\ge 5/2$ while the 
IR dimension of $\W \Geff^{\rm triv}$ is irrelevant. Dangerous contributions 
to the Ward operator are e.g.\ 
\begin{equation}
A^\mu \fdq{}{Z^\mu},\qquad c_A \fdq{}{c_Z},\qquad \fdq{}{\G^b}.
\label{Wdanger1}
\end{equation}
Owing to the choices of the IR dimensions, the operators in (\ref{Wdanger1}) 
lead to insertions with $\dimIR=3$, which are allowed. This explains in 
particular the choices for the $\G$ modes. 

\mytable{
\begin{center}
\vspace*{-12mm}\rule{0pt}{1pt}
\begin{tabular*}{\textwidth}{@{\extracolsep{3mm}}lccccccc}
Field    & $\eps^\al$ & $\xi^\mu$ & $\au$ & $\av$ & $\G^b$,
$\G^R$ & $\Phi, \Phi'$ (remnant) & $\Psi, \Psi'$ \\
$\dimIR$ & $0$ & $0$ & $1$   & $2$   & $1$    & $2$ \phantom{(remnant)} 
& $2$
\end{tabular*}
\end{center}
}{IR dimensions of non-propagating fields}{tab:IRnonprop}

Indeed, it is easy to check that our choice of IR dimensions guarantees that 
all Ward operators lead to IR-finite insertions. In the same way we can show 
the IR finiteness of the ST identity. Our decomposition of $\Gamma$ yields 
three contributions
\begin{equation}
\S(\Gamma) = \S(\Gamma^{\rm dyn}) + \S_{\Gamma^{\rm dyn}} \Geff^{\rm triv}
+ \int \dx \fdq{\Geff^{\rm triv}}{Y_a}\fdq{\Geff^{\rm triv}}{\phi_a},
\label{SGammadecomp}
\end{equation}
the last of which is IR harmless. The first contribution
$\S(\Gamma^{\rm dyn})$ corresponds to an 
insertion%
\footnote{This can be seen by temporarily introducing the generating  
  functional $Z$ of general Green functions by Legendre transformation 
  introducing sources $j_a, j'_b$ and exponentiation. In contrast to the ST 
  identity on $\Gamma$, the insertions of the ST operator on $Z$ have the 
  explicit form
  \begin{equation}
  \S(Z)=\left[\fdq{\Geff}{Y_a} j_a+\s\phi'_b j'_b\right]^\rho_\delta\cdot Z ,
  \end{equation}
  where the required degree of subtraction can be read off.} 
whose IR dimension is determined by the quantum action principle
\cite{QAP}, which  
assigns a degree of $8-\dimIR Y-\dimIR \phi$ to the term
$(\delta \Gamma^{\rm dyn}/\delta Y)(\delta\Gamma^{\rm dyn}/\delta\phi)$ 
and $\dimIR(\s \phi')+4-\dimIR \phi'$ to 
$\s\phi' (\delta \Gamma/\delta \phi')$. 
Owing to the choice of the IR dimensions, $\dimIR(\S(\Gamma^{\rm
dyn}))\ge 5/2$. 

These simple arguments are 
sufficient to show the IR finiteness of $\S(\Gamma^{\rm dyn})$ because 
$\Geff^{\rm dyn}$ contains only contributions with $\dimIR\ge 4$. They
cannot be extended to the complete vertex functional, since we 
have allowed for terms with $\dimIR<4$ in $\Geff^{\rm triv}$. Instead, we discuss the  
second contribution in (\ref{SGammadecomp}) separately. It reads explicitly
\begin{equation}
\S_{\Gamma^{\rm dyn}} \Geff^{\rm triv} = \int \dx \fdq{\Gamma^{\rm dyn}}{Y_a} 
\fdq{\Geff^{\rm triv}}{\phi_a} +  
\int \dx \fdq{\Geff^{\rm triv}}{Y_a} \fdq{\Gamma^{\rm dyn}}{\phi_a} 
+ \int \dx \s \phi_b' \fdq{\Geff^{\rm triv}}{\phi_b'} .
\label{eq:SdynGtriv}
\end{equation}

The third term in (\ref{eq:SdynGtriv}) is trivial, since it involves no loop 
contributions. The first two terms have the form of a Ward identity with 
local, at most linear field transformations
$\tilde\delta Y_a = \delta\Geff^{\rm triv}/\delta\phi_a$, 
$\tilde\delta \phi_a=\delta\Geff^{\rm triv}/\delta Y_a$.
By explicitly inspecting all possible terms in $\Geff^{\rm triv}$ we
find that this leads to an insertion with $\dimIR\ge5/2$.
Thus we 
have shown that our choice of IR dimensions of the fields indeed leads to 
finite Ward and ST identities.

It remains one problem to be solved: for the restoration of the Ward 
and ST identities we might be forced to use IR-forbidden symmetry-restoring 
counterterms in $\Geff$. The complete list of such IR-dangerous counterterms 
involving at least two propagating fields reads
\begin{equation}
A^\mu A_\mu,\quad A^\mu Z_\mu,\quad  c_A\bar c_A,\quad c_A\bar
c_Z,\quad c_Z \bar c_A.
\label{eq:IRforbidden}
\end{equation}
All other IR-dangerous terms are excluded from $\Geff$ by 
lepton-number conservation for terms involving neutrinos, and due to CP 
conservation for terms involving $\G^b$, $\G^R$, like $A^\mu A_\mu \G^b$ and 
$c_A\bar c_A \G^b$.  

None of the terms in (\ref{eq:IRforbidden}) appears in the classical action. 
If these terms can also be excluded from the counterterm part of
$\Geff$, the BPHZL scheme guarantees the IR finiteness 
of the vertex functional $\Gamma$. This is a special 
property of the BPHZL scheme since loop corrections corresponding to terms in 
(\ref{eq:IRforbidden}) vanish automatically at $p=0$ due to the special form 
of the subtraction operator (\ref{eq:subtractions}). In general, IR finiteness 
of $\Gamma$ follows from the conditions
\begin{equation}
\Gamma_{AA}|_{p=0}=
\Gamma_{AZ}|_{p=0}=
\Gamma_{c_A\cbar_A}|_{p=0}=
\Gamma_{c_A\cbar_Z}|_{p=0}=
\Gamma_{c_Z\cbar_A}|_{p=0}=0.
\label{IRNormCond}
\end{equation}
It remains to be shown 
that the IR-dangerous counterterms are not required 
for the restoration of the symmetry identities.

Firstly we discuss the term $A^\mu Z_\mu$. Using the results of
section \ref{sec:fieldparam} we find that among the
invariant counterterms of the first type there are the parameters from field 
reparametrization 
\begin{multline}
\left[ (\deltan \Z_V)_{ZA} A_\mu \fdq{}{Z_\mu} + 
(\deltan \Z_V)_{ZZ} Z_\mu  \fdq{}{Z_\mu} \right] 
\frac{1}{2}\left( M_Z^2 Z^\mu Z_\mu \right) \\
=(\deltan \Z_V)_{ZA}M_Z^2A^\mu Z_\mu 
+ (\deltan \Z_V)_{ZZ}M_Z^2Z^\mu Z_\mu  
\end{multline}
at our disposal. Clearly, $(\deltan \Z_V)_{ZA}$ can be adjusted such that 
$\Gamma_{A^\mu Z^\nu}|_{p=0}$. Similarly, the constants 
$(\deltan \Z_c)_{ZA}$ and $(\deltan \Z_{\bar{c}})_{ZA}$,
corresponding to the field reparametrization of $c$, $\cbar$, can be used to 
cancel the counterterms  $c_A\bar c_Z$, $c_Z \bar c_A$ and equivalently to
establish $\Gamma_{c_A\cbar_Z}|_{p=0}=\Gamma_{c_Z\cbar_A}|_{p=0}=0$.

The remaining conditions in (\ref{IRNormCond}) cannot be established
by a field reparametrization. Instead, we show that they are satisfied
owing to the result of section \ref{sec:photonmassless} that there is
exactly one unbroken gauge symmetry in the MSSM.

We turn to the term $A^\mu A_\mu$. As shown in section
\ref{sec:photonmassless}, the determinant of the gauge-boson self energy
vanishes at $p=0$. Since
$\Gamma_{ZZ}|_{p=0},\Gamma_{W^+W^-}|_{p=0}\ne0$, and since the
off-diagonal elements $\Gamma_{AZ}|_{p=0}$ vanish, the photon is
massless, i.e.\ $\Gamma_{AA}|_{p=0}=0$.
In the same way it is shown that $\Gamma_{c_A \bar c_A}|_{p=0}=0$.

This completes our proof that the conditions (\ref{IRNormCond}) can
be satisfied without destroying the symmetries. Hence, in
the BPHZL scheme the counterterm contributions to the forbidden terms
in (\ref{eq:IRforbidden}) vanish, and the IR finiteness is proven.

In schemes that are different from the BPHZL scheme, the counterterm
contributions to $\Geff$ might be different, but since $\Gamma$ is
a regularization-scheme independent quantity, the infrared finiteness
of $\Gamma$ holds for all schemes. For example, in dimensional
regularization the loop and counterterm contributions to the
IR-dangerous terms need not vanish individually but the net result for
$\Gamma$ is the same as in the BPHZL scheme.

\section{Renormalization scheme} 
\label{sec:renscheme}
\setcounter{equation}{0}

In this section we specify a renormalization scheme for the MSSM. 
We use an on-shell scheme where all normalization conditions are
formulated at the physical particle masses, and where transitions between 
different fields are avoided for on-shell momenta as far as possible. As a 
consequence, up to effects due to the finite widths, every field corresponds 
either to a mass eigenstate or to an unphysical degree of freedom with 
simple formal properties.

Before we start with the discussion of the renormalization scheme, we 
introduce some notations. The complex mass $\M_\phi$ of an unstable particle 
$\phi$ corresponds to the location of the particle pole in the complex 
invariant-mass plane and is decomposed into a real mass $M_\phi$ and a finite 
width $\Gamma_\phi$ as follows
\begin{equation}
\M_\phi^2=M_\phi^2-\i \Gamma_\phi M_\phi.
\end{equation}
Since the complex particle pole is gauge independent, it is a proper quantity
for the definition of normalization conditions. 

\subsection{Normalization conditions}
\label{sec:normcond}

The on-shell normalization conditions are listed separately for the different 
sectors of the MSSM. 

\subsubsection*{Gauge-boson and Higgs sector}

Since the $W^\pm$ and $Z$ bosons are unstable particles, the poles of
their propagators are complex. Since we require that all tadpole contributions 
vanish, the poles $\M_{W,Z}^2$ are determined by the normalization conditions
\begin{align}
\label{eq:detAZ}
\left.{\rm det}\twomat{\Gamma_{AA}^T & \Gamma_{AZ}^T \\
                 \Gamma_{ZA}^T & \Gamma_{ZZ}^T }
\right|_{p^2=\M_Z^2}&=0,&
\Gamma_{W^+W^-}^T|_{p^2=\M_W^2}&=0,
\end{align}
where $\Gamma^T$ is the transverse part of the gauge-boson self energy
defined in (\ref{eq:transverse}).
The real masses $M_W$, $M_Z$ are adopted as the two free parameters in the 
gauge-boson sector. The imaginary parts of $\M_{W,Z}^2$ are determined
as functions of $M_{W,Z}$. To establish the masslessness of the photon and
to avoid mixing of gauge-bosons as far as possible, we 
require the conditions at the real masses:
\begin{subequations}
\begin{align}
\label{eq:IRenforced1}
\Gamma_{AA}^T |_{p^2=0}&=0,& 
\Gamma_{AZ}^T |_{p^2=0}&=0, \\
{\rm Re} \Gamma_{ZA}^T |_{p^2=M_Z^2}&=0.&&
\end{align}
\end{subequations}
The normalization conditions for the residua of the full gauge-boson 
propagators are given by
\begin{subequations}
\begin{align}
\partial_{p^2}\Gamma^T_{AA}|_{p^2=0}&=1,&
{\rm Re}\partial_{p^2}\Gamma^T_{ZZ}|_{p^2=M_Z^2}&=1,\\
{\rm Re}\partial_{p^2}\Gamma^T_{W^+W^-}|_{p^2=M_W^2}&=1.&&
\end{align}
\end{subequations}

In the Higgs sector, the real mass $M_{A^0}$ of the neutral CP-odd Higgs boson 
$A_0$ is chosen as the only free parameter. Accordingly, we require the 
normalization condition
\begin{align}
\left.{\rm det}\twomat{\Gamma_{G^0G^0} & \Gamma_{G^0A^0}\\
                 \Gamma_{A^0G^0} & \Gamma_{A^0A^0}}
\right|_{p^2=\M_{A^0}^2}=0.
\label{eq:AGpole}
\end{align}
with ${\rm Re}\M_{A^0}^2=M_{A^0}^2$.
The masses of the Higgs bosons $H^0$, $h^0$, $H^\pm$ are no free 
parameters of the MSSM. These masses $\M_{H^0}$, $\M_{h^0}$, $\M_{H^\pm}$ are
defined by
\begin{align}
\left.{\rm det}\twomat{\Gamma_{H^0H^0} & \Gamma_{H^0h^0}\\
                 \Gamma_{h^0H^0} & \Gamma_{h^0h^0}}
\right|_{p^2=\M_{H^0}^2,\M_{h^0}^2}&=0,&
\left.{\rm det}\twomat{\Gamma_{G^+G^-} & \Gamma_{G^+H^-}\\
                 \Gamma_{H^+G^-} & \Gamma_{H^+H^-}}
\right|_{p^2=\M_{H^\pm}^2}&=0.
\label{eq:HPMpole}
\end{align}
In this way, the complex masses of the Higgs bosons $H^0$, $h^0$, $H^\pm$ 
can be 
calculated order by order as a function of $M_{A^0}$ and the other MSSM 
parameters (c.f.\ section \ref{sec:physicalparameters}).
The demixing conditions in the Higgs sector are given by
\begin{subequations}
\label{eq:Higgsdemixing}
\begin{align}
\label{BCondition}
{\rm Re}\Gamma_{A^0B_A}|_{p^2=M_{A^0}^2} & = 0,&
{\rm Re}\Gamma_{A^0B_Z}|_{p^2=M_{A^0}^2} & = 0,\\
{\rm Re}\Gamma_{A^0G^0}|_{p^2=M_{A^0}^2} & = 0,&
{\rm Re}\Gamma_{H^+H^-}|_{p^2=M_{H^\pm}^2} & = 0,\\
{\rm Re}\Gamma_{H^+G^-}|_{p^2=M_{H^\pm}^2} & = 0,&
{\rm Re}\Gamma_{H^+B^-}|_{p^2=M_{H^\pm}^2} & = 0,\\
{\rm Re}\Gamma_{H^0h^0}|_{p^2=M_{H^0}^2} & = 0,&
{\rm Re}\Gamma_{h^0H^0}|_{p^2=M_{h^0}^2} & = 0.
\end{align}
\end{subequations}
The demixing between Higgs bosons and longitudinal gauge bosons is 
not realized in (\ref{eq:Higgsdemixing}).
The effects of this mixing are controlled by the ST identity as shown 
later in section \ref{sec:EvaluationNormCond}. 
The residua of the Higgs propagators are fixed by 
\begin{subequations}
\begin{align}
{\rm Re}\partial_{p^2}\Gamma_{A^0A^0}|_{p^2=M_{A^0}^2}&=1,&
{\rm Re}\partial_{p^2}\Gamma_{H^0H^0}|_{p^2=M_{H^0}^2}&=1,\\
{\rm Re}\partial_{p^2}\Gamma_{h^0h^0}|_{p^2=M_{h^0}^2}&=1,&
{\rm Re}\partial_{p^2}\Gamma_{H^+H^-}|_{p^2=M_{H^\pm}^2}&=1,\\
\partial_{p^2}\Gamma_{G^0G^0}|_{p^2=0}&=1,&
\partial_{p^2}\Gamma_{G^+G^-}|_{p^2=0}&=1.
\end{align}
\end{subequations}

\subsubsection*{Chargino and neutralino sector}

In the chargino and neutralino sector, we adopt the three real masses 
$m_{\chi^\pm_1}$, $m_{\chi^\pm_2}$, $m_{\chi^0_1}$ as the free parameters 
and require
\begin{equation}
\label{eq:Mchi}
{\rm det}\left(\Gamma_{f_i\bar{f}_j}\right)|_{p^2=\M_{f_k}^2}=0,
\end{equation}
where the Dirac spinors $f_i$ run either over the charginos $\chi_i^\pm$
with $i=1,2$ or over the neutralinos $\chi_i^0$ with $i=1,2,3,4$. 
By this equation also the imaginary parts of
$\M^2_{\chi^\pm_1,\chi^\pm_2,\chi^0_1}$ and the complex masses
$\M^2_{\chi^0_{2,3,4}}$ are defined as functions of the MSSM input
parameters. 
At the real masses, we require the demixing conditions ($i\ne j$)
\begin{align}
\widetilde{\rm Re}\Gamma_{f_i\bar{f}_j}\, u_{f_j}|_{p^2=m_{f_j}^2}&=0
\end{align}
and normalization conditions for the residua
\begin{align}
\lim_{p^2\to m_{f_i}^2}
\frac{1}{\psl-m_{f_i}}\widetilde{\rm Re}\Gamma_{f_i\bar{f_i}}\,
 u_{f_i}&=u_{f_i},
\end{align}
where the spinors $u_{f_i}$ are solutions of the Dirac equation 
$(\psl-m_{f_i})u_{f_i}=0$ for $p^2=m_{f_i}^2$, 
and $\widetilde{\rm Re}$ takes the real part of the loop integrals but not 
of the couplings and the $\gamma$ matrices.  

\subsubsection*{Matter sector}

Since our matter representation is restricted to one generation, we 
choose the three real masses $m_e$, $m_u$, $m_d$ as the free parameter of the 
quark and lepton sector. As in the chargino and neutralino sector, we require 
the following on-shell conditions:
\begin{equation}
{\rm det}\left(\Gamma_{f\bar{f}}\right)|_{p^2=\M_f^2}=0,
\end{equation}
where $f$ runs over the massive fermions $e,u,d$. 
The residua of the quark and lepton propagators are fixed by
\begin{align}
{\displaystyle\lim_{p^2\to m_{f}^2}}
\frac{1}{\psl-m_{f}}\widetilde{\rm Re}\Gamma_{f\bar f} u_f &=u_f,
\end{align}
where $f$ runs over all Dirac spinors $\nu,e,u,d$, and
the spinors $u_f$ satisfy $(\psl-m_f) u_f=0$ for $p^2=m_f^2$.

In the squark and slepton sector, we choose the real masses of the sfermions
$\tilde{u}_1$, $\tilde{u}_2$, $\tilde{d}_1$, $\tilde{\nu}_1$,
$\tilde{e}_1$ as input parameters. The on-shell conditions for the squark 
and slepton sector read
\begin{align}
\label{eq:onshellfermion}
\left.{\rm det}\left(\Gamma_{\tilde{f}_i\tilde{f}_j^\dagger}\right)
\right|_{p^2=\M_{\tilde{f}_k}}=0
\end{align}
with $\tilde{f}_i=\tilde{u}_i$, $\tilde{d}_i$, $\tilde{\nu}_i$, 
$\tilde{e}_i$ and $i,j,k=1,2$. The masses of the sfermions $\tilde{d}_2$, 
$\tilde{e}_2$ are dependent parameters. The normalization conditions of the 
demixing and the residua of the propagators are given by ($i\ne j$)
\begin{align}
{\rm Re}\Gamma_{\tilde{f}_i\tilde{f}_j^\dagger}
|_{p^2=M_{\tilde{f}_j}^2}&=0, &
{\rm Re}\partial_{p^2}\Gamma_{\tilde{f}_i\tilde{f}_i^\dagger}
|_{p^2=M_{\tilde{f}_i}^2}&=1.
\end{align}

\subsubsection*{Gauge-fixing sector}

The unphysical Goldstone bosons are characterized as the three massless 
directions of the scalar potential:
\begin{subequations}
\begin{align}
\Gamma_{G^0G^0}|_{p^2=0}&=0,&\Gamma_{A^0G^0}|_{p^2=0}&=0,\\
\Gamma_{G^+G^-}|_{p^2=0}&=0,&\Gamma_{G^+H^-}|_{p^2=0}&=0.
\end{align}
\end{subequations}
The on-shell conditions in the ghost sector read
\begin{align}
\left.{\rm det}\twomat{\Gamma_{c_A\cbar_A} & \Gamma_{c_A\cbar_Z}\\
                 \Gamma_{c_Z\cbar_A} & \Gamma_{c_Z\cbar_Z}}
\right|_{p^2=\M_{c_Z}^2}&=0,&
\Gamma_{c^+\cbar^-}|_{p^2=\M_{c^\pm}^2}&=0.
\end{align}
The demixing conditions are split into normalization conditions required 
for the IR finiteness of the MSSM (see section \ref{sec:IR})
\begin{subequations}
\begin{align}
\label{eq:IRenforced2}
\Gamma_{c_A\cbar_A}|_{p^2=0}&=0,&
\Gamma_{c_A\cbar_Z}|_{p^2=0}&=0,\\
\Gamma_{c_Z\cbar_A}|_{p^2=0}&=0,&&
\end{align}
\end{subequations}
and the remaining demixing conditions
\begin{subequations}
\begin{align}
{\rm Re}\Gamma_{c_A\cbar_Z}|_{p^2=M_{c_Z}^2}&=0,&
{\rm Re}\Gamma_{c_Z\cbar_A}|_{p^2=M_{c_Z}^2}&=0.
\end{align}
\end{subequations}
$M_{c_Z}^2$, $M_{c^\pm}^2$ are no free parameters of the MSSM and depend on 
the gauge-fixing parameters. The conditions for the residua of the ghost 
propagators are given by
\begin{subequations}
\begin{align}
\partial_{p^2}\Gamma_{c_A\cbar_A}|_{p^2=0}&=1,&
{\rm Re}\partial_{p^2}\Gamma_{c_Z\cbar_Z}|_{p^2=M_{c_Z}^2}&=1,\\
{\rm Re}\partial_{p^2}\Gamma_{c^+\cbar^-}|_{p^2=M_{c^\pm}^2}&=1.&&
\end{align}
\end{subequations}

\subsubsection*{Comments on the normalization conditions}

The normalization conditions (\ref{eq:IRenforced1}) and (\ref{eq:IRenforced2})
are necessary for the IR finiteness of the MSSM [c.f.\ (\ref{eq:IRforbidden})] 
as discussed in detail in section \ref{sec:IR}. 

Since not all masses of the MSSM are independent parameters, there is a
conceptual difference between the on-shell conditions concerning real 
masses that are free parameters of the MSSM, like $M_{A^0}$ in 
(\ref{eq:AGpole}), and masses that are dependent parameters, like $M_{H^\pm}$
in (\ref{eq:HPMpole}). The former serve as normalization conditions for the 
respective self energies, while the latter are simply definitions for the 
masses $\M_{H^\pm}$ etc.

We have not established the demixing between the physical Higgs fields $A^0$, 
$H^\pm$, the Goldstone bosons $G^0$, $G^\pm$, the longitudinal gauge bosons 
$\partial^\mu A_\mu$, $\partial^\mu Z_\mu$, $\partial^\mu W^{\pm}_\mu$, and 
the auxiliary $B$ fields. Therefore, the propagators in this sector have to 
be determined by inverting the full matrix of the 2-point functions between 
the Higgs, gauge-boson, and $B$ fields. 
It is not obvious whether $\M_{A^0,H^\pm}^2$ as defined by (\ref{eq:AGpole}), 
(\ref{eq:HPMpole}) coincide with the poles of the propagators. That they
actually do is shown in section \ref{sec:EvaluationNormCond} 
using the ST identity.

For unstable particles, a complete demixing is generally not possible due 
to the non-vanishing of the imaginary parts of the self-energy corrections. 
The demixing is realized only up to terms of the order of the finite widths 
of unstable particles. On the other hand, unstable particles appear, in a 
strict sense, only as resonances in scattering processes and do not 
contribute to asymptotic states. 

\subsection{Generating invariant counterterms}
\label{sec:GenerateCTs}

In this section, the consequences of section \ref{sec:CTs} for
generating the invariant counterterms in practice are drawn. 
In section \ref{sec:CTs}, we have seen that all invariant counterterms 
are generated by applying three kinds of renormalization transformations to 
the classical action corresponding to the renormalization 
of the free parameters of the model, the field renormalization of symmetric 
fields, and the finite field reparametrizations of symmetric fields by 
physical fields. Schematically, the transformations read
\begin{subequations}
\label{eq:RenTransGeneral}
\begin{align}
g_i & \to g_{i{\rm bare}}=g_i + \delta g_i,\\
\label{eq:ztrans}
\phi^{\rm sym} & \to \sqrt{z}\phi^{\rm sym},\\
\phi^{\rm sym} & = R \phi^{\rm phys}.
\end{align}
\end{subequations}
where $g_i$ stands for an arbitrary invariant counterterm.
Clearly, there is an overlap between the renormalization of the
symmetric fields by the counterterm $\sqrt{z}$ and the reparametrization in 
terms of physical fields involving the demixing matrices $\Z$. 
In $\Gamma$ and $\Geff$ the counterterms $\sqrt{z}$ and $\Z$ appear
only in the combination $\sqrt{z}\Z$. Both counterterms $\sqrt{z}$ 
and $\Z$ can be simply combined to one general matrix valued field 
renormalization.

Applying the transformations (\ref{eq:RenTransGeneral}) on the classical 
action yields a bare action,
\begin{equation}
\Gcl \to \Gamma_{\rm bare},
\end{equation}
$\Gamma_{\rm bare}$ is introduced only as an auxiliary quantity to generate 
invariant counterterms and does not necessarily 
coincide with $\Geff$ (\ref{eq:defGeff}) from which the Feynman rules are 
derived. The invariant counterterms of order $\hbar^n$ can be obtained by
expanding all counterterms in $\Gamma_{\rm bare}$,  
i.e.\ $g_{i{\rm bare}}=g_i+\sum_n\deltan g_i$, omitting all contributions 
involving products of counterterms, and taking only the pure 
$\deltan$ contributions:
\begin{equation}
\Ginv^{(n)}\equiv\deltan \Gamma=
\Gamma_{\rm bare}|_{\rm 1st\ order\ in\ \deltan}.
\end{equation}
The structure of $\Ginv^{(n)}$ is identical for all $n$ and, in particular,
coincides with the one of $\Ginv^{(1)}$.

Two possibilities exist how the classical action can be parametrized. Either
$\Gcl$ can be formulated in terms of the symmetric fields or in terms 
of physical fields:
\begin{equation}
\Gcl=\Gamma_{\rm cl}^{\rm sym}(g_i,\phi^{\rm sym})
=\Gamma_{\rm cl}^{\rm phys}(g_i,\phi^{\rm phys}).
\end{equation}
Both parametrizations are related by
\begin{equation}
\Gamma_{\rm cl}^{\rm phys}(g_i,\phi^{\rm phys})=  
\Gamma_{\rm cl}^{\rm sym}(g_i,\Z^{(0)}\phi^{\rm phys}),
\label{eq:GclPhSym}
\end{equation}
where $\Z^{(0)}$ denotes the lowest-order demixing matrix.
Depending on whether $\Gamma_{\rm cl}^{\rm sym}$ or 
$\Gamma_{\rm cl}^{\rm phys}$ is used as a starting point, different but 
equivalent parametrizations of $\Ginv$ are obtained. We sketch in the 
following both possibilities in the general case before applying them to 
the individual sectors of the MSSM.

It is closer to section \ref{sec:CTs} to start with 
$\Gamma_{\rm cl}^{\rm sym}(g_i,\phi^{\rm sym})$. The transformations 
(\ref{eq:RenTransGeneral}) yield
\begin{equation}
\Gamma_{\rm bare} = 
\Gamma_{\rm cl}^{\rm sym}(g_{i{\rm bare}}, \sqrt{z}R\phi^{\rm phys}).
\label{eq:GBare1}
\end{equation}
As a result only the combination $\sqrt{z}R$ appears in 
$\Gamma_{\rm cl}^{\rm sym}$. It will be useful to split off the 
lowest-order part and define
\begin{equation}
\sqrt{Z}_{\rm 1st\ param.} = (\Z^{(0)})^{-1} \sqrt{z} \Z,
\label{eq:firstparam}
\end{equation}
where $\sqrt{Z}$ is a general field renormalization matrix.
This kind of parametrization for the $Z$ factors will be used in the
sfermion sector and in the neutralino-chargino sector.

An equivalent but different parametrization can be obtained by using 
$\Gamma_{\rm cl}^{\rm phys}$ as the starting point. For the fields 
$\phi^{\rm phys}$ the renormalization transformation 
(\ref{eq:RenTransGeneral}) yields 
\begin{equation}
\phi^{\rm phys}=(\Z^{(0)})^{-1}\phi^{\rm sym} \to 
(\Z^{(0)}_{\rm bare})^{-1}\sqrt{z} \Z \phi^{\rm phys},
\end{equation}
where $\Z^{(0)}_{\rm bare}$ is obtained from $\Z^{(0)}$, which is
a function of the parameters $g_i$, by replacing $g_i\to g_{i{\rm bare}}$. 
In agreement with (\ref{eq:GclPhSym}) and (\ref{eq:GBare1}), the bare action 
can be written as 
\begin{align}
\Gamma_{\rm bare} & = \Gamma_{\rm cl}^{\rm phys}\left(g_{i{\rm bare}}, 
(\Z^{(0)}_{\rm bare})^{-1} \sqrt{ z}R\phi^{\rm phys}\right).
\end{align}
In this case, the combination 
\begin{equation}
\sqrt{Z}_{\rm 2nd\ param.} = (\Z^{(0)}_{\rm bare})^{-1}\sqrt{z} \Z
\label{eq:ZParam2}
\end{equation}
appears naturally, and thus the renormalization transformations amount
simply to 
\begin{subequations}
\begin{align}
g_i & \to g_i + \delta g_i,\\
\phi^{\rm phys} & \to \sqrt{Z}\phi^{\rm phys}.
\end{align}
\end{subequations}
The renormalization constants $\sqrt Z$ introduced in both parametrizations 
differ. The two parametrizations can be related by
\begin{equation}
\sqrt{Z}_{\rm 2nd\ param.} = (\Z^{(0)}_{\rm bare})^{-1}
    \Z^{(0)}\sqrt{Z}_{\rm 1st\ param.}.
\end{equation}
The second parametrization is most useful in cases in which the analytic
form of $\Z^{(0)}$ is explicitly known, and where the form of
$\Gamma_{\rm cl}^{\rm phys}$ has a simple structure. 
We will use it in the gauge-boson, Higgs, and ghost sector.

\subsection{Evaluation of the normalization conditions}
\label{sec:EvaluationNormCond}

The general structure of the invariant counterterms is discussed in 
detail in sections \ref{sec:CTs} and \ref{sec:GenerateCTs}. As the net result, 
there are counterterms to all independent parameters as well as general 
matrix valued field renormalizations. 

Assuming that the vertex functional is already renormalized up to order 
$\hbar^{n-1}$, the normalization condition have to be fulfilled by a proper
choice of the genuine invariant counterterms of order $\hbar^n$, since 
the lower-order counterterms are already fixed. In this section, 
we want to verify whether the normalization conditions can be satisfied.

\subsubsection*{Gauge-boson sector}

The invariant counterterms of the gauge-boson sector are obtained from the 
classical action by performing the following renormalization transformations:
\begin{align}
M_W &\to M_{W{\rm bare}} = M_W + \delta M_W,\\
M_Z &\to M_{Z{\rm bare}} = M_Z + \delta M_Z,\\
W^\pm_\mu &\to \sqrt{\ZW} W^\pm_\mu,\\
\twovect{A_\mu}{Z_\mu} &\to \sqrt{\ZV} \twovect{A_\mu}{Z_\mu}.
\label{eq:AZRenTransform}
\end{align}
The field-renormalization matrix $\sqrt{\ZV}$ reads
\begin{align}
\sqrt{\ZV}&=\twomat{ \cos\theta_{W{\rm bare}} & \sin\theta_{W{\rm bare}} \\
                    -\sin\theta_{W{\rm bare}} & \cos\theta_{W{\rm bare}}}
\twomat{\sqrt{z_{V'}} & 0 \\ 0 & \sqrt{z_V}} \Z_V,
\end{align}
where the second parametrization (\ref{eq:ZParam2}) is used. According to 
(\ref{eq:ThetaWDef}), the bare weak mixing angle is defined by 
\begin{equation}
\cos\theta_{W{\rm bare}} = \frac{M_{W{\rm bare}}}{M_{Z{\rm bare}}}.
\end{equation} 
Working out the pure $\deltan$ part of $\Gamma_{\rm bare}$, we obtain 
the invariant counterterms of order $\hbar^n$:
\begin{eqnarray}
\Ginv^{(n)}|_{\rm gauge}&=& 
\int\dx\bigg\{
-\frac14(A^{\mu\nu},Z^{\mu\nu})\deltan \ZV\twovect{A_{\mu\nu}}{Z_{\mu\nu}}
-\frac12 \deltan \ZW W^{+\mu\nu}W^-_{\mu\nu}
\nonumber\\&&{}
+\frac12 (A^\mu,Z^\mu)
 \left[\twomat{0 & 0 \\ 0 & M_Z^2}\deltan \ZV
      +\twomat{0 & 0 \\ 0 & \deltan M_Z^2}
 \right]
\twomat{A_\mu \\ Z_\mu}
\nonumber\\&&{}
+ \left(\deltan \ZW M_W^2 + \deltan M_W^2\right) W^{+\mu} W^-_\mu
\bigg\},
\end{eqnarray}
where $A^{\mu \nu}, Z^{\mu \nu}, W^{\pm \mu \nu}$ are defined as in 
(\ref{eq:GclBil}).
This yields the following contributions to the vertex functions
appearing in the normalization conditions:
\begin{subequations}
\label{eq:renvector}
\begin{align}
\deltan\Gamma^T_{AA}|_{p^2=0} & = 0,
\label{eq:GammaAA}
\\
\deltan\Gamma^T_{AZ}|_{p^2=0}& = - \tfr12(\deltan\ZV)_{ZA}M_Z^2,
\label{eq:GammaAZ}
\\
\deltan\Gamma^T_{ZZ}|_{p^2=M_Z^2}& = - \deltan M_Z^2,
\\
\deltan\Gamma^T_{ZA}|_{p^2=M_Z^2}& = \tfr12(\deltan\ZV)_{AZ}M_Z^2,
\\
\deltan\Gamma^T_{W^+W^-}|_{p^2=M_W^2}& = - \deltan M_W^2,
\\
\partial_{p^2}\deltan\Gamma^T_{AA}|_{p^2=0}& = (\deltan\ZV)_{AA},  
\\
\partial_{p^2}\deltan\Gamma^T_{ZZ}|_{p^2=M_Z^2}& = (\deltan\ZV)_{ZZ}, 
\\
\partial_{p^2}\deltan\Gamma^T_{W^+W^-}|_{p^2=M_W^2}& = \deltan \ZW.
\end{align}
\end{subequations}
The notation $\deltan \Gamma$ is used for the contributions of the invariant
counterterms of order $\hbar^n$. The counterterm contribution to the 
determinant appearing in (\ref{eq:detAZ}) is not given explicitly, but it is 
obvious that (\ref{eq:detAZ}) can be satisfied by adjusting $\deltan M_{W,Z}$.
The structure of (\ref{eq:renvector}) shows 
that the demixing and residua conditions fix the counterterms $\deltan\ZW$, 
$\deltan\ZV$. However, there is no counterterm available for the on-shell 
condition of the photon. As already discussed in section \ref{sec:IR}, 
(\ref{eq:GammaAA}) is a direct consequence of the ST identity and of the 
demixing normalization condition (\ref{eq:GammaAZ}). 

As a result, all normalization conditions in the gauge-boson sector can be
satisfied.

\subsubsection*{Ghost sector}

The complete field-renormalization transformations in the ghost sector read
\begin{align}
\twovect{c_A}{c_Z} &\to \sqrt{\Zc}\twovect{c_A}{c_Z},&
\twovect{\cbar_A}{\cbar_Z} & \to \sqrt{\Zcbar}\twovect{\cbar_A}{\cbar_Z},\\
c^\pm &\to \sqrt{z_c} c^\pm,&
\cbar^\pm &\to \sqrt{\ZW}^{-1} \cbar^\pm
\end{align} 
and the renormalization of the gauge-fixing parameters are given by
\begin{equation}
\zeta^{\phi_i \phi_j} \to
\zeta^{\phi_i \phi_j}+\deltan\zeta^{\phi_i \phi_j}.
\end{equation}
As in the gauge-boson sector, the second parametrization of section 
\ref{sec:GenerateCTs} is used:
\begin{align}
\sqrt{\Zc}&=\twomat{ \cos\theta_{W{\rm bare}} & \sin\theta_{W{\rm bare}} \\
                    -\sin\theta_{W{\rm bare}} & \cos\theta_{W{\rm bare}}}
\twomat{\sqrt{z_{V'}} & 0 \\ 0 & \sqrt{z_c}} \Z_c,\\
\sqrt{\Zcbar}&=\twomat{ \cos\theta_{W{\rm bare}} & \sin\theta_{W{\rm bare}} \\
                       -\sin\theta_{W{\rm bare}} & \cos\theta_{W{\rm bare}}}
\twomat{\sqrt{z_{V'}}^{-1} & 0 \\ 0 & \sqrt{z_V}^{-1}} \Z_{\bar{c}}.
\end{align}
The result for the invariant counterterms of order $\hbar^n$ reads
\begin{eqnarray}
\lefteqn{\Ginv^{(n)}|_{\rm gh}=
\int\dx\bigg\{-(\cbar_A,\cbar_Z) \bigg[\Box 
\frac{\deltan\Zcbar^T + \deltan\Zc}{2}
+\frac12\deltan\Zcbar\twomat{0 & \zeta^{AG^0} \\ 0 & \zeta^{ZG^0}} M_Z^2
}\nonumber\\&&{}
+\twomat{0 & \zeta^{AG^0} \\ 0 & \zeta^{ZG^0}}\frac12\deltan\Zc M_Z^2
+\twomat{0 & \zeta^{AG^0} \\ 0 & \zeta^{ZG^0}}\deltan M_Z^2
+M_Z^2 \twomat{0 & \deltan\zeta^{AG^0} \\ 0 & \deltan\zeta^{ZG^0}}\bigg]
\twovect{c_A}{c_Z}
\nonumber\\&&{}
-(\cbar^-,\cbar^+)\bigg[
\left(\Box+\zeta^{G^\pm}M_W^2\right)\frac{\deltan z_c-\deltan \ZW}{2}
 + \deltan\zeta^{G^\pm} M_W^2 + \zeta^{G^\pm} \deltan M_W^2 \bigg]
 \twovect{c^+}{c^-}
\bigg\}.
\nonumber\\&&{}
\end{eqnarray}
Using the classical value $\zeta^{AG^0}=0$ corresponding to the $R_\xi$ gauge 
(see section \ref{sec:GammaCl}) the contributions to the self energies 
appearing in the normalization conditions read
\begin{subequations}
\begin{align}
\label{eq:cAnorm}
\deltan\Gamma_{c_A\cbar_A}|_{p^2=0} & = 0,\\
\deltan\Gamma_{c_A\cbar_Z}|_{p^2=0} & = 
- \tfr12  \zeta^{ZG^0} M_Z^2 (\deltan\Zc)_{ZA},\\
\deltan\Gamma_{c_Z\cbar_A}|_{p^2=0} & = 
- \tfr12 \zeta^{ZG^0} M_Z^2 (\deltan\Zcbar)_{ZA}
- \deltan\zeta^{AG^0} M_Z^2, \\
\label{eq:cZ}
\deltan\Gamma_{c_Z\cbar_Z}|_{p^2=M_{c_Z}^2} & = 
- \zeta^{ZG^0}\deltan M_Z^2 
- \deltan\zeta^{ZG^0}M_Z^2,\\
\deltan\Gamma_{c_A\cbar_Z}|_{p^2=M_{c_Z}^2} & = 
\tfr12 (\deltan\Zcbar)_{AZ} \zeta^{ZG^0} M_Z^2,\\
\deltan\Gamma_{c_Z\cbar_A}|_{p^2=M_{c_Z}^2} & = 
\tfr12 (\deltan\Zc)_{AZ} \zeta^{ZG^0} M_Z^2
- \deltan\zeta^{AG^0} M_Z^2,\\
\label{eq:cpm}
\deltan\Gamma_{c^+\cbar^-}|_{p^2=M_{c^\pm}^2} & = 
- \deltan \zeta^{G^\pm} M_W^2 
- \zeta^{G^\pm} \deltan M_W^2, \\
\partial_{p^2}\deltan\Gamma_{c_A\cbar_A}|_{p^2=0} & = 
\tfr12 \left[(\deltan\Zc)_{AA} + (\deltan\Zcbar)_{AA}\right],\\
\partial_{p^2}\deltan\Gamma_{c_Z\cbar_Z}|_{p^2=M_{c_Z}^2} & = 
\tfr12 \left[(\deltan\Zc)_{ZZ}  + (\deltan\Zcbar)_{ZZ} \right],\\
\partial_{p^2}\deltan\Gamma_{c^+\cbar^-}|_{p^2=M_{c^\pm}^2} & = 
\tfr12 \left(\deltan z_c - \deltan\ZW \right).
\end{align}
\end{subequations}

As in the case of the photon, there 
is no invariant counterterm that can be adjusted in order to
satisfy the normalization condition (\ref{eq:cAnorm}). However, 
$\Gamma_{c_A\cbar_A}|_{p^2=0}=0$ holds automatically as a consequence 
of the ST identity if $\Gamma_{c_A\cbar_Z}|_{p^2=0}=0$ and 
$\Gamma_{c_Z\cbar_A}|_{p^2=0}=0$.

The gauge parameters $\zeta^{AG^0}$, $\zeta^{ZG^0}$, $\zeta^{G^\pm}$ and
their counterterms are fixed by the gauge-fixing conditions
(\ref{eq:gaugecondition}). Since there are no free counterterms 
available in (\ref{eq:cZ}) and (\ref{eq:cpm}), the values of $\M_{c_Z}^2$, 
$\M_{c^\pm}^2$ are dependent parameters. They are determined via the 
conditions $\Gamma_{c_Z\cbar_Z}|_{p^2=\M_{c_Z}^2}=0$ and
$\Gamma_{c^+\cbar^-}|_{p^2=\M_{c^\pm}^2}=0$.

In this way all desired normalization conditions in the ghost sector can 
be satisfied.

\subsubsection{Scalar and longitudinal gauge-boson sector}
\label{sec:A0longmixing}

Furthermore, we consider the normalization conditions in the sector of 
neutral, CP-odd, and charged Higgs bosons, of longitudinal gauge bosons, 
and of $B$ fields. 

Using the renormalization transformations
\begin{align}
M_{A^0}^2 &\to M_{A^0}^2+\delta M_{A^0}^2,\\
\twovect{G^0}{A^0} &\to \sqrt{Z_{A^0}} \twovect{G^0}{A^0},\\
\twovect{G^{\pm}}{H^{\pm}} &\to \sqrt{Z_{H^\pm}}\twovect{G^\pm}{H^\pm}
\end{align}
in the classical action, we obtain the invariant counterterms 
of order $\hbar^n$. 
We use the second parametrization (\ref{eq:ZParam2}):
\begin{align}
\sqrt{Z_{A^0,H^\pm}}&=\twomat{ \cos\beta_{\rm bare} & \sin\beta_{\rm bare} \\
                              -\sin\beta_{\rm bare} & \cos\beta_{\rm bare}}
\twomat{\sqrt{z_{H_1}} & 0 \\ 0 & \sqrt{z_{H_2}}} \Z_{A^0,H^\pm}.
\end{align}
Dividing the present sector into a pure Higgs-boson part 
$\Ginv|_{HH}$, a Higgs--gauge-boson part $\Ginv|_{HV}$, and a 
Higgs--$B$-field part $\Ginv|_{HB}$, we obtain for the respective 
counter\-term contributions:
\begin{eqnarray}
\lefteqn{\Ginv^{(n)}|_{HH} = \int\dx\bigg\{
-\frac12 (G^0,A^0)\Box \deltan Z_{A^0}\twovect{G^0}{A^0}}
\nonumber\\&&{}
-(G^0,A^0)\bigg[\twomat{0 & 0 \\ 0 & M_{A^0}^2}\deltan Z_{A^0}
+\twomat{0 & 0 \\ 0 & \deltan M_{A^0}^2}\bigg]
\twovect{G^0}{A^0} 
\nonumber\\&&
+(G^-,\ H^-)\bigg[-\Box \frac{\deltan Z_{H^\pm}^T+\deltan Z_{H^\pm}}{2}
-\twomat{0 & 0 \\ 0 & \deltan M_{H^\pm}^2}
\nonumber\\&&
-\frac12\bigg(\deltan Z_{H^\pm}^T
 \twomat{ 0 & 0 \\ 0 & M_{H^\pm}^{2(0)}}
+\twomat{ 0 & 0 \\ 0 & M_{H^\pm}^{2(0)}} \deltan Z_{H^\pm}
\bigg)\bigg]
\twovect{G^+}{H^+}\bigg\}
\nonumber\\&&
+\mbox{tadpoles},
\\
\lefteqn{\Ginv^{(n)}|_{HV} = 
\int\dx\bigg\{- (\partial_\mu G^0,\partial_\mu A^0)
\bigg[\frac{\deltan Z_{A^0}^T}{2} 
\twomat{0 & M_Z\\ 0 & 0} }
\nonumber\\&&{}
+ \twomat{ 0 & M_Z\\ 0 & 0} \frac{\deltan Z_V}{2}
+ \twomat{ 0 & \deltan M_Z \\ 0& 0} \bigg]
  \twovect{A^\mu}{Z^\mu} 
\nonumber\\&&
+\bigg[-\i (\partial_\mu G^+,\partial_\mu H^+)\bigg[
\frac{\deltan Z_{H^\pm}^T+\deltan \ZW}{2} \twovect{M_W}{0}
+ \twovect{
\deltan M_W}{0}
\bigg] W^{-\mu}
 + {\rm c.c.}\bigg]\bigg\}
\nonumber\\&&{}
+\mbox{tadpoles}.
\end{eqnarray}
The part $\Ginv|_{HB}$ involving $B$ fields will be discussed
separately later. 

The tadpole contributions contain the counterterms $\deltan t_1$, 
$\deltan t_2$, which are fixed by the conditions $\delta\Gamma/\delta H^0=0$ 
and $\delta\Gamma/\delta h^0= 0$. Since tadpole contributions contain no 
other adjustable counterterms, they are not written explicitly.

The constant $\deltan M_{H^\pm}^2{}^{(0)}$ is not a free parameter, but
it is determined by the lowest-order 
relation $M_{H^\pm}^2{}^{(0)}=M_{A^0}^2+M_W^2$ (\ref{eq:MHpmDef}) and hence 
$\deltan M_{H^\pm}^2{}^{(0)}=\deltan M_{A^0}^2 + \deltan  M_W^2$ 
(see section \ref{sec:physicalparameters}). 

The counterterms contribute in the following way to the vertex
functions involving the physical Higgs bosons $A^0$, $H^\pm$ 
(up to tadpoles):
\begin{subequations}
\begin{align}
\deltan\Gamma_{A^0A^0}|_{p^2=M_{A^0}^2} & = - \deltan M_{A^0}^2,\\
\deltan \Gamma_{A^0G^0}|_{p^2=M_{A^0}^2} & = 
 \tfr12 M_{A^0}^2(\deltan Z_{A^0})_{AG},\\
\deltan \Gamma_{A^0 A_\mu}|_{p^2=M_{A^0}^2} & = 0,\\
\deltan \Gamma_{A^0 Z_\mu}|_{p^2=M_{A^0}^2} & =
-\i  p^\mu \tfr12 (\deltan Z_{A^0})_{AG} M_Z,\\
\deltan \Gamma_{H^+H^-}|_{p^2=M_{H^\pm}^2} & = 
- \deltan M_{H^\pm}^2,\\
\deltan \Gamma_{H^+G^-}|_{p^2=M_{H^+}^2} & = 
\tfr12 (\deltan Z_{H^\pm})_{HG} M_{H^\pm}^2{}^{(0)} ,\\
\deltan \Gamma_{H^+ W^-_\mu}|_{p^2=M_{H^\pm}^2} & = 
p^\mu\tfr12 (\deltan Z_{H^\pm})_{HG} M_W .
\end{align}
\end{subequations}
The contributions to the vertex functions involving would-be  Goldstone bosons
read
\begin{subequations}
\begin{align}
\deltan\Gamma_{G^0G^0}|_{p^2=0} & = 0,\\
\deltan\Gamma_{A^0G^0}|_{p^2=0} & = 
- M_{A^0}^2 (\deltan Z_{A^0})_{GH},\\
\deltan\Gamma_{G^+G^-}|_{p^2=0} & = 0,\\
\deltan\Gamma_{G^+H^-}|_{p^2=0} & = 
- M_{H^\pm}^2{}^{(0)} (\deltan Z_{H^\pm})_{GH} .
\end{align}
\end{subequations}

Clearly, the demixing conditions for $\Gamma_{A^0G^0}$, $\Gamma_{H^+G^-}$ 
can be satisfied by a suitable choice of the non-diagonal elements of  
$\deltan Z_{A^0,H^\pm}$, and the on-shell condition for $\Gamma_{A^0A^0}$ 
can be satisfied by choosing $\deltan M_{A^0}^2$ appropriately.

However, in the sector of the charged scalars $H^\pm$, $G^\pm$, the 
counterterm $\deltan M_{H^\pm}^2$ is not independent as discussed above,
and hence $M_{H^\pm}^2$ is not a free parameter.

We turn to an important question mentioned already at the end of section
\ref{sec:normcond}. It is posed by the mixings between the scalars
and the longitudinal gauge bosons: between $A^0$ and $A^\mu$, $A^0$ and 
$Z^\mu$, and $H^\pm$ and $W^{\pm\mu}$. 
No counterterms are available to
cancel these mixings. Therefore, the Higgs propagators are not obtained 
from inverting only the Higgs self-energy matrix but by inverting the 
larger matrix
\begin{eqnarray}
\GG{HVB} & = & \left(\begin{array}{c|c}
\begin{array}{cc}\GG{h_ih_j} & \GG{h_iV^b}\\
                 \GG{V^ah_j} & \GG{V^aV^b}
\end{array}
 & \begin{array}{c}\GG{h_iB^b} \\ \GG{V^aB^b} \end{array}
\\
\hline
\begin{array}{cc}
\GG{B^ah_j} & \GG{B^aV^b} \end{array}
& \GG{B^aB^b} 
\end{array}\right),
\label{GammaHVB}
\end{eqnarray}
where for simplicity $h_i$, $V^a$, $B^a$ have been written for all
Higgs, gauge-boson, and $B$-field degrees of freedom. It is not obvious 
whether $\M_{A^0}^2$, $\M_{H^\pm}^2$ as defined by (\ref{eq:AGpole}),
(\ref{eq:HPMpole}) coincide with the poles of the propagators, 
i.e.\ with the zeros of ${\rm det}(\Gamma_{HVB})$. The ST
identity answers this question (c.f.\ \cite{Gambino:1999ai}). We
need the ingredients
\begin{eqnarray}
0 & = & 
\GG{c^c Y_{V}^b}\GG{V^a V^b} + \GG{c^c Y_{h_j}}\GG{V^a h_j}
\label{GZSTIV},\\
0 & = &
\GG{c^c Y_{V}^b}\GG{h_i V^b} + \GG{c^c Y_{h_j}}\GG{h_i h_j}
\label{GZSTIH},
\end{eqnarray}
and the fact that $\GG{c^c Y_{V^b}}$ is invertible for $p\ne0$ since  
$\GG{c^c Y_V^{b\mu}}(p,-p) = -\i p_\mu\delta_{cb}+\O(\hbar)$. 
The relations (\ref{GZSTIV}), (\ref{GZSTIH}) follow from the ST 
identities $0 = \delta^2\S(\Gamma)/(\delta V^a \delta c^c)$ and 
$0 = \delta^2\S(\Gamma)/(\delta h_i \delta c^c)$, respectively.

Firstly we consider the restrictions of gauge invariance on the
submatrix $\Gamma_{HV}$, defined in analogy to (\ref{GammaHVB}). From
(\ref{GZSTIV}), (\ref{GZSTIH}) we obtain
\begin{eqnarray}
\Gamma_{HV}d^{(c)} & = & 0\quad \mbox{with}\quad 
d^{(c)}= \twovect{\GG{c^c Y_{h_j}}}{ \GG{c^c Y_V^b}},
\end{eqnarray}
so for every generator of the gauge group there is a zero mode of
$\Gamma_{HV}$ for any $p^2$. 
However, if we suppose $p^2$ is chosen such that ${\rm det}(\GG{h_ih_j})=0$,
then there are coefficients $d^{(0)}_j$ with $\GG{h_ih_j}d^{(0)}_j=0$.
Using (\ref{GZSTIH}), $\GG{h_ih_j}d^{(0)}_j=0$ implies also 
$\GG{V^ah_j}d^{(0)}_j=0$. Thus
\begin{eqnarray}
\Gamma_{HV}d^{(0)} & = & 0\quad \mbox{with}\quad
d^{(0)} = \twovect{d^{(0)}_j}{0},
\end{eqnarray}
i.e.\ there is an additional zero mode of $\Gamma_{HV}$. Owing to
the invertibility of $\GG{c^c Y_{V^b}}$, the $d^{(c)}$  with
$c=0,\dots,n$ constitute $n+1$ linearly independent vectors, where $n$
is the number of gauge-group generators.

For $\Gamma_{HVB}$ this implies that there are $n+1$ linearly
independent vectors with 
\begin{equation}
\Gamma_{HVB}
\left(\begin{array}{c}d^{(c)}\\ \hline 0\end{array}\right)
=
\left(\begin{array}{c}0\\ \hline x^a\end{array}\right),
\end{equation} 
where the non-zero entries $x^a$ are confined to the
lower $n$ rows corresponding to the $B^a$ [c.f.\ (\ref{GammaHVB})]. 
Hence, there must be a non-trivial linear combination of the $d^{(c)}$ 
which is annihilated by $\Gamma_{HVB}$, so that ${\rm det}(\Gamma_{HVB})=0$ 
holds as was to be shown. 

Thus, ${\rm det}(\GG{h_ih_j})=0$   for $p^2\ne0$ implies ${\rm
det}(\GG{HVB})=0$. Therefore, the conditions (\ref{eq:AGpole}),
(\ref{eq:HPMpole}) really characterize $\M_{A^0,H^\pm}^2$ as poles of
the Higgs propagators. It is noteworthy that in this argument the
form of the gauge fixing and thus the $B$-dependent terms are
irrelevant.

In the Goldstone sector there are also no free renormalization
constants available to satisfy the diagonal normalization conditions 
 $\GG{G^0G^0}=\GG{G^+G^-}=0$ at $p=0$. 
Again these conditions hold automatically as a consequence of the ST
identity. As shown in section \ref{sec:photonmassless},  the
eigenvalue zero of the Higgs self-energy matrix at $p=0$ is threefold
degenerate, corresponding to the three spontaneously broken gauge
symmetries. Since also $\Gamma_{G^0A^0}=\Gamma_{G^+H^-}=0$ at $p=0$, 
this implies that $\Gamma_{G^0G^0}=\Gamma_{G^+G^-}=0$ at $p=0$. The three 
fields $G^{0,\pm}$ thus correspond to the three massless directions of the 
scalar potential.
In appendix \ref{app:GoldstQuant} the characterization of the 
Goldstone modes is discussed in more detail, and explicit results for the 
derivatives $\delta/\delta G^a$ are given.

The conditions for the residues of the propagators can be satisfied by
a suitable choice of the diagonal field-renormalization constants:
\begin{subequations}
\begin{align}
\deltan\partial_{p^2}\Gamma_{A^0A^0}|_{p^2=M_{A^0}^2} & = 
(\deltan Z_{A^0})_{GG},\\
\deltan\partial_{p^2}\Gamma_{H^+H^-}|_{p^2=M_{H^\pm}^2} & = 
(\deltan Z_{H^\pm})_{GG},\\
\deltan\partial_{p^2}\Gamma_{G^0G^0}|_{p^2=0} & = (\deltan Z_{A^0})_{AA},\\
\deltan\partial_{p^2}\Gamma_{G^+G^-}|_{p^2=0} & = (\deltan Z_{H^\pm})_{HH}.
\end{align}
\end{subequations}

We come back to the part of the vertex functional involving the $B$ fields 
$\Gamma_{HB}$. There are no loop contributions to this part of the
vertex functional, and this part is completely governed by the 
gauge-fixing conditions (\ref{eq:gaugecondition}) and the form of the 
gauge-fixing functions (\ref{eq:gaugefixingfct}). Clearly, the 
demixing requirements involving $B$ fields,
$\GG{H^+B^-}=\GG{A^0B_A}=\GG{A^0B_A}=0$, have to be compatible with
the gauge-fixing functions. 

In fact, we have anticipated these normalization conditions in section
\ref{sec:gaugefixing} by requiring that $\F^{a,\prime}|_{\Phi=0}$
depend only on the Goldstone fields but not on the physical Higgs
fields $H^\pm$, $A^0$. This requirement forced us to introduce the
background Higgs multiplets $\Phi_i^{a,\prime}$ and the construction
of section \ref{sec:gaugefixing}. On the other hand, it guarantees the demixing
conditions $\GG{H^+B^-}=\GG{A^0B_A}=\GG{A^0B_A}=0$ so that the
$H^\pm$, $A^0$ fields are as close to mass eigenstates as possible.

\subsubsection*{CP-even Higgs bosons}

In the former sectors the ST identity is always required 
for the proof that all normalization conditions can be satisfied. In the 
remaining sectors, there are enough counterterms at our disposal, and 
the ST identity need not be considered. Therefore, we discuss this sections
only briefly and emphasize the common structure of all sectors. 

We continue with the CP-even part of the Higgs sector. The renormalization 
transformation reads
\begin{equation}
\twovect{H^0}{h^0} \to \sqrt{Z_{H^0}} \twovect{H^0}{h^0}
\end{equation}
with
\begin{equation}
\sqrt{Z_{H^0}}=\twomat{ \cos\alpha_{\rm bare} & \sin\alpha_{\rm bare} \\
                        -\sin\alpha_{\rm bare} & \cos\alpha_{\rm bare}}
\twomat{\sqrt{z_{H_1}} & 0 \\ 0 & \sqrt{z_{H_2}}} \Z_{H^0}.
\end{equation}
The generated invariant counterterms are
\begin{eqnarray}
\Ginv^{(n)}|_{Hh}  & = & \int\dx \frac12(H^0,h^0)\bigg[
-\Box \deltan Z_{H^0} - 
\twomat{\deltan M_{H^0}^2 & 0 \\ 0 &\deltan M_{h^0}^2}
\nonumber\\&&{}
+ \twomat{M_{H^0}^2{}^{(0)} & 0 \\ 0 & M_{h^0}^2{}^{(0)}} \deltan Z_{H^0} 
\bigg] \twovect{H^0}{h^0} 
\nonumber\\&&{}+\mbox{tadpoles}.
\label{eq:CPevenHiggs}
\end{eqnarray}
The mass parameters appearing in (\ref{eq:CPevenHiggs}) are not free parameter
of the MSSM. Thus the on-shell conditions 
${\rm Re}\Gamma_{H^0H^0}|_{p^2=\M_{H^0}^2}=0$ 
and ${\rm Re}\Gamma_{h^0h^0}|_{p^2=\M_{h^0}^2}=0$ determine 
the values of $\M_{H^0}^2$, $\M_{h^0}^2$. The remaining on-shell 
conditions in this sector can be satisfied by adjusting the 
field-renormalization matrix $\deltan Z_{H^0}$. 

\subsubsection*{Charginos and neutralinos}

In the chargino-neutralino sector, it is more convenient to apply the first 
kind of field renormalization as discussed in section \ref{sec:GenerateCTs}. 
Thus, we use the parametrization of the classical action in terms of 
symmetric fields and parameters and perform the following renormalization 
transformations:
\begin{align}
{\cal X} &\to {\cal X}+\delta {\cal X},&
{\cal Y} &\to {\cal Y}+\delta {\cal Y},\\
\twovect{\lambda^+_\alpha}{h_{2\alpha}^1}&\to
\twovect{\sqrt{z_\lambda}\lambda^+_\alpha}{\sqrt{z_{h_2}}h_{2\alpha}^1},&
\twovect{\lambda^+_\alpha}{h_{2\alpha}^1} &= 
\Z_{\chi^+} \twovect{\chi_{1\alpha}^+}{\chi_{2\alpha}^+},\\
\twovect{\lambda^-_\alpha}{h_{1\alpha}^2}&\to
\twovect{\sqrt{z_\lambda}\lambda^-_\alpha}{\sqrt{z_{h_1}}h_{1\alpha}^2},&
\twovect{\lambda^-_\alpha}{h_{1\alpha}^2} &= 
\Z_{\chi^-} \twovect{\chi_{1\alpha}^-}{\chi_{2\alpha}^-},\\
\left(\! \begin{array}{c} \lambda'_\alpha \\ \lambda^3_\alpha \\ 
      h_{1\alpha}^1 \\ h_{2\alpha}^2 \end{array}\!\right) &\to 
\left(\! \begin{array}{c} \sqrt{z_{V'}}\lambda'_\alpha \\
      \sqrt{z_\lambda}\lambda^3_\alpha \\ \sqrt{z_{h_1}}h_{1\alpha}^1 \\ 
      \sqrt{z_{h_2}}h_{2\alpha}^2 \end{array}\!\right) ,&
\left(\! \begin{array}{c} \lambda'_\alpha \\ \lambda^3_\alpha \\ 
      h_{1\alpha}^1 \\ h_{2\alpha}^2 \end{array}\!\right) 
&= \Z_{\chi^0} \left( \!\begin{array}{c} \chi_{1\alpha}^0 \\  
                     \chi_{2\alpha}^0 \\ \chi_{3\alpha}^0 \\\chi_{4\alpha}^0 
                     \end{array}\!\right).
\end{align}
This transformations reflect directly the results of section
\ref{sec:CTs}. The matrices ${\cal X}$, ${\cal Y}$ are defined in (\ref{YDef})
and (\ref{XDef}). Combining both field transformations, we can write in 
agreement with (\ref{eq:firstparam})
\begin{align}
\twovect{\lambda^+_\alpha}{h_{2\alpha}^1} &\to
\Z_{\chi^+}^{(0)}\sqrt{Z_{\chi^+}}
 \twovect{\chi_{1\alpha}^+}{\chi_{2\alpha}^+},\\
\twovect{\lambda^-_\alpha}{h_{1\alpha}^2} &\to 
\Z_{\chi^-}^{(0)} \sqrt{Z_{\chi^-}}
 \twovect{\chi_{1\alpha}^-}{\chi_{2\alpha}^-},\\
\left(\! \begin{array}{c} \lambda'_\alpha \\ \lambda^3_\alpha \\ 
      h_{1\alpha}^1 \\ h_{2\alpha}^2 \end{array}\!\right) 
&\to \Z_{\chi^0}^{(0)} \sqrt{Z_{\chi^0}}
 \left(\!\begin{array}{c} \chi_{1\alpha}^0 \\  \chi_{2\alpha}^0 \\ 
       \chi_{3\alpha}^0 \\\chi_{4\alpha}^0 \end{array}\!\right).
\end{align}
In the following ${\cal U}$, ${\cal V}$, ${\cal N}$ denote the inverse 
lowest-order demixing matrices $(\Z_{\chi^{+,-,0}}^{(0)})^{-1}$
 [c.f.\ (\ref{eq:UVNDef})]. Applying the transformations to the classical 
action yield the following invariant counterterms in the 
chargino-neutralino sector:
\begin{eqnarray}
\Ginv^{(n)}|_{\chi^\pm} &=& \int\dx \bigg\{ 
\sum_{i,j=1}^2 \bar{\chi}^-_i \i \gamma^\mu \partial_\mu 
\frac{1}{2} \left(\deltan Z_{\chi^+} + \deltan Z_{\chi^+}^T\right)_{ij} 
P_L \chi^+_j
\nonumber\\&&{}
+\bar{\chi}^+_i \i \gamma^\mu \partial_\mu 
\frac{1}{2}\left(\deltan Z_{\chi^-} + \deltan Z_{\chi^-}^T\right)_{ij} 
P_R \chi^-_j 
\nonumber\\&&{}
- \bigg(
  \bar\chi^-_i \left[{\cal U}^* \deltan {\cal X} {\cal V}^\dagger
    + \frac{1}{2}\left(\deltan Z_{\chi^-}^T M_{\chi^\pm} + M_{\chi^\pm} \deltan
      Z_{\chi^+}\right) \right]_{ij}
  P_L \chi^+_j + {\rm c.c.}\bigg)\bigg\},
\nonumber\\&&
\\
\Ginv^{(n)}|_{\chi^0} &=& \int\dx \bigg\{
\frac12\bar{\chi}^0_i
\i{\gamma}^\mu\partial_\mu \frac{1}{2}\left(\deltan Z_{\chi^0} + \deltan
  Z_{\chi^0}^T\right)_{ij}  \chi^0_j 
\nonumber\\&&{}
- \bar\chi^0_i\left[{\cal N}^* \deltan {\cal Y} {\cal N}^\dagger
   + \frac{1}{2}\left(\deltan Z_{\chi^0}^T M_{\chi^0} + M_{\chi^0} \deltan
     Z_{\chi^0}\right) \right]_{ij}
P_L \chi^0_j + {\rm c.c.}\bigg\}.\nonumber\\
\end{eqnarray}
The left- and right-handed projectors are denoted by 
$P_{L,R}=(1\pm\gamma_5)/2$.
The diagonal mass matrices are  
$M_{\chi^0}={\cal N}^* {\cal Y} {\cal N}^\dagger
={\rm diag}(m_{\chi^0_1},\ldots,m_{\chi^0_4})$,
$M_{\chi^\pm}={\cal U}^* {\cal X} {\cal V}^\dagger
={\rm diag}(m_{\chi^\pm_1},m_{\chi^\pm_2})$.

In $\deltan {\cal X}$ and $\deltan {\cal Y}$ there appear on the one hand the 
three independent renormalization constants $\deltan\mu$, $\deltan M_1$,
$\deltan M_2$. On the other hand, $\deltan {\cal X}$ and $\deltan {\cal Y}$ 
depend on further renormalization
constants, namely $\deltan M_{W,Z}$ and $\deltan \tan\beta$. The
latter constants are fixed in other sectors --- $\deltan M_{W,Z}$ have
been fixed in the gauge-boson sector, and $\deltan \tan\beta$ has not been
fixed yet, but as it is a physical parameter we choose not to dispose
of it via an on-shell or demixing condition.

Hence,
the three on-shell conditions for the poles at $\M_{\chi^+_{1,2},\chi^0_1}^2$
can be satisfied by choosing
$\deltan \mu$, $\deltan M_1$, $\deltan M_2$ appropriately.
In this way, all masses in the chargino-neutralino sector are 
determined.

The remaining normalization conditions in this sector are the demixing
conditions and the conditions for the residua. It is easy to see that
these conditions can all be satisfied by one unique choice of the
$\deltan Z_{\chi^{+,-,0}}$.

\subsubsection{Quarks and leptons}

The renormalization of the quarks and the charged leptons is
identical, so we consider the up-quark $u$ as an example. 
The relevant renormalization transformations are 
\begin{eqnarray}
m_u &\to & m_u+\delta m_u,\\
u & \to &\sqrt{Z_{u_L}}P_L u + \sqrt{Z_{u_R}}P_R u,\\
\sqrt{Z_{u_L}} & = & \sqrt{z_q}\Z_{u_L},\\
\sqrt{Z_{u_R}} & = & \sqrt{z_u}.
\end{eqnarray}
These transformations result in the following counterterms of order $\hbar^n$:
\begin{eqnarray}
\Ginv^{(n)}|_{u} & = & \int\dx \bar{u}
\bigg[\i\gamma^\mu\partial_\mu (\deltan Z_{u_L}P_L+\deltan Z_{u_R}P_R)
\nonumber\\&&{}
-\deltan m_u - m_u\frac{\deltan Z_{u_L}+\deltan Z_{u_R}}{2}
\bigg]u.
\end{eqnarray}
Obviously, the on-shell condition for the mass and for the residue can
be satisfied by adjusting the renormalization constants $\deltan m_u$
and $\deltan Z_{u_{L,R}}$.

\subsubsection{Squarks and sleptons}

In the squark-slepton sector we apply the first kind of
parametrization discussed in section \ref{sec:GenerateCTs}, like in
the chargino-neutralino sector. The renormalization transformations
read
\begin{align}
M_{\tilde{f}}^2 &\to M_{\tilde{f}}^2 + \delta
M_{\tilde{f}}^2,
\qquad \tilde{f}=\tilde{e},\tilde{d},\tilde{u},\\
\twovect{Q^1}{U^\dagger} & \to  
\Rsu^{(0)}\sqrt{Z_{\tilde{u}}}\twovect{\tilde{u}_1}{\tilde{u}_2},\\
\twovect{Q^2}{D^\dagger} & \to  
\Rsd^{(0)}\sqrt{Z_{\tilde{d}}}\twovect{\tilde{d}_1}{\tilde{d}_2},\\
\twovect{L^2}{R^\dagger} & \to 
\Rse^{(0)}\sqrt{Z_{\tilde{e}}}\twovect{\tilde{e}_1}{\tilde{e}_2},\\
L^1 & = \sqrt{z_{L}} \tilde{\nu}_1,
\end{align}
where 
\begin{align}
\Rsu^{(0)}\sqrt{Z_{\tilde{u}}}&
=\twomat{\sqrt{z_{Q}} & 0 \\ 0 & \sqrt{z_{U}}} \Rsu,\\
\Rsd^{(0)}\sqrt{Z_{\tilde{d}}}&
=\twomat{\sqrt{z_{Q}} & 0 \\ 0 & \sqrt{z_{D}}} \Rsd ,\\
\Rse^{(0)}\sqrt{Z_{\tilde{e}}}&
=\twomat{\sqrt{z_{L}} & 0 \\ 0 & \sqrt{z_{R}}} \Rse
\end{align}
in agreement with the general formula (\ref{eq:ZParam2}).
Applying these transformations to the classical action, the following 
counterterm contributions are obtained:
\begin{eqnarray}
\Ginv^{(n)}|_{\tilde{f}} &= & 
\int\dx \sum_{\tilde{f}=\tilde{u},\tilde{d},\tilde{e},\tilde{\nu}}
-(\tilde{f}_1^\dagger,\tilde{f}_2^\dagger)
\bigg[\frac{1}{2}(\deltan Z_{\tilde{f}}^T+\deltan Z_{\tilde{f}})\Box
\nonumber\\&&{}
     + \Z_{\tilde{f}}^{(0)}{}^T \deltan M_{\tilde{f}}^2
     \Z_{\tilde{f}}^{(0)} 
     +\frac{1}{2}(\deltan Z_{\tilde{f}}^T M_{\tilde{f}}^2
     + M_{\tilde{f}}^2  \deltan Z_{\tilde{f}})\bigg]
\twovect{\tilde{f}_1}{\tilde{f}_2},
\end{eqnarray}
where in the case of the sneutrino $\tilde{\nu}$ the matrix
$\deltan Z_{\tilde{f}}$ degenerates to $\deltan Z_{\tilde{\nu}}=\deltan z_L$ 
and $\deltan M_{\tilde{\nu}}^2$ has to be replaced by 
$\deltan m_{\tilde{\nu}}^2$. 

The counterterms, which are already determined  in other sectors, 
are $\deltan \mu$, $\deltan m_{e,d,u}$, $\deltan M_{W,Z}$, 
$\deltan \tan\beta$, and the $A$ parameters $\deltan A_{e,d,u}$. 
The remaining five counterterms appearing in $\deltan M_{\tilde{f}}^2$ are 
$\deltan m_{\tilde{l},\tilde{e},\tilde{q},\tilde{d},\tilde{u}}^2$. 
These free mass counterterms are sufficient to satisfy the five on-shell 
conditions (\ref{eq:onshellfermion})
for the sfermions 
$\tilde{f}=\tilde{u}_1,\tilde{u}_2,\tilde{d}_1,\tilde{\nu},\tilde{e}_1$.
After fixing these five counterterms, all masses in this sector are
determined. The on-shell conditions for $\tilde{f}=\tilde{d}_2,\tilde{e}_2$ 
can only be fulfilled by choosing 
$\M_{\tilde{d}_2}^2$, $\M_{\tilde{e}_2}^2$ appropriately.

The demixing conditions and the conditions for the residua fix
uniquely the values of the field renormalization constants 
$\deltan Z_{\tilde{u},\tilde{d},\tilde{e}}$, $\deltan z_L$.

\subsubsection{Summary}

The evaluation of the normalization conditions shows that all 
normalization conditions of section \ref{sec:normcond} can be satisfied.
However, two non-trivial aspects of the normalization conditions are
remarkable.

For the IR and Goldstone normalization conditions 
$\Gamma_{AA}^T=\Gamma_{c_A\cbar_A}=\GG{G^0G^0}=\GG{G^+G^-}=0$ at $p=0$,
there are no free invariant counterterms available. 
Rather, these conditions hold automatically 
as a consequence of the ST identity and the corresponding 
demixing conditions at $p=0$. Furthermore, owing to the scalar-vector
mixing, the definition of $\M_{A^0,H^\pm}^2$ by itself does not
guarantee that $\M_{A^0,H^\pm}^2$ correspond to the poles of the Higgs
propagators. Again the ST identity has been used to prove this
correspondence.

As a result of the normalization conditions, the counterterms to the
following quantities are 
determined: $M_{W,Z}^2$, $M_{A^0}^2$, $t_{1,2}$, $\mu$, $M_{1,2}$,
$m_{e,d,u}$, $m_{\tilde{l},\tilde{e},\tilde{q},\tilde{d},\tilde{u}}$,  
all field renormalization constants $\sqrt{Z_\phi}$, $\sqrt{z_\phi}$ 
(see section \ref{sec:CTs}) except for certain combinations of 
$\sqrt{\Zc}$, $\sqrt{\Zcbar}$, $\sqrt{z_c}$.

The normalization conditions of section \ref{sec:normcond} do not determine 
the counterterms to the parameters $e$, $\tan\beta$, $A_{e,d,u}$. These 
physical parameters can be fixed via five additional normalization conditions.
  
\section*{Discussion and outlook}

In this paper we have presented the renormalization of the MSSM. The
main results are the formulation of the Ward and ST identities, the
$R_\xi$ gauge conditions and the on-shell normalization conditions.
We have shown that all these conditions can be satisfied
simultaneously and that the MSSM is multiplicatively renormalizable
and infrared finite.

More in detail, the elements of the construction are the
following. The basic structure of the symmetry identities is similar
to the one in the Standard Model \cite{Kraus:1997bi}, supersymmetry
being treated as in
\cite{Maggiore:1996gg,Hollik:1999xh,Hollik:2000pa,Maggiore:1995gr}.
Supersymmetry and gauge invariance are formulated in terms of a ST
identity that involves supersymmetry and translational ghosts in
addition to the usual Faddeev-Popov ghosts; the Abelian subgroup is
specified by a local Ward identity, and soft breaking is introduced
via an external chiral supermultiplet. In order to account for the
spontaneous breaking of gauge invariance, we reparametrize all fields
in terms of physical fields with a more direct correspondence to mass
eigenstates. The vacuum expectation values of the Higgs fields
themselves are extended to external fields $(\Phi+\mbv)$, since
otherwise the $R_\xi$ gauge fixing would explicitly break global gauge
invariance. 

However, unlike in the simpler models studied in
\cite{Kraus:1997bi,Kraus:1995jk}, it is impossible in the MSSM to 
use external fields with the multiplet structure of the dynamical
Higgs fields, i.e.\ two isospin doublets $(\Phi+\mbv)_{1,2}$. The
reason is the mixing of the Goldstone fields 
$G^{0,\pm}$ with the physical Higgs fields $A^0,H^\pm$, which would
inevitably lead to the appearance of $A^0,H^\pm$ in the gauge-fixing
term. Instead, we choose two 8-component external fields transforming
according to the product of the adjoint and the fundamental
representation.

Further complications arise in the proof of the infrared finiteness
and the discussion of the normalization conditions. In order to prove
the infrared finiteness, we use on the one hand the ST identity like in
\cite{Kraus:1997bi} to constrain the appearance of the massless fields
in the Green functions. On the other hand, in the symmetry identities
also the external fields could cause infrared problems, and we have to
use a special  reparametrization to obtain the optimal assignment of
the infrared dimensions. In the case of the normalization conditions,
the ST identity is necessary to show that the on-shell conditions at
$p^2=0$ can be satisfied. Owing to the mixing of the physical Higgs
fields with unphysical Goldstone and longitudinal gauge fields, it is
not directly obvious that the on-shell condition for $M_{A^0}^2$ really
corresponds to the pole of the $A^0$ propagator. This correspondence
can again be proven via the ST identity. 

There is an interesting difference between the on-shell conditions in
the SM and the ones in the MSSM. In the SM all particle 
masses can be chosen as free input parameters and,
conversely, all parameters except for the electric charge are fixed
via the on-shell conditions. In the MSSM, however, the masses of $H^0$,
$h^0$, $H^\pm$, $\chi^0_{2,3,4}$, and of $\tilde{d}_2$, $\tilde{e}_2$
are not free. They are fixed via the symmetries as functions of the
free input parameters. Moreover, the free parameters
$\tan\beta$ and $A_{u,d,e}$ are usually not fixed via on-shell
conditions but via other conditions, e.g.\ in the interaction sector.

On the practical side, the results of the present paper yield the
basis for the determination of the counterterms in the MSSM. Since no
consistent symmetry-preserving regularization is known, in general
symmetry-restoring counterterms are needed. They are determined by the
symmetry identities and can be calculated using the same methods as in
\cite{Hollik:1999xh,Hollik:2001cz} in the case of supersymmetric QED and QCD. 
The invariant counterterms are determined by the normalization
conditions presented in section \ref{sec:renscheme}.

In the present approach to the renormalization of the MSSM the
non-renormalization theorems \cite{NRT1} and the relations  between the
renormalization constants of soft-breaking and supersymmetric parameters
\cite{NRT2} escaped the algebraic treatment since they are --- as it is well 
known --- not a direct consequence of supersymmetry. 
In order to derive these improved
renormalization properties in the framework of algebraic renormalization
one has to use the methods of \cite{NRT3,NRT4} with an extended version of the
classical action. The non-renormalization theorems of the
MSSM would yield stringent tests of concrete loop calculations, and
they would make visible the improved renormalization behaviour --- one
of the main motivations to study supersymmetry at all --- in terms of
explicit equations.

\begin{appendix}

\section{Notations and conventions} 
\label{ap:conventions}

We use the following conventions 
for the metric tensor $g^{\mu \nu}$ and the generators of the SU(2) gauge 
group $T^a$:
\begin{equation}
g^{\mu \nu}=\mathrm{diag}(1,-1,-1,-1) , \qquad
\T^a=\frac{1}{2} \sigma^a,
\end{equation}
where the Pauli matrices are denoted by $\sigma^a$, $a=1,2,3$.
Hence, the generators of the SU(2) gauge group obey the following relations:
\begin{equation}
[\T^a,\T^b]=\i \epsilon^{a b c} \T^c ,\qquad
2 \tr (\T^a \T^b) = \delta^{ab}
\end{equation}
with $\epsilon^{123}=+1$. Spinorial derivatives are defined as follows
\begin{equation}
\frac{\partial}{\partial \theta^\alpha} \theta^\beta=
{\delta_\alpha}^\beta,  \qquad
\frac{\partial}{\partial \bar\theta^{\dot{\alpha}}}
\bar\theta^{\dot{\beta}} =
{\delta_{\dot{\alpha}}}^{\dot{\beta}}
\end{equation}
and the superspace integrations read
\begin{equation}
\int\dV =\int \dx \D^2 \bar \D^2, \qquad
\int \dS =\int \dx \D^2
\end{equation}
with
\begin{equation}
\D_\alpha=\frac{\partial}{\partial \theta^\alpha}
-\i \sigma^\mu_{\alpha \dot{\alpha}} \bar\theta^{\dot{\alpha}} 
\partial_\mu, \qquad
\bar \D_{\dot{\alpha}}=- \frac{\partial}{\partial \bar\theta^{\dot{\alpha}}}
+\i \theta^\alpha \sigma^\mu_{\alpha \dot{\alpha}} \partial_\mu.
\end{equation}
Furthermore, we use the following notations:
\begin{equation}
\sigma^{\mu \nu}_{\alpha \beta}=\frac{\i}{2}
({\sigma^\mu}_{\alpha \dot{\alpha}} 
\bar{\sigma}^{\nu \dot{\alpha}}{}_\beta-
{\sigma^\nu}_{\alpha \dot{\alpha}} 
\bar{\sigma}^{\mu \dot{\alpha}}{}_\beta), \qquad
\bar\sigma^{\mu \nu}_{\dot{\alpha} \dot{\beta}}=\frac{\i}{2}
(\bar\sigma_{\dot{\alpha}}^{\mu \alpha} 
\sigma^\nu_{\alpha \dot{\beta}}
-\bar\sigma_{\dot{\alpha}}^{\nu \alpha} 
\sigma^\mu_{\alpha \dot{\beta}}).
\end{equation}

The field-strength tensors and covariant derivatives are defined by
\begin{subequations} 
\begin{align}
F_{\mu \nu}&=\partial_\mu V_\nu-\partial_\nu V_\mu+\i g[V_\mu,V_\nu] ,\\
F_{\mu \nu}^\prime&=\partial_\mu V_\nu^\prime-\partial_\nu V_\mu^\prime ,\\
\D_\mu \phi &=\left\{
\begin{array}{ll}
\partial_\mu \phi +\i g [V_\mu,\phi ] & 
\quad \mbox{adjoint representation}, \\
\partial_\mu \phi+\i g V_\mu \phi 
+\i g^\prime V^\prime_\mu \frac{Y}{2} \phi 
& \quad \mbox{isospin doublet}, \\
\partial_\mu \phi+\i g^\prime V^\prime_\mu \frac{Y}{2} \phi &
\quad \mbox{isospin singlet}.
\end{array}
\right.
\end{align}
\end{subequations}
In our conventions, the rules for complex conjugation are
\begin{subequations}
\begin{align}
\overline{\s \phi} &= \s \overline{\phi} &&\text{ for bosonic $\phi$,} \\
\overline{\s \phi} &= - \s \overline{\phi} && \text{ for fermionic $\phi$.}
\end{align}
\end{subequations}

\section{BRS transformations}\label{app:BRS}

The BRS transformations of the field components of the MSSM read:

\noindent Gauge multiplets:
\begin{subequations}
\begin{align}
\s V_\mu&= \D_\mu c 
+\eps^\alpha \sigma_{\mu \alpha \dot{\alpha}} 
 \bar\lambda^{\dot{\alpha}}
+\lambda^{\alpha} \sigma_{\mu \alpha \dot{\alpha}}
 \bar\eps^{\dot{\alpha}}
-\i \xi^\nu \partial_\nu V_\mu, \label{BRS_V}\\
\s \lambda_\alpha &=
 \frac{1}{2} \eps^\beta \sigma^{\mu \nu}_{\beta \alpha} F_{\mu \nu}
-\i g \{\lambda_\alpha,c\}
+\eps_\alpha D_V
-\i \xi^\mu \partial_\mu \lambda_\alpha,\\
\s \bar\lambda_{\dot{\alpha}}&=
-\frac{1}{2} \bar\sigma^{\mu \nu}_{\dot{\alpha}\dot{\beta}} 
 \bar\eps^{\dot{\beta}}F_{\mu \nu}
-\i g \{\bar\lambda_{\dot{\alpha}},c\}
-\bar\eps_{\dot{\alpha}} D_V
-\i \xi^\mu \partial_\mu \bar\lambda_{\dot{\alpha}},\\
\s D_V&=
 \i \D_\mu \lambda^\alpha \sigma^\mu_{\alpha \dot{\alpha}} 
 \bar\eps^{\dot{\alpha}}
-\i \D_\mu \eps^\alpha \sigma^\mu_{\alpha \dot{\alpha}} 
 \bar\lambda^{\dot{\alpha}}
+\i g [D_V,c]
-\i \xi^\mu \partial_\mu D_V,\\
\s V_\mu^\prime&= 
 \partial_\mu c^\prime 
+\eps^\alpha \sigma_{\mu \alpha \dot{\alpha}} 
 \bar\lambda^{\prime \dot{\alpha}}
+\lambda^{\prime \alpha} \sigma_{\mu \alpha \dot{\alpha}}
 \bar\eps^{\dot{\alpha}}
-\i \xi^\nu \partial_\nu V_\mu^\prime,\\
\s \lambda_\alpha^\prime &=
 \frac{1}{2}\eps^\beta \sigma^{\mu \nu}_{\beta \alpha} F_{\mu \nu}^\prime
+\eps_\alpha D_V^\prime
-\i \xi^\mu \partial_\mu \lambda_\alpha^\prime,\\
\s \bar\lambda_{\dot{\alpha}}^\prime&=
-\frac{1}{2}\bar\sigma^{\mu \nu}_{\dot{\alpha}\dot{\beta}} 
\bar\eps^{\dot{\beta}} F_{\mu \nu}^\prime
-\bar\eps_{\dot{\alpha}} D_V^\prime
-\i \xi^\mu \partial_\mu \bar\lambda^\prime_{\dot{\alpha}},\\
\s D_V^\prime&=
 \i \partial_\mu \lambda^{\prime \alpha} \sigma^\mu_{\alpha \dot{\alpha}} 
 \bar\eps^{\dot{\alpha}}
-\i \partial_\mu \eps^\alpha \sigma^\mu_{\alpha \dot{\alpha}} 
 \bar\lambda^{\prime \dot{\alpha}}
-\i \xi^\mu \partial_\mu D_V^\prime.
\end{align}
\end{subequations}
Matter Fields (The Higgs and quark fields transform analogously to 
$L$, $l$, $F_L$):
\begin{subequations}
\begin{align}
\s L&=- \i g c L
- \i g^\prime \frac{Y}{2} c^\prime L
+ \sqrt{2} \eps^\alpha l_\alpha
- \i \xi^\mu \partial_\mu L,\\
\s l_\alpha &=
- \i g c l_\alpha
-\i g^\prime \frac{Y}{2} c^\prime l_\alpha
+\sqrt{2} \eps_\alpha F_L
+\i \sqrt{2} \sigma^\mu_{\alpha \dot{\alpha}} 
 \bar\eps^{\dot{\alpha}}\D_\mu L
-\i \xi^\mu \partial_\mu l_\alpha,\\
\s F_L&=
 - \i g c F_L
-\i g^\prime \frac{Y}{2} c^\prime F_L
-2 g \bar\eps_{\dot{\alpha}} \bar\lambda^{\dot{\alpha}} L
-g^\prime Y \bar\eps_{\dot{\alpha}} 
 \bar\lambda^{\prime \dot{\alpha}} L
+\i \sqrt{2} \D_\mu l^\alpha \sigma^\mu_{\alpha \dot{\alpha}} 
 \bar\eps^{\dot{\alpha}} \nonumber \\
&\quad
-\i \xi^\mu \partial_\mu F_L,\\
\s R&=
-\i g^\prime \frac{Y}{2} c^\prime R
+\sqrt{2} \eps^\alpha r_\alpha
-\i \xi^\mu \partial_\mu R,\\
\s r_\alpha &=
-\i g^\prime \frac{Y}{2} c^\prime r_\alpha
+\sqrt{2} \eps_\alpha F_R
+\i \sqrt{2} \sigma^\mu_{\alpha \dot{\alpha}} 
 \bar\eps^{\dot{\alpha}}\D_\mu R
-\i \xi^\mu \partial_\mu r_\alpha,\\
\s F_R&=
-\i g^\prime \frac{Y}{2} c^\prime F_R
-g^\prime Y \bar\eps_{\dot{\alpha}} 
 \bar\lambda^{\prime \dot{\alpha}} R
+ \i \sqrt{2} \D_\mu r^\alpha \sigma^\mu_{\alpha \dot{\alpha}} 
 \bar\eps^{\dot{\alpha}}- \i \xi^\mu \partial_\mu F_R. 
\end{align}
\end{subequations}
Ghosts, Antighosts, and Lagrange multipliers:
\begin{subequations}
\begin{align}
\label{eq:ghostBRSstart}
\s c&=
-\i g c c 
+2 \i \eps^\alpha \sigma^\mu_{\alpha \dot{\alpha}} 
 \bar\eps^{\dot{\alpha}} V_\mu
-\i \xi^\mu \partial_\mu c,\\
\s \bar{c}&= B
-\i \xi^\mu \partial_\mu \bar{c},\\
\s B&= 2 \i \eps^\alpha \sigma^\mu_{\alpha \dot{\alpha}} 
 \bar{\eps}^{\dot{\alpha}} \partial_\mu \bar{c}
-\i \xi^\mu \partial_\mu B,\\
\s c^\prime&=
 2 \i \eps^\alpha \sigma^\mu_{\alpha \dot{\alpha}} 
 \bar\eps^{\dot{\alpha}} V_\mu^\prime
-\i \xi^\mu \partial_\mu c^\prime,\\
\s \bar{c}^\prime&= B^\prime
-\i \xi^\mu \partial_\mu \bar{c}^\prime,\\
\s B^\prime&= 2 \i \eps^\alpha \sigma^\mu_{\alpha \dot{\alpha}} 
 \bar{\eps}^{\dot{\alpha}} \partial_\mu \bar{c}^\prime
-\i \xi^\mu \partial_\mu B^\prime,\\
\s \xi^\mu &=
 2 \eps^\alpha \sigma^\mu_{\alpha \dot{\alpha}} 
 \bar\eps^{\dot{\alpha}},\\
\s \eps_\alpha&= 0,\\
\s \bar\eps_{\dot{\alpha}}&= 0. 
\label{BRS_epsdot}
\end{align}
\end{subequations}
$Y$ represents the hypercharge of the respective field as given in 
table \ref{ta:fields}. 

The commutation relations of the generators are encoded in the
transformations of the ghosts in (\ref{eq:ghostBRSstart})--(\ref{BRS_epsdot}): 
$c c$ in $\s c$ contains the commutator $[T^a,T^b]$ of SU(2) generators due 
to the anticommuting nature of $c$. The additional term 
$\eps^\alpha\sigma^\mu_{\alpha\dot{\alpha}}\bar\eps^{\dot{\alpha}} V_\mu$
in $\s c$ corresponds to the supersymmetry anticommutator $\{Q_\al, \bar
Q_\da\}$ and reflects the fact that in the Wess-Zumino gauge supersymmetry
transformations are accompanied by field dependent gauge transformations. 
The usual well-known part of the supersymmetry anticommutator is found in 
$\s \xi^\mu$. 

\section{Explicit results for the classical action} \label{ap:explicit}
\setcounter{equation}{0}

The results of the classical action in terms of component fields
are listed in the following:
\begin{subequations}
\begin{align}
& {} \nonumber 
\frac{1}{16}\int\dV
\bar{\hat{L}} \e^{2 g \hat{V}+ g^\prime Y \hat{V}^\prime} \hat{L}=
\int \dx \bigg(\overline{\D_\mu L} \D^\mu L
+\i l^{\alpha} \sigma^\mu_{\alpha \dot{\alpha}} \overline{\D_\mu l}{}^{\dot{\alpha}}
+\bar{F}_L F_L
+g \bar{L} D_V L
\\ & {} \qquad {}
-\sqrt{2} g \bar l_{\dot{\alpha}} \bar{\lambda}^{\dot{\alpha}}L   
-\sqrt{2} g \bar{L} \lambda^\alpha l_{\alpha}
+g^\prime \frac{Y}{2} \bar{L} D_V^\prime L
-g^\prime \frac{Y}{2} \sqrt{2} \bar l_{\dot{\alpha}} 
 \bar{\lambda}^{\prime\dot{\alpha}}L   
-g^\prime \frac{Y}{2} \sqrt{2} \bar{L} \lambda^{\prime \alpha} l_{\alpha}
\bigg), \\ & {} \nonumber
\frac{1}{16}\int \dV 
\bar{\hat{R}} \e^{g^\prime Y \hat{V}^\prime} \hat{R}=
\int \dx \bigg(\overline{\D_\mu R} \D^\mu R
+\i r^{\alpha} \sigma^\mu_{\alpha \dot{\alpha}} \overline{\D_\mu r}{}^{\dot{\alpha}}
+\bar{F_R} F_R
+g^\prime \frac{Y}{2} \bar{R} D_V^\prime R
\\ & {} \qquad {}
-g^\prime \frac{Y}{2} \sqrt{2} \bar r_{\dot{\alpha}} 
 \bar{\lambda}^{\prime\dot{\alpha}}R   
-g^\prime \frac{Y}{2} \sqrt{2} \bar{R} \lambda^{\prime \alpha} r_{\alpha}
\bigg), \\ & {}
-\frac{1}{512 g^2} \int \dS
2 \tr[ \hat{F}^\alpha \hat{F}_\alpha]=
 \int \dx 2 \tr\bigg[
-\frac{1}{4} F_{\mu \nu} F^{\mu \nu}
+\i \lambda^\alpha \sigma^\mu_{\alpha \dot{\alpha}} 
 \D_\mu \bar\lambda^{\dot\alpha}
+\frac{1}{2}  D_V^2 
\bigg], \\ & {} \nonumber
-\frac{1}{512 g^2} \int \dS 
\hat{A} 2 \tr[ \hat{F}^\alpha \hat{F}_\alpha]=
\int \dx 2 \tr\bigg[
-\au \eps^\alpha \lambda_\alpha D_V
+\frac{1}{2} \au \lambda^\alpha 
 {{\sigma^{\mu \nu}}_\alpha}^\beta \eps_\beta F_{\mu \nu} 
\nonumber \\ & {} \qquad {}
+\frac{1}{2} (\av+v_A) \lambda^\alpha \lambda_\alpha 
\bigg], \\ & {} 
-\frac{1}{128 g^{\prime 2}} \int \dS  
\hat{F}^{\prime \alpha} \hat{F}^\prime_{\alpha}=
\int \dx \bigg(
-\frac{1}{4}\hat{F}^\prime_{\mu \nu} \hat{F}^{\prime\mu \nu}
+\i \lambda^{\prime \alpha} \sigma^\mu_{\alpha \dot{\alpha}} 
  \partial_\mu \bar\lambda^{\prime \dot\alpha}
+\frac{1}{2} D_V^{\prime 2}
\bigg), \\ & {} \nonumber
-\frac{1}{128 g^{\prime 2}} \int \dS 
\hat{A} \hat{F}^{\prime \alpha} \hat{F}_\alpha^\prime=
\int \dx \bigg[-\au \eps^\alpha \lambda_\alpha^\prime D_V^\prime
+\frac{1}{2} \au \lambda^{\prime \alpha} 
 {{\sigma^{\mu \nu}}_\alpha}^\beta \eps_\beta F_{\mu \nu}^\prime 
\\ & {} \qquad {}
+\frac{1}{2} (\av+v_A) \lambda^{\prime \alpha} \lambda_\alpha^\prime 
\bigg], \\ & {} \nonumber
-\frac{1}{4} \int \dS 
\hat{H}_1^T (\i \sigma_2) \hat{L} \hat{R}=
\int \dx \bigg[
-H_1^T (\i \sigma_2) l r 
+H_1^T (\i \sigma_2) F_L R
+H_1^T (\i \sigma_2) L F_R
\\ & {} \qquad {}
-h_1^{T \alpha} (\i \sigma_2) l_\alpha R
-h_1^{T \alpha} (\i \sigma_2) L r_\alpha
+F_1^T (\i \sigma_2) L R 
\bigg], \\ & {} 
\frac{1}{4} \int \dS 
\hat{H}_1^T (\i \sigma_2) \hat{H}_2=
\int \dx \bigg[
-H_1^T (\i \sigma_2) F_2 
-F_1^T (\i \sigma_2) H_2
+h_1^{T \alpha} (\i \sigma_2) h_{2 \alpha}\bigg]
\end{align}
\end{subequations}
with
\begin{subequations}
\begin{align}
\hat{F}^\alpha
&=4 g \e^{-\i \theta \sigma^\mu \bar\theta \partial_\mu} 
\left( - 2 \lambda^\alpha - 2 \theta^\alpha D_V
+\sigma^{\mu \nu \alpha \beta} \theta_\beta F_{\mu \nu}
-2 \i {\sigma^{\mu \alpha}}_{\dot{\alpha}} 
 \D_\mu \bar\lambda^{\dot\alpha} \theta^2
\right), \\
\hat{F}^{\prime \alpha}
&=2 g^\prime  \e^{-\i \theta \sigma^\mu \bar\theta \partial_\mu}
\left( - 2 \lambda^{\prime \alpha} - 2 \theta^\alpha D_V^\prime
+\sigma^{\mu \nu \alpha \beta} \theta_\beta F_{\mu \nu}^\prime
-2 \i {\sigma^{\mu \alpha}}_{\dot{\alpha}} \D_\mu^\prime
 \bar\lambda^{\prime \dot\alpha} \theta^2
\right).
\end{align}
\end{subequations}
The remaining terms of the classical action can be derived by obvious 
substitutions or by the fact, that a product of chiral superfields
is again a chiral superfield.
For instance, the results involving a chiral superfield 
$\hat{A} \hat{L}$ can be simply obtained from the corresponding result with 
a chiral field  $\hat{L}$ if we substitute
\begin{align}
L&\to0, &
l_\alpha&\to \sqrt{2} \eps_\alpha \au L, & 
F_L&\to (\av+v_A)L-\sqrt{2} \au \eps^\alpha l_\alpha.
\end{align}

\section{Explicit form of Ward operators} \label{ap:operators}
\setcounter{equation}{0}

\begin{subequations}
\begin{align}
\W(\omega)&=\int \dx \left\{
2 \tr\left[
-\i g [\omega,V^\mu] \frac{\delta}{\delta V^\mu}
-\i g [\omega,\lambda^\alpha]  
 \frac{\delta}{\delta \lambda^\alpha}
-\i g [\omega,\bar\lambda_{\dot{\alpha}}] 
 \frac{\delta}{\delta \bar\lambda_{\dot{\alpha}}}
-\i g [\omega,c] \frac{\delta}{\delta c}
\right.\right.
\nonumber \\ &{} \qquad \qquad {}
+\i g [\omega,Y_V^\mu] \frac{\delta}{\delta Y_V^\mu}
+\i g [\omega,y_{\lambda}^\alpha]  
 \frac{\delta}{\delta y_{\lambda}^\alpha}
+\i g [\omega,\bar{y}_{\lambda \dot{\alpha}}] 
 \frac{\delta}{\delta \bar{y}_{\lambda \dot{\alpha}}}
+\i g [\omega,Y_c] \frac{\delta}{\delta Y_c}
\nonumber \\ &{} \qquad \qquad {}
\left.
-\i g [\omega,B] \frac{\delta}{\delta B}
-\i g [\omega,\bar{c}] \frac{\delta}{\delta \bar{c}}
\right]
\nonumber \\ &{} \qquad {}
- \i g \left[\omega ( \Phi_i^a +\mbv_i^a) 
  +(\i \epsilon^{bca}) \omega^b ( \Phi_i^c + \mbv_i^c )\right]^T
  \frac{\delta}{\delta \Phi_i^{aT}}
- \i g\, \omega (\Phi'_i + \mbv'_i)^T \frac{\delta}{\delta \Phi_i^{\prime T}}
\nonumber \\ &{} \qquad {}
+ \i g \left[\omega ( \bar\Phi_i^a +\bar\mbv_i^a) 
  +(\i \epsilon^{bca}) \omega^b ( \bar\Phi_i^c + \bar\mbv_i^c )\right]
  \frac{\delta}{\delta \bar\Phi_i^{a}}
+ \i g\, \omega (\bar\Phi'_i + \bar\mbv'_i) 
  \frac{\delta}{\delta \bar\Phi_i'}
\nonumber \\ &{} \qquad {}
- \i g \left[\omega \Psi_i^a 
  +(\i \epsilon^{bca}) \omega^b \Psi_i^c \right]^T
  \frac{\delta}{\delta \Psi_i^{aT}}
- \i g\, \omega \Psi_i^{\prime T} \frac{\delta}{\delta \Psi_i^{\prime T}}
\nonumber \\ &{} \qquad {}
+ \i g \left[\omega \bar\Psi_i^a 
  +(\i \epsilon^{bca}) \omega^b \bar\Psi_i^c \right]
  \frac{\delta}{\delta \bar\Psi_i^{a}}
+ \i g\, \omega \bar\Psi'_i \frac{\delta}{\delta \bar\Psi_i'}
\nonumber \\ &{} \qquad {}
-\i g (\omega L)^T \frac{\delta}{\delta L^T}
+\i g \omega \bar{L} \frac{\delta}{\delta \bar{L}}
-\i g (\omega l^{\alpha})^T \frac{\delta}{\delta l^{\alpha T}}
-\i g \omega \bar{l}_{\dot{\alpha}} 
 \frac{\delta}{\delta \bar{l}_{\dot{\alpha}}}
\nonumber \\ &{} \qquad {}
+\i g (\omega Y_L)^T \frac{\delta}{\delta Y_L^T}
-\i g \omega \bar{Y}_L \frac{\delta}{\delta \bar{Y}_L}
+\i g (\omega y_l^{\alpha})^T \frac{\delta}{\delta y_l^{\alpha T}}
+\i g \omega \bar{y}_{l \dot{\alpha}} 
 \frac{\delta}{\delta \bar{y}_{l \dot{\alpha}}}
\nonumber \\ &{} \qquad {}
-\i g \omega R \frac{\delta}{\delta R}
+\i g \omega \bar{R} \frac{\delta}{\delta \bar{R}}
-\i g \omega r^{\alpha} \frac{\delta}{\delta r^{\alpha}}
-\i g \omega \bar{r}_{\dot{\alpha}} 
 \frac{\delta}{\delta \bar{r}_{\dot{\alpha}}}
\nonumber \\ &{} \qquad {}
\left. 
+\i g \omega Y_R \frac{\delta}{\delta Y_R}
-\i g \omega \bar{Y}_R \frac{\delta}{\delta \bar{Y}_R}
+\i g \omega y_r^{\alpha} \frac{\delta}{\delta y_r^{\alpha}}
+\i g \omega \bar{y}_{r \dot{\alpha}} 
 \frac{\delta}{\delta \bar{y}_{r \dot{\alpha}}}
\right\}
\nonumber \\ &{} \qquad {}
+\mbox{analogous quark and Higgs terms},
\label{eq:Wexpl}
\\
\W'(\omega')&= \int \dx \left[
-\i g' \omega' \frac{Y}{2} (\Phi_i^a + \mbv_i^a)^T 
 \frac{\delta}{\delta \Phi_i^{aT}}
-\i g' \omega' \frac{Y}{2} (\Phi_i' + \mbv_i')^T 
 \frac{\delta}{\delta \Phi_i^{\prime T}}
\right.
\nonumber \\ &{} \qquad {}
+\i g' \omega' \frac{Y}{2} (\bar\Phi_i^a + \bar\mbv_i^a) 
 \frac{\delta}{\delta \bar\Phi_i^a}
+\i g' \omega' \frac{Y}{2} (\bar\Phi_i' + \bar\mbv_i') 
 \frac{\delta}{\delta \bar\Phi_i'}
\nonumber \\ &{} \qquad {}
-\i g' \omega' \frac{Y}{2} \Psi_i^{aT} 
 \frac{\delta}{\delta \Psi_i^{aT}}
-\i g' \omega' \frac{Y}{2} \Psi_i^{\prime T} 
 \frac{\delta}{\delta \Psi_i^{\prime T}}
+\i g' \omega' \frac{Y}{2} \bar\Psi_i^a 
 \frac{\delta}{\delta \bar\Psi_i^a}
+\i g' \omega' \frac{Y}{2} \bar\Psi_i' 
 \frac{\delta}{\delta \bar\Psi_i'}
\nonumber \\ &{} \qquad {}
-\i g^\prime \frac{Y}{2} \omega^\prime L^T \frac{\delta}{\delta L^T}
+\i g^\prime \frac{Y}{2} \omega^\prime \bar{L} 
 \frac{\delta}{\delta \bar{L}}
-\i g^\prime \frac{Y}{2} \omega^\prime l^{\alpha T} 
 \frac{\delta}{\delta l^{\alpha T}}
-\i g^\prime \frac{Y}{2} \omega^\prime \bar{l}_{\dot{\alpha}} 
 \frac{\delta}{\delta \bar{l}_{\dot{\alpha}}}
\nonumber \\ &{} \qquad {}
+\i g^\prime \frac{Y}{2} \omega^\prime Y_L^T \frac{\delta}{\delta Y_L^T}
-\i g^\prime \frac{Y}{2} \omega^\prime \bar{Y}_L 
 \frac{\delta}{\delta \bar{Y}_L}
+\i g^\prime \frac{Y}{2} \omega^\prime y_l^{\alpha T} 
 \frac{\delta}{\delta y_l^{\alpha T}}
+\i g^\prime \frac{Y}{2} \omega^\prime \bar{y}_{l \dot{\alpha}} 
 \frac{\delta}{\delta \bar{y}_{l \dot{\alpha}}}
\nonumber \\ &{} \qquad {}
-\i g^\prime \frac{Y}{2} \omega^\prime R \frac{\delta}{\delta R}
+\i g^\prime \frac{Y}{2} \omega^\prime \bar{R} \frac{\delta}{\delta \bar{R}}
-\i g^\prime \frac{Y}{2} \omega^\prime r^{\alpha} 
 \frac{\delta}{\delta r^{\alpha}}
-\i g^\prime \frac{Y}{2} \omega^\prime \bar{r}_{\dot{\alpha}} 
 \frac{\delta}{\delta \bar{r}_{\dot{\alpha}}}
\nonumber \\ &{} \qquad {}
\left. 
+\i g^\prime \frac{Y}{2} \omega^\prime Y_R \frac{\delta}{\delta Y_R}
-\i g^\prime \frac{Y}{2} \omega^\prime \bar{Y}_R 
 \frac{\delta}{\delta \bar{Y}_R}
+\i g^\prime \frac{Y}{2} \omega^\prime y_r^{\alpha} 
 \frac{\delta}{\delta y_r^{\alpha}}
+\i g^\prime \frac{Y}{2} \omega^\prime \bar{y}_{r \dot{\alpha}} 
 \frac{\delta}{\delta \bar{y}_{r \dot{\alpha}}}
\right]
\nonumber \\ &{} \qquad {}
+\mbox{analogous quark and Higgs terms},
\label{eq:Wpexpl}
\\
\W_L&=\int \dx \left[
 L^{T} \frac{\delta}{\delta L^{T}}
-\bar{L} \frac{\delta}{\delta \bar{L}}
+l^{\alpha T} \frac{\delta}{\delta l^{\alpha T}}
+\bar{l}_{\dot{\alpha}} \frac{\delta}{\delta \bar{l}_{\dot{\alpha}}}
\right.
\nonumber \\ &{} \qquad {}
\left. 
-Y_L^{T} \frac{\delta}{\delta Y_L^{T}}
+\bar{Y}_L \frac{\delta}{\delta \bar{Y}_L}
-y_l^{\alpha T} \frac{\delta}{\delta y_l^{\alpha T}}
-\bar{y}_{l \dot{\alpha}} 
 \frac{\delta}{\delta \bar{y}_{l \dot{\alpha}}}
\right]
\nonumber \\ &{} \qquad {}
+\mbox{analogous terms for right-handed leptons},
\label{eq:WLexpl}
\\
\W_Q&=\int \dx \left[
 Q^{T} \frac{\delta}{\delta Q^{T}}
-\bar{Q} \frac{\delta}{\delta \bar{Q}}
+q^{\alpha T} \frac{\delta}{\delta q^{\alpha T}}
+\bar{q}_{\dot{\alpha}} \frac{\delta}{\delta \bar{q}_{\dot{\alpha}}}
\right.
\nonumber \\ &{} \qquad {}
\left. 
-Y_Q^{T} \frac{\delta}{\delta Y_Q^{T}}
+\bar{Y}_Q \frac{\delta}{\delta \bar{Y}_Q}
-y_q^{\alpha T} \frac{\delta}{\delta y_q^{\alpha T}}
-\bar{y}_{q \dot{\alpha}} 
 \frac{\delta}{\delta \bar{y}_{q \dot{\alpha}}}
\right]
\nonumber \\ &{} \qquad {}
+\mbox{analogous terms for right-handed quarks},
\label{eq:WQexpl}
\\
\W^R &=\int \dx \left\{
2 \tr\left[
-\i \lambda^\alpha \frac{\delta}{\delta \lambda^\alpha}
-\i g \bar\lambda_{\dot{\alpha}} 
 \frac{\delta}{\delta \bar\lambda_{\dot{\alpha}}}
+\i y_{\lambda}^\alpha \frac{\delta}{\delta y_{\lambda}^\alpha}
+\i \bar{y}_{\lambda \dot{\alpha}} 
 \frac{\delta}{\delta \bar{y}_{\lambda \dot{\alpha}}}
\right]\right.
\nonumber \\ &{} \qquad {}
-\i \lambda'{}^\alpha \frac{\delta}{\delta \lambda'{}^\alpha}
-\i g \bar\lambda'_{\dot{\alpha}} 
 \frac{\delta}{\delta \bar\lambda'_{\dot{\alpha}}}
+\i y_{\lambda'}^\alpha \frac{\delta}{\delta y_{\lambda'}^\alpha}
+\i \bar{y}_{\lambda' \dot{\alpha}} 
 \frac{\delta}{\delta \bar{y}_{\lambda' \dot{\alpha}}}
\nonumber \\ &{} \qquad {}
-2 \i \av \frac{\delta}{\delta \av}
+2 \i \bar\av \frac{\delta}{\delta \bar\av}
\nonumber \\ &{} \qquad {}
+ \i (\Phi_i^a +\mbv_i^a) \frac{\delta}{\delta \Phi_i^{aT}}
+ \i (\Phi'_i + \mbv'_i)^T \frac{\delta}{\delta \Phi_i^{\prime T}}
- \i (\bar\Phi_i^a +\bar\mbv_i^a) \frac{\delta}{\delta \bar\Phi_i^{a}}
- \i (\bar\Phi'_i + \bar\mbv'_i) \frac{\delta}{\delta \bar\Phi_i'}
\nonumber \\ &{} \qquad {}
+ \i \Psi_i^a \frac{\delta}{\delta \Psi_i^{aT}}
+ \i \Psi_i^{\prime T} \frac{\delta}{\delta \Psi_i^{\prime T}}
- \i \bar\Psi_i^a \frac{\delta}{\delta \bar\Psi_i^{a}}
- \i \bar\Psi'_i \frac{\delta}{\delta \bar\Psi_i'}
\nonumber \\ &{} \qquad {}
+\i H_1^T \frac{\delta}{\delta H_1^T}
-\i \bar{H}_1 \frac{\delta}{\delta \bar{H}_1}
+\i H_2^T \frac{\delta}{\delta H_2^T}
-\i \bar{H}_2 \frac{\delta}{\delta \bar{H}_2}
\nonumber \\ &{} \qquad {}
-\i Y_1^T \frac{\delta}{\delta Y_1^T}
+\i \bar{Y}_1 \frac{\delta}{\delta \bar{Y}_1}
-\i Y_2^T \frac{\delta}{\delta Y_2^T}
+\i \bar{Y}_2 \frac{\delta}{\delta \bar{Y}_2}
\nonumber \\ &{} \qquad {}
+\frac{\i}{2} L^T \frac{\delta}{\delta L^T}
-\frac{\i}{2} \bar{L} \frac{\delta}{\delta \bar{L}}
-\frac{\i}{2} l^{\alpha T} \frac{\delta}{\delta l^{\alpha T}}
-\frac{\i}{2} \bar{l}_{\dot{\alpha}} 
 \frac{\delta}{\delta \bar{l}_{\dot{\alpha}}}
\nonumber \\ &{} \qquad {}
-\frac{\i}{2} Y_L^T \frac{\delta}{\delta Y_L^T}
+\frac{\i}{2} \bar{Y}_L \frac{\delta}{\delta \bar{Y}_L}
+\frac{\i}{2} y_l^{\alpha T} \frac{\delta}{\delta y_l^{\alpha T}}
+\frac{\i}{2} \bar{y}_{l \dot{\alpha}} 
 \frac{\delta}{\delta \bar{y}_{l \dot{\alpha}}}
\nonumber \\ &{} \qquad {}
+\frac{\i}{2} R \frac{\delta}{\delta R}
-\frac{\i}{2} \bar{R} \frac{\delta}{\delta \bar{R}}
-\frac{\i}{2} r^\alpha \frac{\delta}{\delta r^\alpha}
-\frac{\i}{2} \bar{r}_{\dot{\alpha}} 
 \frac{\delta}{\delta \bar{r}_{\dot{\alpha}}}
\nonumber \\ &{} \qquad {}
\left.
-\frac{\i}{2} Y_R \frac{\delta}{\delta Y_R}
+\frac{\i}{2} \bar{Y}_R \frac{\delta}{\delta \bar{Y}_R}
+\frac{\i}{2} y_r^\alpha \frac{\delta}{\delta y_r^\alpha}
+\frac{\i}{2} \bar{y}_{r \dot{\alpha}} 
 \frac{\delta}{\delta \bar{y}_{r \dot{\alpha}}}
\right\}
\nonumber \\ &{} \qquad {}
-\i \eps^\alpha \frac{\partial}{\partial \eps^\alpha}
-\i \bar{\eps}_{\dot{\alpha}} 
 \frac{\partial}{\partial \bar{\eps}_{\dot{\alpha}}}
\nonumber \\ &{} \qquad {}
+\mbox{analogous quark terms}.
\label{eq:WRexpl}
\end{align}
\end{subequations}

\section{Cohomologically trivial invariant counterterms}
\label{app:invCT}

In this section we sketch the determination of the most general form
of the cohomologically trivial contribution $\B\hat\Gamma_{\rm
ct,inv,1b}$ to the invariant counterterms. This contribution satisfies
the first of the conditions in (\ref{eq:GinvConditions})
automatically since $\B$ is nilpotent. The remaining conditions in
(\ref{eq:GinvConditions}) restrict the form of $\B\hat\Gamma_{\rm
inv}{}_2$. In addition, the counterterms have to depend on the field $A_1$
only through the combination $a_\alpha=\sqrt2\epsilon_\alpha A_1$.

\begin{itemize}
\item Gauge-fixing sector: this sector contains the field monomials
depending on $\bar{c}$, $\bar{c}^\prime$, $B$, and $B^\prime$. These
terms are fixed to all orders by the gauge-fixing condition.
\item Field monomials involving the translational ghost $\xi^\mu$:
these monomials are fixed to all orders 
by the translation ghost equation.
\item Field renormalization: counterterms corresponding to field
renormalization arise as
\begin{equation}
\B \int \dx Y_{\phi_i} \phi_i =\pm \int \dx 
\left(\phi_i \frac{\delta}{\delta \phi_i}
-Y_{\phi_i} \frac{\delta}{\delta Y_{\phi_i}}\right) \Gcl,
\end{equation}
where $\phi_i, Y_{\phi_i}$ run through all pairs of fields and
corresponding $Y$ fields. 
The $\pm$ signs hold for bosonic and fermionic $\phi_i$, respectively. 
In order not to invalidate the gauge-fixing conditions, the field 
renormalizations of the gauge bosons and Higgs fields have to be 
supplemented by suitable renormalizations of the gauge-fixing terms.
\item Renormalization of $x_1,x_2,x_1',x_2'$: 
these renormalizations are generated by the following $\B$ variations:
\begin{equation}
\B\int\dx Y_i^T \left(\hat\Phi_i+\hat\mbv_i\right) =\int \dx 
\left[\left(\hat\Phi_i+\mbv_i\right)^T \frac{\delta}{\delta {H}_i}
-Y_i^T \hat\Psi_i\right] \Gcl
\end{equation}
and similar for $(\hat\Phi_i+\hat\mbv_i)\to(\Phi'_i+\mbv'_i)$.
\item Field monomials including spurion fields:
only combinations of the following invariant counterterms obey the 
requirement that $\au$ appears only in the combination 
$\au \epsilon_\alpha$. 
\begin{itemize}
\item
Counterterms related to the soft-breaking parameters
$c_{HH}$, $c_{V'}$, $c_{V}$, $c_{HLR}$, $c_{HQD}$, $c_{HQU}$:
\begin{align}
&\B\int\dx [\au [H^{\rm eff}_1{}^T (\i \sigma_2) H^{\rm eff}_2],&
&\B\int\dx \, [\au \lambda'{}^\alpha \lambda'_\alpha],\\
&\B\int\dx \, [\au \tr(\lambda^\alpha \lambda_\alpha)],&
&\B\int\dx \, [\au H^{\rm eff}_1{}^T (i \sigma_2) L R],\\
&\B\int\dx \, [\au H^{\rm eff}_1{}^T (i \sigma_2) Q D],&
&\B\int\dx \, [\au H^{\rm eff}_2{}^T (i \sigma_2) Q U].
\end{align}
\item
Counterterms causing field and BRS redefinitions:
\begin{align}
\label{eq:Higgsct}
&\B\int\dx \left[ \au y_l^{T \alpha} \epsilon_\alpha L \right] ,&
&\B\int\dx \left[ \au y_r^\alpha \epsilon_\alpha R \right] , \\
&\B\int\dx \left[ \au y_q^{T \alpha} \epsilon_\alpha Q \right] , &
&\B\int\dx \left[ \au y_u^\alpha \epsilon_\alpha U \right], \\
&\B\int\dx \left[ \au y_d^\alpha \epsilon_\alpha D \right],&
&\B\int\dx \left[ \au y_i^{T \alpha} \epsilon_\alpha H^{\rm eff}_i \right],\\
\label{eq:newct}
&\B\int\dx \left[ \au y_i^{T \alpha} \epsilon_\alpha 
(\hat{\Phi}_i+\hat\mbv_i) \right] , &
&\B\int\dx \left[ \au y_i^{T \alpha} \epsilon_\alpha 
(\Phi'_i+\mbv'_i) \right].
\end{align}
These counterterms correspond to the renormalization of
$c_{H_i1}$, $c_{L1}$, $c_{R1}$, $c_{Q1}$, $c_{U1}$, $c_{D1}$ apart
from the counterterms (\ref{eq:newct}).
The counterterms (\ref{eq:newct}) are defined analogous to the 
counterterm (\ref{eq:Higgsct}) with $H^{\rm eff}_i$ replaced by the $\Phi$
fields, but do not correspond to parameters in the classical action. 
All of these counterterms (\ref{eq:newct}) are of the type of the
$u_3$ counterterms discussed in \cite{Hollik:2000pa}.
\item Additional unphysical counterterm:
\begin{eqnarray}
\label{eq:irrelevantCT}
\B\int\dx \left[ \au \epsilon^\alpha \sigma^\mu_{\alpha \dot{\alpha}} 
\tr(V_\mu \bar{y}_\lambda^{\dot{\alpha}}) \right].
\end{eqnarray}
This counterterm corresponds to the $v_3$ counterterms discussed in
\cite{Hollik:2000pa}. An analogous counterterm involving the abelian 
gauge field $V'_\mu$ is not invariant since it violates the nilpotency 
condition $\wl_\mu\Ginv=0$.
\item Counterterms corresponding to the soft-breaking parameters
$c_{H_i2}$, $c_{L2}$, $c_{R2}$, $c_{Q2}$, $c_{U2}$, $c_{D2}$:
\begin{eqnarray}
&&\B\int\dx  \left\{\aub \bar{H}^{\rm eff}_i 
\left[(\av+v_A) H^{\rm eff}_i - \au \sqrt{2} \epsilon^\alpha h_{i \alpha} \right] 
\right\} , \\
&&\B\int\dx  \left\{\aub \bar{L} 
\left[(\av+v_A) L - \au \sqrt{2} \epsilon^\alpha l_\alpha \right] 
\right\} , \\
&&\B\int\dx  \left\{\aub \bar{R} 
\left[(\av+v_A) R - \au \sqrt{2} \epsilon^\alpha r_\alpha \right] 
\right\} , \\
&&\B\int\dx  \left\{\aub \bar{Q} 
\left[(\av+v_A) Q - \au \sqrt{2} \epsilon^\alpha q_\alpha \right] 
\right\} , \\
&&\B\int\dx  \left\{\aub \bar{U} 
\left[(\av+v_A) U - \au \sqrt{2} \epsilon^\alpha u_\alpha \right] 
\right\} , \\
&&\B\int\dx  \left\{\aub \bar{D} 
\left[(\av+v_A) D - \au \sqrt{2} \epsilon^\alpha d_\alpha \right] 
\right\}.
\end{eqnarray}
\end{itemize}
An explicit term-by-term analysis shows that no further
invariant counterterm exists that involves at least one dynamical
field. The remaining counterterms consist purely of external fields
and are therefore unphysical.
\end{itemize}

\section{Characterization of the Goldstone Modes}
\label{app:GoldstQuant}

In the physical part of the classical action, only the combination $H_i +
x_i \hat \Phi_i + x_i' \Phi'_i$ occurs. In this section we investigate the
relation between 2-point functions of $H_i$, $\hat\Phi_i$ and $\Phi'_i$ in
higher orders and 
demonstrate how the definition of Goldstone fields has to be adjusted.

Classically, the Goldstone modes correspond to the flat directions of
the potential and can be algebraically characterized as 
\begin{equation}
\sum_{i=1}^2 \left(- \i g \fdq{}{H_i} T^a \vecv_i + {\rm c.c.}\right)
\sim \fdq{}{G^a}.
\label{Goldstdef1}
\end{equation}
This characterization implies $\Gamma_{G^a X}|_{p=0}=0$ for all fields
$\phi\ne B$ at the classical level.

On the quantum level, a characterization of the Goldstone modes can be
derived from the Ward and ST identities.

The rigid Ward identity, differentiated w.r.t.\ a field $\phi$ ($\phi\ne B$), 
yields at $p=0$
\begin{equation}
\sum_{a=1,2,3,'} \left(\delta_b \mbv_i^{aj} \Gamma_{\phi \Phi_i^{aj}}
 + \delta_b \bar{\mbv}_i^{aj} \Gamma_{\phi \bar{\Phi}_i^{aj}}\right)=0,
\label{rigidWIbasic}
 \end{equation}
where $\delta_\omega \mbv_i^{aj}=(\delta_\omega \Phi_i^{aj})|_{\Phi=0}$,
$\delta_\omega \mbv'_i{}^j= (\delta_\omega \Phi'_i{}^j)|_{\Phi'=0}$.
Using the $\Phi$ fields defined in eq.\ (\ref{BackgrGoldstDef}) this results 
in
\begin{equation}
\Gamma_{\phi \G^b}|_{p=0}=0,
\end{equation}
i.e.\ the modes $\G^b$, whose definition (\ref{BackgrGoldstDef})
contains no higher-order contributions, correspond to flat directions
of the scalar potential.

(\ref{rigidWIbasic}) does not characterize the dynamical
Goldstone modes. To obtain such a characterization, it has to be 
supplemented with a relation of 2-point functions involving $\Phi$ fields 
and the corresponding ones involving the Higgs fields $H_{1,2}$. Such a 
relation can be derived from the ST identity.

Differentiating the ST identity w.r.t.\ $\Psi^a$ ($a=1,2,3,'$) and an
arbitrary physical  field $\phi$ ($\phi\ne B$) we obtain
\begin{equation}
\Gamma_{\Psi_i^{a j}  Y_r^s}\Gamma_{\phi H_r^s} 
+ \Gamma_{\Psi_i^{a j}  \bar{Y}_r^s}\Gamma_{\phi \bar{H}_r^s} 
+ \Gamma_{\phi \Phi_i^{a j}}=0. \label{STIqXprime}
\end{equation}
A similar identity can be derived for $\Gamma_{\phi \bar{\Phi}_i^{a j}}$.

Using these relations of the 2-point functions in (\ref{rigidWIbasic}) 
yields at $p=0$
\begin{eqnarray}
0 & = & 
\sum_{a=1,2,3,'} \left[\delta_b \mbv_i^{aj} 
\left(\Gamma_{\Psi_i^{a j}  Y_r^s}\Gamma_{\phi H_r^s} 
+ \Gamma_{\Psi_i^{a j}  \bar{Y}_r^s}\Gamma_{\phi \bar{H}_r^s} \right) \right.
\nonumber\\&&{}
\left. + \delta_b  \bar{\mbv}_i^{aj} 
\left(\Gamma_{\bar{\Psi}_i^{a j}  Y_r^s}\Gamma_{\phi H_r^s} 
+ \Gamma_{\bar{\Psi}_i^{a j}
  \bar{Y}_r^s}\Gamma_{\phi \bar{H}_r^s} \right)\right].
\label{rigidWIresult}
\end{eqnarray}
This identity characterizes the Goldstone modes as
\begin{eqnarray}
\fdq{}{G^b}& \sim &
\sum_{a=1,2,3,'} \left[
\left.\left( \delta_b \mbv_i^{aj} \Gamma_{\Psi_i^{a j}  Y_r^s}
+ \delta_b  \bar{\mbv}_i^{aj} \Gamma_{\bar{\Psi}_i^{a j}  Y_r^s}
\right)\right|_{p=0} \fdq{}{H_r^s} \right.
\nonumber\\&&{}
\left.
+ \left.\left( \delta_b \mbv_i^{aj} \Gamma_{\Psi_i^{a j}  \bar{Y}_r^s}
+ \delta_b  \bar{\mbv}_i^{aj} \Gamma_{\bar{\Psi}_i^{a j}  \bar{Y}_r^s}
\right)\right|_{p=0} \fdq{}{\bar{H}_r^s} \right].
\label{rigidWIresult2}
\end{eqnarray}
With this characterization, we obtain
\begin{equation}
\Gamma_{G^b \phi}|_{p=0} = 0,
\label{eq:GoldstCharacterization}
\end{equation}
so that the $G^b$ indeed correspond to the flat directions of the
potential. For $\phi=B$ (\ref{STIqXprime}) is replaced by
\begin{equation}
\Gamma_{\Psi_i^{aj}  Y_r^s}\Gamma_{B^b H_r^s} 
+ \Gamma_{\Psi_i^{aj}  \bar{Y}_r^s}\Gamma_{B^b \bar{H}_r^s} 
+\Gamma_{B^b  \Phi_i^{aj}} +
 \Gamma_{\Psi_i^{aj} \bar c^b}=0, 
\label{STIqB}
\end{equation}
which results in
\begin{equation}
\Gamma_{B^b G^a}|_{p=0} \ne 0.
\end{equation}

In section \ref{sec:EvaluationNormCond} we have seen that there are
exactly three linearly independent modes $G^b=G^0, G^\pm$ which
satisfy (\ref{eq:GoldstCharacterization}). (\ref{rigidWIresult2}) 
supplements this result with an explicit formula for the Goldstone modes 
and correspondingly with the first column of the matrices $\Z_{A^0}$, 
$\Z_{H^\pm}$. However, in contrast to the characterization of the 
$\Phi$ fields, (\ref{rigidWIresult2}) contains contributions of higher 
orders.

Finally, we evaluate the characterization (\ref{rigidWIresult2}) 
at the classical level. In lowest-order, we have
\begin{align}
\Gamma_{\Psi_i^{aj}  Y_r^s}&= - 2 x_i \delta_{ir} T^a{}^s{}_{j},&
\Gamma_{\Psi_i^{\prime j} Y_r^s}&= -  x'_i  \delta_{ir}\delta^{js}
\end{align}
and thus obtain
\begin{equation}
\fdq{}{G^b} \sim
\delta_b\left[2 x_i (T^a \mbv^a_i)^s + x'_i \mbv'_i{}^s\right]
\fdq{}{H_i^s} + {\rm c.c.}
\end{equation}
Owing to the decomposition (\ref{eq:decomposition}) for $\vec{v}_i$,
this is identical to equation (\ref{Goldstdef1}).

\end{appendix}

\end{document}